\documentclass{aa}
\usepackage{times}
\usepackage{graphics}
\usepackage{xspace}
\usepackage{epsfig}
\usepackage{rotating}
\usepackage{dcolumn}
\usepackage{longtable}

\makeatletter
\@ifundefined{chapter}{\def\thebibliography#1{
\section*{References}                          
  \list
  {\relax}{\setlength{\labelsep}{0em}
        \setlength{\itemindent}{-\bibhang}
        \setlength{\itemsep}{\parskip}          
        \setlength{\parsep}{0pt}
        \setlength{\leftmargin}{\bibhang}}
    \def\newblock{\hskip .11em plus .33em minus .07em}
    \sloppy\clubpenalty4000\widowpenalty4000
    \sfcode`\.=1000\relax}}%
{\def\thebibliography#1{
  \list
  {\relax}{\setlength{\labelsep}{0em}
        \setlength{\itemindent}{-\bibhang}
        \setlength{\itemsep}{\parskip}          
        \setlength{\parsep}{0pt}
        \setlength{\leftmargin}{\bibhang}}
    \def\newblock{\hskip .11em plus .33em minus .07em}
    \sloppy\clubpenalty4000\widowpenalty4000
    \sfcode`\.=1000\relax}}

\newlength{\bibhang}
\let\@internalcite\cite
\def\cite{\@ifstar{\citey}{\citefull}}
\def\citefull{\def\astroncite##1##2{##1\ ##2}\@internalcite}
\def\citey{\def\astroncite##1##2{##1\ (##2)}\@internalcite}
\def\citeyear{\def\astroncite##1##2{##2}\@internalcite}
\def\citename{\def\astroncite##1##2{##1}\@internalcite}
\def\@citex[#1]#2{\if@filesw\immediate\write\@auxout{\string\citation{#2}}\fi
  \def\@citea{}\@cite{\@for\@citeb:=#2\do
    {\@citea\def\@citea{; }\@ifundefined       
       {b@\@citeb}{{\bf ??}\@warning              
       {Citation `\@citeb' on page \thepage \space undefined}}%
{\csname b@\@citeb\endcsname}}}{#1}}

\def\@cite#1#2{#1\if@tempswa #2\fi}       
\def\@biblabel#1{}

\def\astroncite#1#2{#1\ #2}
\makeatother

\newcommand{\newrule}{\rule[-0.05cm]{0.cm}{0.4cm}}

\setlongtables

\begin{document}

\thesaurus{06(%
08.06.1; 
13.25.5; 
08.12.1)} 

\title{X-ray flares on zero-age- and pre-main sequence stars in Taurus-Auriga-Perseus}

\author{B. Stelzer\inst {1}, R. Neuh\"auser\inst {1} \and V. Hambaryan\inst {2}} 

\institute{Max-Planck-Institut f\"ur extraterrestrische Physik,
  Giessenbachstr.~1,
  D-85740 Garching,
  Germany \and
  Astrophysikalisches Institut Potsdam,
  An der Sternwarte 16,
  D-14482 Potsdam,
  Germany} 

\offprints{B. Stelzer}
\mail{B. Stelzer, stelzer@xray.mpe.mpg.de}
\titlerunning{X-ray flares in Taurus-Auriga-Perseus}

\date{Received $<$6 April 1999$>$ / Accepted $<$3 February 2000$>$ } 
\maketitle
 
\begin{abstract}

We present the results of a systematic search for X-ray flares 
on young stars observed during 
{\em ROSAT} PSPC observations of the Taurus-Auriga-Perseus sky region. 
All pointed PSPC observations currently available from the 
{\em ROSAT} Public Data Archive 
with known pre-main sequence T Tauri Stars or young Pleiads
or Hyads in the field of view are analyzed. A study of the activity
of late-type stars 
of different ages provides information on the evolution of their
coronal activity, which may be linked to their angular momentum.

We develop a criterion for the detection of flares based on the 
shape of the X-ray lightcurve. Applying our detection method to all 
104 PSPC pointings from the archive we find 52 flares. 
Among them 15 are detected on T Tauri Stars,
20 on Pleiads, and 17 on Hyads. 
Only the 38 events which can definitely be attributed to late-type
stars (i.e. stars of spectral type G and later) are considered in the
statistical analysis of the properties of flaring stars. We 
investigate the influence of stellar parameters such as age, rotation and
multiplicity on individual flare parameters and flare frequency. 

From the total exposure time 
falling to the share of each sample and the duration of the individual
flares we compute a flare rate. 
We take into account that the detection sensitivity for
large X-ray flares depends on the S/N and hence on the stellar distance.
The values we derive for the flare rates are 
$0.86 \pm 0.16$\% for T Tauri Stars, 
$0.67 \pm 0.13$\% for Pleiads and $0.32 \pm 0.17$\% for Hyads.
The flare rate of classical T Tauri Stars 
may be somewhat higher than that of weak-line T Tauri Stars 
($F_{\rm c} = 1.09 \pm 0.39$\% versus
$F_{\rm w} = 0.65 \pm 0.16$\%).

Hardness ratios are used to track the heating
that takes place during stellar flares. Hardness ratios are evaluated for 
three distinct phases of the flare: 
the rise, the decay, and the quiescent (pre- and post-flare) stage.
In most cases the hardness increases
during the flares as compared to the quiescent state. 
During both quiescence and flare phase TTSs 
display the largest hardness ratios, and
the Hyades stars show the softest spectrum.

\end{abstract}

\section{Introduction}\label{sect:intro}

The Taurus-Auriga-Perseus region offers the opportunity to study 
the X-ray emission of young stars at several evolutionary stages. The 
youngest stars observed 
by {\em ROSAT} in this portion of the sky are the T Tauri Stars (TTSs) of
the Taurus-Auriga and Perseus star forming regions, 
late-type pre-main sequence (PMS) stars of $M \leq 3 {\rm M}_\odot$ 
with an estimated age of $10^5-10^7\,{\rm yrs}$. 
Two young star clusters, the Pleiades and Hyades, are also located 
in this region of the sky at 
age of $10^8\,{\rm yrs}$ and $6~10^8\,{\rm yrs}$, respectively.
They consist mostly of zero-age main-sequence (ZAMS) stars, 
except for some higher mass post-main sequence stars and
brown dwarfs, which are not studied here. 

From the early observations by the {\em Einstein} satellite 
it was concluded that the X-ray emission of young 
stars arises in an optically thin, hot plasma at temperatures above
$10^6\,{\rm K}$
(\cite{Feigelson81.1}). The emission region has been associated with the
stellar corona where the X-rays are produced --- more or less 
analogous to the solar X-ray emission --- through a stellar 
$\alpha$-$\Omega$-dynamo. 
The dynamo is driven by the combination of rotation and 
convective motions. 
Correlations between
the X-ray emission of late-type stars and the stellar rotation 
support the notion that 
dynamo-generated magnetic fields are responsible for heating the coronae
(\cite{Pallavicini81.1}). 
But successful direct measurements of the magnetic fields of TTSs 
have been performed only recently (see e.g. \cite{Guenther99.2}).
The details of the heating mechanism are still not well understood.

The correlation between stellar rotation and X-ray emission of
late-type stars suggests that the rotational evolution of young stars 
determines the development of stellar activity.
The rotational evolution of low-mass PMS stars partly depends on the
circumstellar environment. While classical TTSs (hereafter cTTSs) are
surrounded by a circumstellar disk, inferred from IR dust emission 
(\cite{Bertout88.1}, \cite{Strom89.1}, and \cite{Beckwith90.1}) and 
more recently from direct imaging (e.g. \cite{McCaughrean96.1}), 
weak-line TTSs (wTTSs) lack such a disk, or at least the disk is not
optically thick. 
Owing to contraction wTTSs spin up as they approach the main
sequence. For cTTSs, on the other hand, coupling between the 
disk and the star may prevent spin-up (\cite{Bouvier93.1}).
The period observed on the ZAMS depends on the time the star has spent
in the cTTS phase.
After the main-sequence is reached, the rotation rate decreases again 
(see \cite{Bouvier97.1}).
As a consequence of their slower rotation, stars on the
ZAMS and main sequence (MS) should on average show less X-ray activity 
than PMS stars. 

Earlier investigations of X-ray observations of young late-type stars
were mostly concerned with the quiescent emission  
(see \cite{Neuhaeuser95.1}, \cite{Stauffer94.1}, \cite{Gagne95.1}, 
\cite{Hodgkin95.1}, \cite{Micela96.1}, 1999, \cite{Pye94.1}, and
\cite{Stern94.1}).
In contrast to these studies we focus on the occurrence of X-ray flares.
Furthermore we discuss a larger sample than most of the previous 
studies by using {\em all} currently available observations from the
{\em ROSAT} Public Data Archive that contain any TTS, Pleiad or Hyad
in the field of view.

X-ray flares may be used as a diagnostics of stellar activity.
They are thought to originate in magnetic loops.
In contrast to findings from quasi-static loop
modeling, the only direct determination of the size of a flaring
region (\cite{Schmitt99.1}) shows that the emitting region is
very compact.
In the loops which confine the coronal plasma 
magnetic reconnection suddenly frees large
amounts of energy which is dissipated into heat and thus leads to a
temporary enhancement of the X-ray emission. 
The decay of the lightcurve is accompanied
by a corresponding (exponential) decay of the temperature and emission
measure, which are obtained from  
one- or two-temperature 
spectral models for an optically thin, thermal plasma 
(\cite{Raymond77.1}, \cite{Mewe85.1}, 1986).

The most powerful X-ray flares have been observed on the
youngest objects, notably a flare on the infrared 
Class I protostar YLW~15 in 
$\rho$ Oph which has been presented by \citey{Grosso97.1}.
X-ray flares on TTSs observed so far (see \cite{Montmerle83.1},
\cite{Preibisch93.1}, \cite{Strom94.1}, 
\cite{Preibisch95.1}, \cite{Gagne95.1}, \cite{Skinner97.1}, \cite{Tsuboi98.1})
exceed the maximum emission observed
from solar flares by a factor of $10^3$ and more. Some extreme
events have shown X-ray luminosities of 
$\sim L_{\rm x} = 10^{33}\,{\rm erg/s}$.
Although some of the strongest X-ray flares ever observed were detected
on TTSs 
to date no systematic search for TTS flares was undertaken. 

This paper is devoted to a study of the relation between X-ray flare 
activity and other stellar parameters, such as age, rotation rate, and
multiplicity.
For this purpose 
we perform a statistical investigation of {\em ROSAT} observations. 
We develop a method for the flare
detection based on our conception of the typical shape of a flare 
lightcurve,
where the term `typical shape' refers to the characteristics of
the X-ray lightcurve described above, i.e. a significant
rise and subsequent decay of the lightcurve to the previous emission
level. 
The database and source detection is
described in Sect.~\ref{sect:data}. 
In Sect.~\ref{sect:lcs} we describe how the lightcurves are
generated. Our flare detection algorithm is explained 
in Sect.~\ref{sect:detect}, where
we also present all flare parameters derived from the X-ray lightcurves. 
Then we describe the influence of observational restrictions on the data 
analysis and how the related biases can be overcome 
(Sect.~\ref{sect:bias}).
In Sect.~\ref{sect:statcomp} we compare the flare characteristics of 
different samples of flaring
stars selected by their age, rotation rate, and multiplicity. We
present luminosity functions for TTSs, Pleiads, 
and Hyads during flare and quiescence.
Luminosity functions of the non-active state of these stars
have been presented before (see e.g. \cite{Pye94.1}, \cite{Hodgkin95.1}, 
\cite{Neuhaeuser97.3})
and some of the flares discussed here have been discussed 
by \citey{Gagne95.1}, \citey{Strom94.1}, and \citey{Preibisch93.1}. 
However,
 this is the first statistical evaluation of flare luminosities. 
Flare rates comparing stellar subgroups with different properties
(such as age, $v\,\sin{i}$, and stellar multiplicity)
are compiled in Sect.~\ref{sect:rate}. 
Because of lack of sufficient statistics for a 
detailed spectral analysis, hardness ratios are used to describe the
spectral properties of the flares.
In Sect.~\ref{sect:hr} we present the observed relations between 
hardness ratios measured during different activity phases and between
hardness and X-ray luminosity.
Finally, we discuss and summarize 
our results in Sect.~\ref{sect:discussion} and Sect.~\ref{sect:conclusions}.

\section{Database and data reduction}\label{sect:data}

In this section we introduce the stellar sample 
and explain the analysis of the raw data. 
Details about our membership lists for TTSs,
Pleiads, and Hyads are given below
(Sect.~\ref{subsect:member}). We have retrieved all pointed
{\em ROSAT} PSPC observations from the archive 
that contain at least one of the stars from these lists in their field.
The observations are listed in Table~\ref{tab:pids}. 
After performing source detection on all of these pointings, we have 
cross correlated the membership lists with the detected X-ray sources and
identified individual TTSs, Pleiads, and Hyads in the X-ray image.
The process of source detection and identification is described in 
Sect.~\ref{subsect:soudet}.

\subsection{The stellar sample}\label{subsect:member}

The analysis presented here is 
confined to the Taurus-Auriga-Perseus region.
This portion of the sky includes the Taurus-Auriga complex, the MBM\,12
cloud, and 
the Perseus molecular clouds with the reflection nebula NGC\,1333 
and the young cluster IC\,348. Two open clusters containing mostly
ZAMS stars are found nearby the above mentioned star forming
regions, the Pleiades and the Hyades.
The choice of this specific sky region thus
enables us to compare the X-ray emission of young stars at different ages.

\begin{figure}
\begin{center}
\resizebox{8cm}{!}{\includegraphics{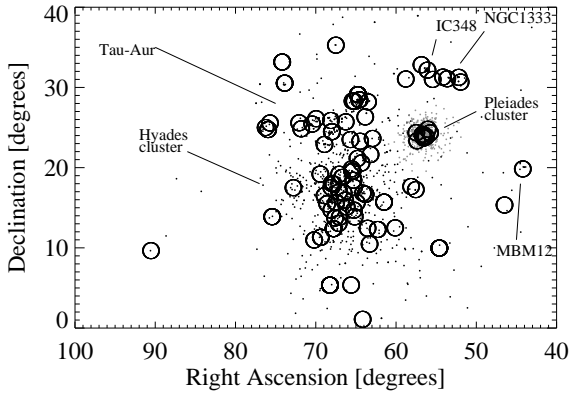}}
\caption{Sky map of the Tau-Aur-Per region showing the pointing positions of
the {\em ROSAT} PSPC observations. On display are all PSPC pointings
selected from the {\em ROSAT} Public Data Archive which contain at least
one TTS, Pleiad, or Hyad from the membership lists given in the text. In
addition all stars from the membership lists are plotted.}
\label{fig:skymap}
\end{center}
\end{figure}

Our sample of low-mass PMS stars 
in and around Taurus consists of all TTSs
which are either on or very close to the Taurus star forming clouds
or off the clouds at locations where they can still be linked with the 
Taurus clouds (see e.g. \cite{Neuhaeuser97.2}).
We restrict our Taurus sample to objects between 
$\alpha _{2000} = 2h$ and $5h$
and $\delta _{2000} = - 10^{\circ}$ and $40 ^{\circ}$.
The TTSs in Taurus comprise those 
listed in the Herbig-Bell catalog (\cite{Herbig88.1}; HBC), 
in \citey{Neuhaeuser95.1} or in \citey{Kenyon95.1}.
In addition we include TTSs newly identified either as
counterparts to previously unidentified {\em ROSAT} sources
(\cite{Strom94.1}, \cite{Wichmann96.1}, \cite{Magazzu97.1},
\cite{Neuhaeuser97.2}, \cite{Zickgraf98.1}, 
\cite{Li98.1}, \cite{Briceno98.1})
or by other means 
(\cite{Torres95.1}, \cite{Oppenheimer97.1}, \cite{Briceno98.1}, 
\cite{Reid99.1}, \cite{Gizis99.1}).
TTSs in the molecular cloud MBM\,12 (see \cite{Hearty00.1}) 
are also in the examined sky region.

In addition to TTSs from Tau-Aur we include those from the Perseus molecular
cloud complex, mainly IC\,348 and
NGC\,1333, in our analysis.
Our list of TTSs in IC\,348 comprises X-ray detections 
identified with H$\alpha$ emission
stars or with proper motion members (Tables~4~and~5 in
\cite{Preibisch96.1}), and emission line stars from \citey{Herbig98.1},
and \citey{Luhman99.1}. TTS members of NGC\,1333 are listed in 
\citey{Preibisch97.3}. 

All objects with low lithium strength are excluded,
because it is dubious whether they are young.
We accept only those objects as PMS stars 
which show more lithium than Pleiades stars of the same spectral type,
i.e. we exclude all those with $W_{\lambda}$(Li) lower than $0.2$\,\AA~for 
F- and G-type stars and lower than $0.3$\,\AA~for K-type stars.
When applying this criterion, we always use the spectrum
with the best resolution and best S/N, i.e. the high-resolution
spectra from Wichmann et al. (in preparation). 
If no high-resolution spectra are 
available, we use the medium-resolution spectra from \citey{Martin99.1},
\citey{Neuhaeuser97.2}, or \citey{Magazzu97.1}.

Members of the Pleiades and Hyades clusters are 
selected from the Open Cluster database 
compiled by C. Prosser and collegues 
(available at http://cfa-www.harvard.edu/ $\sim$stauffer/opencl/index.html). 
The tables collected in Prosser's database 
provide a summary of membership classification based on different methods,
such as photometry, spectra, radial velocity and H$\alpha$ emission. In 
addition a final membership determination is given 
which we use to define our membership lists.

Fig.~\ref{fig:skymap} shows a sky map with the positions 
of the selected PSPC observations.

\subsection{Source detection and identification}\label{subsect:soudet}

Source detection is performed on all observations given 
in Table~\ref{tab:pids} and shown in Fig.~\ref{fig:skymap} 
using a combined local and map source detection algorithm based on a
maximum likelihood method (\cite{Cruddace88.1}). All detections with
$ML \geq 7.4$ (corresponding to $\sim 3.5$\,Gaussian $\sigma$ 
determined as best choice by \cite{Neuhaeuser95.1}) 
are written to a source list, which is subsequently cross-correlated 
with the membership lists introduced above. 
The maximum distance $\Delta$ between
optical and X-ray position to be allowed in this identification process
depends on the off-axis angle of the source because the
positional accuracy of the PSPC is worse at larger distances from the
center due to broader point spread function (PSF). From distributions of 
the normalized cumulative number of identifications versus offset $\Delta$ 
for different off-axis ranges we have determined the optimum
cross-correlation radius for all detector positions (similar to
\cite{Neuhaeuser95.1}). A detailed
description of this process together with a table providing $\Delta$ for
different off-axis ranges will be given in Stelzer et al. (in preparation).

Observations which are characterized by strong background variations 
(200020, 200008-0, and 200442), as well as observations consisting of two 
very short intervals separated by a gap of $> 100\,{\rm h}$ 
(200068-0, 200914) are omitted from the flare detection and flare 
analysis. Furthermore, 
we neglect observations with a total duration of less than 1000\,s.
All observations which have been ignored in the analysis presented in this 
paper are marked with an asterisk in Table~\ref{tab:pids}.

\begin{table}
\caption{Complete list of {\em ROSAT} PSPC pointings available from the
{\em ROSAT} Public Data Archive in October 1998, which include at least one
TTS, Pleiad or Hyad from our membership lists (see text) in the field of
view. All observations {\em not} marked by an asterisk have been analyzed
in this work. The total exposure time is given in Column~3. The 
remaining columns give the number of detected and
undetected TTSs, Pleiads and Hyads 
(D -- detection, N -- non-detection). If more than one star
was identified with an X-ray source all identifications are
listed. For multiple systems each component is counted as one entry.}
\label{tab:pids}
\scriptsize
\begin{tabular}{rlrr@{~/~}rr@{~/~}rr@{~/~}r} \\ \hline
No. & {\em ROSAT}    & exp\,(s) & \multicolumn{2}{c}{TTS} &
\multicolumn{2}{c}{Pleids} & \multicolumn{2}{c}{Hyads} \\ 
    & Pointing ID    &       & D & N & D & N & D & N \\ \hline
   1     &       180185p         &        8897   &         16    &          7
 &          0    &          0    &          0    &          0     \\ 
   2     &       200001p--1      &        1520   &          0    &          2
 &          0    &          0    &          0    &          0     \\ 
   3     &       200001p-0       &        4486   &         23    &         12
 &          0    &          0    &          0    &          0     \\ 
   4     &       200001p-1       &       25591   &         26    &          9
 &          0    &          0    &          0    &          0     \\ 
   5     &       200008p--2$^*$      &         696   &          0    &          6
 &          2    &        162    &          0    &          0     \\ 
   6     &       200008p--4$^*$      &         121   &          0    &          5
 &          0    &        220    &          0    &          0     \\ 
   7     &       200008p--5$^*$      &         722   &          0    &          4
 &          1    &        110    &          0    &          2     \\ 
   8     &       200008p-0$^*$       &        5936   &          2    &          4
 &         86    &        165    &          1    &          0     \\ 
   9     &       200008p-2       &        7049   &          4    &          3
 &         88    &        161    &          1    &          0     \\ 
  10     &       200020p$^*$         &       39879   &          0    &          2
 &          0    &          0    &         13    &          8     \\ 
  11     &       200068p--1      &        1307   &          3    &          2
 &         49    &        184    &          0    &          0     \\ 
  12     &       200068p-0$^*$       &       12849   &          3    &          2
 &         93    &        155    &          1    &          0     \\ 
  13     &       200068p-1       &       27071   &          3    &          2
 &        101    &        145    &          1    &          0     \\ 
  14     &       200073p         &        2376   &          0    &          1
 &          0    &          0    &          6    &          3     \\ 
  15     &       200082p-1$^*$       &         814   &          0    &          0
 &          0    &          0    &          0    &          2     \\ 
  16     &       200083p         &        2799   &          0    &          0
 &          0    &          0    &         16    &          5     \\ 
  17     &       200107p--1      &       27692   &          0    &          0
 &          0    &          0    &          2    &          1     \\ 
  18     &       200107p-0       &        3923   &          0    &          0
 &          0    &          0    &          2    &          1     \\ 
  19     &       200402p         &       10469   &          3    &          0
 &          0    &          0    &          0    &          0     \\ 
  20     &       200441p         &       10987   &          0    &          3
 &          0    &          0    &         15    &          8     \\ 
  21     &       200442p         &       19948   &          2    &          0
 &          0    &          0    &          6    &          4     \\ 
  22     &       200443p         &       20074   &         15    &          8
 &          0    &          0    &         12    &          2     \\ 
  23     &       200444p         &       14593   &          0    &          1
 &          0    &          0    &          2    &          5     \\ 
  24     &       200547p         &       28359   &          0    &          3
 &          0    &          0    &          0    &          0     \\ 
  25     &       200553p         &       10961   &          0    &          0
 &          0    &          0    &          8    &          3     \\ 
  26     &       200556p         &       22456   &          5    &          0
 &         67    &         81    &          3    &          0     \\ 
  27     &       200557p         &       27648   &          7    &          0
 &         94    &         96    &          0    &          0     \\ 
  28     &       200576p         &        1521   &          0    &          0
 &          0    &          0    &         12    &          2     \\ 
  29     &       200677p$^*$         &         650   &          3    &         13
 &          0    &          0    &          3    &          0     \\ 
  30     &       200694p         &        1987   &         17    &         13
 &          0    &          0    &          1    &          1     \\ 
  31     &       200694p-1       &        5395   &         16    &         12
 &          0    &          0    &          2    &          0     \\ 
  32     &       200775p         &        4096   &          0    &          2
 &          0    &          0    &          3    &          9     \\ 
  33     &       200776p         &       22995   &          0    &          1
 &          0    &          0    &         10    &          5     \\ 
  34     &       200777p         &       16296   &          0    &          1
 &          0    &          0    &         13    &          8     \\ 
  35     &       200778p         &        1900   &          0    &          1
 &          0    &          0    &         15    &         11     \\ 
  36     &       200911p         &       17460   &          0    &          0
 &          0    &          0    &          7    &          4     \\ 
  37     &       200911p-1       &       13755   &          0    &          0
 &          0    &          0    &          7    &          4     \\ 
  38     &       200912p         &        1656   &          0    &          2
 &          0    &          0    &          3    &          6     \\ 
  39     &       200912p-1       &       23710   &          1    &          1
 &          0    &          0    &          5    &          4     \\ 
  40     &       200913p         &       25341   &          2    &          0
 &          0    &          0    &          7    &          2     \\ 
  41     &       200914p$^*$         &        4098   &          0    &          0
 &          0    &          0    &          1    &          1     \\ 
  42     &       200915p         &        3504   &          0    &          0
 &          0    &          0    &          2    &          2     \\ 
  43     &       200942p         &        7367   &          0    &          0
 &          0    &          0    &          4    &          6     \\ 
  44     &       200945p         &        4099   &          0    &          0
 &          0    &          0    &          7    &          4     \\ 
  45     &       200949p         &        6098   &         17    &          3
 &          0    &          0    &          0    &          0     \\ 
  46     &       200980p         &       10649   &          0    &          0
 &          0    &          0    &          3    &          3     \\ 
  47     &       200980p-1       &        4014   &          0    &          0
 &          0    &          0    &          3    &          3     \\ 
  48     &       200981p         &        4491   &          2    &          1
 &          0    &          0    &          8    &          7     \\ 
  49     &       200982p         &        7724   &          0    &          0
 &          0    &          0    &         10    &          3     \\ 
  50     &       201012p         &        7492   &          2    &          1
 &          0    &          0    &          0    &          0     \\ 
  51     &       201013p         &        4826   &          3    &          2
 &          0    &          0    &          0    &          0     \\ 
  52     &       201013p-1       &        4484   &          2    &          4
 &          0    &          0    &          0    &          0     \\ 
  53     &       201014p         &        9910   &          1    &          1
 &          0    &          0    &          1    &          1     \\ 
  54     &       201015p         &       10058   &          0    &          1
 &          0    &          0    &          1    &          1     \\ 
  55     &       201016p         &       10576   &          6    &          6
 &          0    &          0    &          0    &          0     \\ 
  56     &       201017p         &        8452   &          7    &          9
 &          0    &          0    &          0    &          0     \\ 
  57     &       201023p         &        3058   &          0    &          0
 &          0    &          0    &          0    &          3     \\ 
  58     &       201025p         &        5448   &         11    &         17
 &          0    &          0    &          0    &          0     \\ 
  59     &       201097p         &       10287   &          0    &          0
 &          0    &          0    &          1    &          0     \\ 
  60     &       201278p         &        1358   &          4    &          0
 &          0    &          0    &          0    &          0     \\ 
  61     &       201278p-1       &        4028   &          4    &          0
 &          0    &          0    &          0    &          0     \\ 
  62     &       201305p         &       23643   &         94    &         96
 &          0    &          0    &          0    &          0     \\ 
  63     &       201312p         &        2800   &          8    &         26
 &          0    &          0    &          0    &          0     \\ 
  64     &       201313p         &        4027   &         22    &         12
 &          0    &          0    &          7    &          6     \\ 
  65     &       201314p         &        2691   &          0    &          0
 &          0    &          0    &         12    &          4     \\ 
  66     &       201314p-1       &        1397   &          0    &          0
 &          0    &          0    &          9    &          6     \\ \hline
\end{tabular}
\end{table}

\setcounter{table}{0}

\begin{table}
\caption{{\em continued}}
\scriptsize
\begin{tabular}{rlrr@{/}rr@{/}rr@{/}r} \\ \hline
No. & {\em ROSAT}    & exp\,(s) & \multicolumn{2}{c}{T} & \multicolumn{2}{c}{P} & \multicolumn{2}{c}{H} \\ 
    & Pointing ID    &       & D & N & D & N & D & N \\ \hline
  67     &       201315p$^*$         &         645   &          4    &          1
 &          0    &          0    &          1    &          9     \\ 
  68     &       201315p-1       &        1544   &          5    &          0
 &          0    &          0    &          4    &          7     \\ 
  69     &       201315p-2       &        2298   &          5    &          0
 &          0    &          0    &          4    &          6     \\ 
  70     &       201316p         &        4156   &          1    &          1
 &          0    &          0    &          3    &          3     \\ 
  71     &       201317p         &        1731   &          0    &          0
 &          0    &          0    &          1    &          1     \\ 
  72     &       201319p         &        1811   &          4    &          0
 &          0    &          0    &          3    &          1     \\ 
  73     &       201368p         &       16594   &          0    &          1
 &          0    &          0    &         17    &          9     \\ 
  74     &       201369p         &       15538   &          0    &          0
 &          0    &          0    &         24    &          4     \\ 
  75     &       201370p         &        4975   &          0    &          1
 &          0    &          0    &          5    &          4     \\ 
  76     &       201370p-1       &       13652   &          1    &          0
 &          0    &          0    &          5    &          2     \\ 
  77     &       201484p         &        7673   &          1    &         10
 &          0    &          0    &          0    &          0     \\ 
  78     &       201485p         &        2429   &          0    &          3
 &          0    &          0    &          0    &          0     \\ 
  79     &       201485p-1       &        1904   &          1    &          2
 &          0    &          0    &          0    &          0     \\ 
  80     &       201504p         &      109534   &          2    &          0
 &          0    &          0    &          4    &          0     \\ 
  81     &       201519p         &        5889   &          2    &          0
 &          0    &          0    &          0    &          0     \\ 
  82     &       201532p         &       10183   &          9    &          6
 &          0    &          0    &          1    &          0     \\ 
  83     &       201533p         &       10799   &          4    &          6
 &          0    &          0    &          0    &          0     \\ 
  84     &       201534p         &        6213   &          0    &          1
 &          0    &          0    &          3    &          1     \\ 
  85     &       201598p         &        5652   &          7    &         14
 &          0    &          0    &          0    &          0     \\ 
  86     &       201599p         &        6175   &          6    &         15
 &          0    &          0    &          0    &          0     \\ 
  87     &       201600p         &        5747   &          6    &         15
 &          0    &          0    &          0    &          0     \\ 
  88     &       201601p         &        5817   &          5    &         16
 &          0    &          0    &          0    &          0     \\ 
  89     &       201602p         &        5582   &          8    &         13
 &          0    &          0    &          0    &          0     \\ 
  90     &       201747p         &       19600   &          0    &          1
 &          0    &          0    &         16    &          2     \\ 
  91     &       201748p         &       16999   &          0    &          0
 &          0    &          0    &          2    &          3     \\ 
  92     &       201749p$^*$         &         924   &          0    &          0
 &          0    &          0    &          3    &          1     \\ 
  93     &       201749p-1       &        1540   &          0    &          0
 &          0    &          0    &          3    &          1     \\ 
  94     &       300178p         &         710   &          0    &          7
 &          0    &          0    &          0    &          0     \\ 
  95     &       400312p         &       10735   &          0    &          0
 &          0    &          0    &          1    &          0     \\ 
  96     &       700044p         &        4611   &          5    &          0
 &          0    &          0    &          6    &          4     \\ 
  97     &       700063p         &        1701   &          0    &          0
 &          0    &          0    &          1    &          0     \\ 
  98     &       700825p         &        1435   &          0    &          0
 &          0    &          0    &          0    &          2     \\ 
  99     &       700825p-1       &       15744   &          0    &          0
 &          0    &          0    &          1    &          1     \\ 
 100     &       700913p         &        2096   &          0    &          0
 &          0    &          0    &          1    &          0     \\ 
 101     &       700916p         &        7161   &          0    &          0
 &          0    &          0    &          1    &          0     \\ 
 102     &       700919p         &        2004   &          0    &          0
 &          0    &          0    &          1    &          0     \\ 
 103     &       700945p         &        2516   &          0    &          0
 &          0    &          0    &          1    &          0     \\ 
 104     &       701055p         &        9039   &          0    &          0
 &          0    &          0    &          0    &          0     \\ 
 105     &       701253p         &        5417   &          0    &          0
 &          0    &          0    &          0    &          2     \\ 
 106     &       800051p-0       &        1470   &          1    &          0
 &          0    &          0    &          0    &          0     \\ 
 107     &       800051p-1       &        3347   &          1    &          0
 &          0    &          0    &          0    &          0     \\ 
 108     &       800051p-2       &        3303   &          1    &          0
 &          0    &          0    &          0    &          0     \\ 
 109     &       800083p         &       10220   &          1    &          0
 &          0    &          0    &          0    &          0     \\ 
 110     &       800104p         &        7524   &          1    &          0
 &          0    &          0    &          0    &          0     \\ 
 111     &       800193p         &       21969   &          0    &          0
 &          0    &          0    &          3    &          3     \\ 
 112     &       900138p         &       24899   &          6    &          0
 &          0    &          0    &          0    &          0     \\ 
 113     &       900154p         &       28791   &          0    &          2
 &          0    &          0    &          0    &          0     \\ 
 114     &       900193p         &        9128   &          3    &         22
 &          0    &          0    &          0    &          0     \\ 
 115     &       900353p         &        7718   &         18    &         11
 &          0    &          0    &         12    &          4     \\ 
 116     &       900371p         &        4227   &          1    &          0
 &          0    &          0    &          2    &          0     \\ 
 117     &       900371p-1       &        7718   &          0    &          0
 &          0    &          0    &          2    &          0     \\ \hline
\end{tabular}
\end{table}

\section{Lightcurves}\label{sect:lcs}

Using the arrival time information of the photons counted within a 
pre-defined source circle lightcurves are generated for each of the 
{\em ROSAT} sources 
that have been identified with a TTS, Pleiad or Hyad 
from the membership lists. 

For the source extraction radius we have used the 
99\% quantile of the Point Spread Function (PSF) at 1\,keV, 
i.e. the radius containing 99\% of the 1\,keV 
photons at the respective off-axis angle. In contrast to the
standard EXSAS source radius of 2.5 FWHM, which becomes unreasonably 
large for off-axis sources 
due to the extended wings of the PSF, this
choice of extraction radius limits the source size. 
Close to the detector center, some bright sources slightly overshine the
nominal 99\% quantile of the PSF probably due to small deviations from the
assumed 1\,keV spectrum. We have therefore checked all images
for such bright sources and determined a larger source radius for these
cases based on visual inspection.
In crowded regions, where the 
PSF of several sources overlap, 
the measured counts are upper limits to the actual emission of the sources.
None of the overlapping sources showed a flare, however, 
such that no further attention is drawn to the overestimation of the count
rate in these cases.

The events measured within the circular source region are 
binned into 400\,s intervals. 
Since the typical duration of a flare is less than one hour,
significantly longer integration times would lead to a loss of 
information about the structure of the lightcurve, while for shorter
bin lengths the lightcurves are dominated by the low statistics.
Furthermore, 
the choice of 400\,s integration time guarantees that no additional
variability is introduced by the telescope motion (wobble).

Due to the earth eclipses the data stream is interrupted at
periodical time intervals. Depending on the phase used
for the time integration, at the beginning and/or end of each data segment
the 400\,s intervals are only partly exposed. For the flare detection only bins
with full 400\,s of exposure are used.
To gain independence of the
binning we generate lightcurves with different phasing of the 400\,s 
intervals: First, in order to divide the given observing time into as 
many 400\,s exposures as possible, a lightcurve is binned in such a way that
a new 400\,s interval starts after each observation gap. Thus, data are
lost only at the end of each data segment, because the last bin remains
uncomplete. Secondly, lightcurves are built by simply splitting the total 
observing time into 400\,s intervals beginning from the start of 
the observation regardless of data gaps. 
In this case data are rejected at the beginning {\em and} 
end of each data segment. 

The number of background counts falling in the source circle is determined
from the smoothed background image which is created by cutting out the 
detected sources and then performing a spline fit to the resulting image. 
This method of background acquisition is of advantage in
crowded fields where an annulus around the source position -- the most widely 
used method for estimating the background -- likely is contaminated by 
other sources. The background count rate is found by dividing the
number of background counts in the source circle 
through the exposure time extracted from the standard {\em ROSAT} exposure map.
To take account of possible time variations in the background count rate, the
background is determined separately for each data segment and subtracted 
from the measured count rate in the respective data interval. 
(When referring to `data segments'
we mean parts of the lightcurve that are separated from each other by gaps of
at least 0.5\,h.)

\section{Flare detection}\label{sect:detect}

\subsection{The method}\label{subsect:method_det}

One of the major elements of a flare by customary definition is 
a significant increase in count rate,
after which the initial level of intensity is reached again. Therefore,
our flare detection is based on the deviation of the count rate from the
 (previously determined) mean quiescent level of the source. 
To ensure that the quiescent count rate contains no contribution from 
flares, in the first step, we determine mean count rates for all data
segments of each lightcurve and define the quiescent level as the lowest mean 
measured in any of these data segments.

We define a flare as an event which is characterized by two or more 
consecutive time bins that constitute a sequence of either rising or falling 
count rates, corresponding to rise and decay phase of the flare. 
In addition, to ensure the significance of our
flare detections, we define the upper standard deviation of the quiescent
level as a point of reference and require that
(a) all bins which are part of the flare
are characterized by count rates higher than this level, and that (b)
the sum of the deviations of all these bins is more than 5\,$\sigma$ from 
this level. 
A rise immediately followed by a
decay is counted as {\em one} flare. 
Since the shape of a lightcurve is influenced to some degree by the binning
used, we accept only flares that are detected in lightcurves with both
bin phasings (see Sect.~\ref{sect:lcs}). 

Detections of more than one flare in a single lightcurve are possible. 
To estimate the contribution of each event properly,
after detection the decay of the first flare in each lightcurve 
is modeled by an exponential function, and a new lightcurve is generated
by subtracting the fit function from the data. Having removed the
first flare, we search for further flares in the reduced, 
`flare-subtracted' lightcurve using the same criteria as before. 
This procedure is repeated until no additional flares are detected. 

Since many of the investigated sources are highly variable X-ray emitters
on timescales shorter than resolvable by our method, the
mean count rate used until now in some cases is not a 
good estimate for the quiescent emission. With the knowledge 
obtained about the times at which flares have occurred
we therefore redetermine the 
quiescent count rate taking the mean from the remaining data 
after removal of
all flare contributions. Using this new mean count rate we repeat the 
flare detection procedure.

\subsection{Flare Parameters}\label{subsect:flarepar}

With the detection procedure described in the previous subsection 
we have found 52 flares. We have always identified the nearest optical
position with the X-ray source. In one flare, however, two 
possible optical counterparts, DD\,Tau and CZ\,Tau, are closeby (at
6$^{\prime\prime}$ and 24$^{\prime\prime}$ respectively), so that we can
not be sure which star flared. 
Fifteen events were observed on TTSs, 20 on 
Pleiads, and 17 on Hyads. On two TTSs (RXJ\,0437.5+1851 and T\,Tau) 
and two Hyads (VA\,334 and VB\,141) two flares occurred in the
same observation. VB\,141 showed a third event during a
different {\em ROSAT} exposure.

Hyades stars above $2\,{\rm M}_\odot$, that have already
evolved off the main-sequence, are not considered
in the statistical analysis if they showed a flare. Brown dwarfs in 
the Hyades and Pleiades are not on the main-sequence per definition, but
they are also too faint for X-ray detection (\cite{Neuhaeuser99.1}). 
Thus we discuss only the ZAMS from the Pleiades and Hyades.

A complete list of all TTSs, Pleiads, and Hyads
on which at least one flare was detected 
is given in Table~\ref{tab:opt_par}. Column~1 gives the
designation of the flaring star. Column~2 is the distance estimate
used for the count-to-energy-conversion. For TTSs 
in Taurus-Auriga we adopt a 
value of $140\,{\rm pc}$ (\cite{Elias78.1}, \cite{Wichmann98.1}), while
the TTSs in MBM\,12 are located at $65\,{\rm pc}$ (\cite{Hearty00.1}), and
those in Perseus are located at $350\,{\rm pc}$ (NGC\,1333;
\cite{Herbig88.1}) and $300\,{\rm pc}$ (IC\,348; \cite{Cernicharo85.1}).
Pleiads are assumed to be at a distance of $116\,{\rm pc}$, the value
derived by \citey{Mermilliod97.1}. Finally, we use the individual
Hipparcos parallaxes for Hyades stars if available, and otherwise the mean
value of $46\,{\rm pc}$ (\cite{Perryman98.1}). We give
spectral type, $v\,\sin{i}$, multiplicity, and binary separation 
of the stars and their respective references in columns~3 -- 9. For TTSs 
additional columns specify whether the star is a cTTS or a wTTS.

\begin{figure*}
\begin{center}
\parbox{16cm}{
\parbox{4.5cm}{\resizebox{5.5cm}{!}{\includegraphics{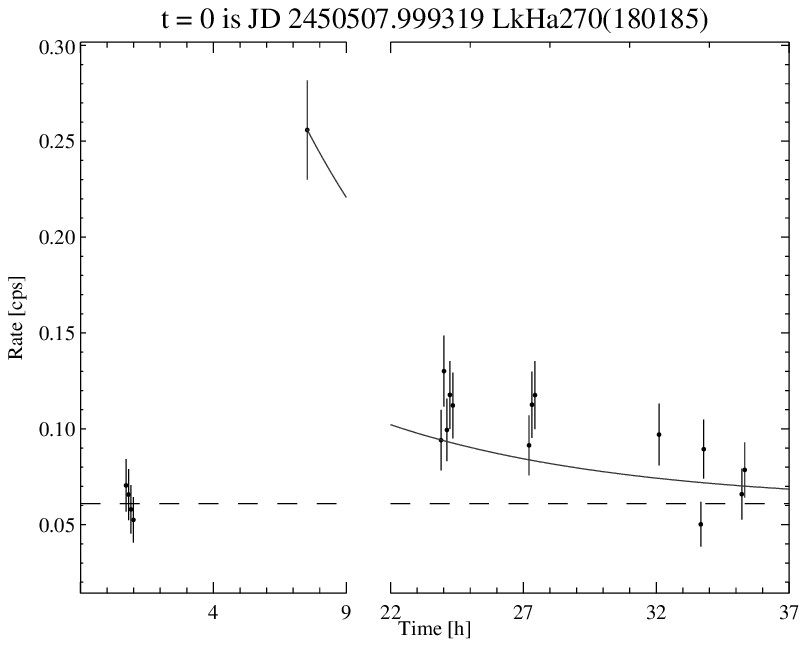}}}
\parbox{1cm}{\hspace*{1.cm}}
\parbox{4.5cm}{\resizebox{5.5cm}{!}{\includegraphics{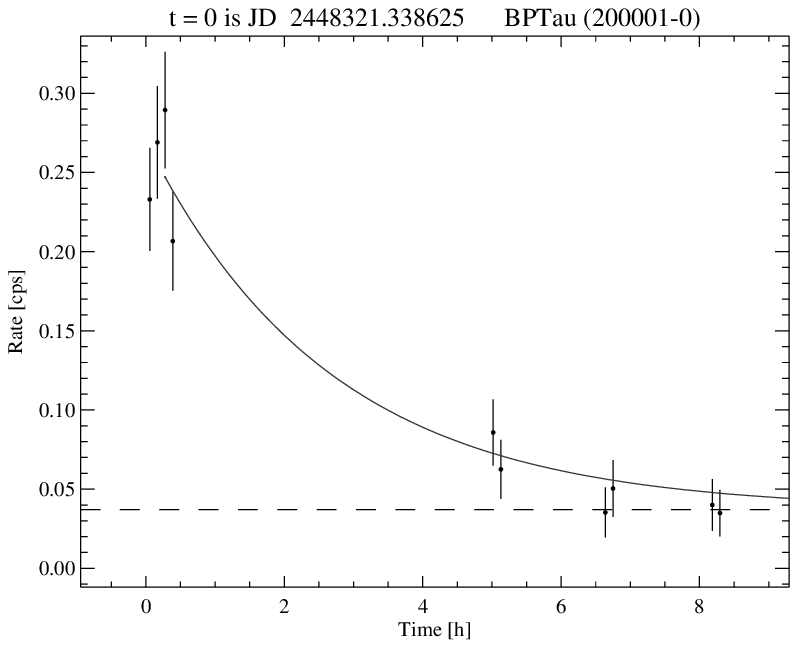}}}
\parbox{1cm}{\hspace*{1.cm}}
\parbox{4.5cm}{\resizebox{5.5cm}{!}{\includegraphics{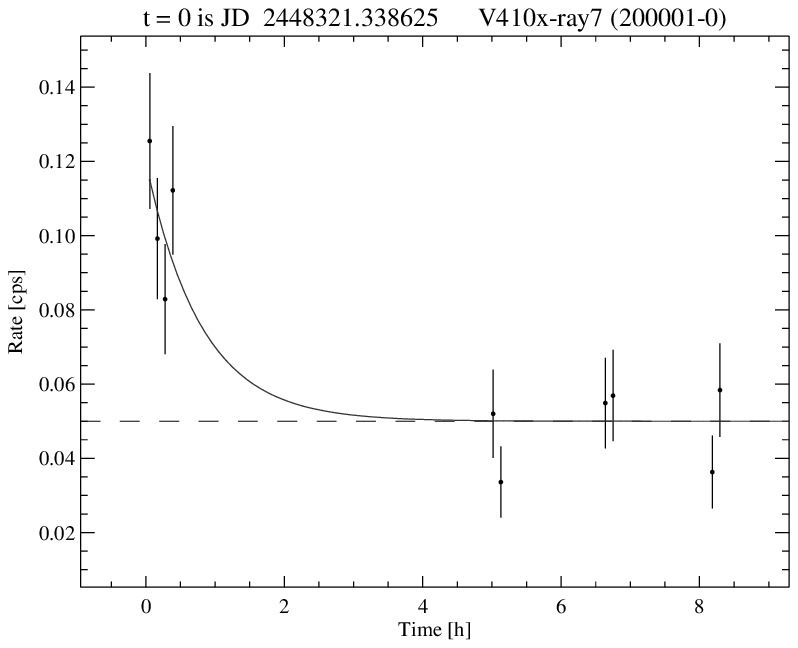}}}
}
\parbox{16cm}{
\parbox{4.5cm}{\resizebox{5.5cm}{!}{\includegraphics{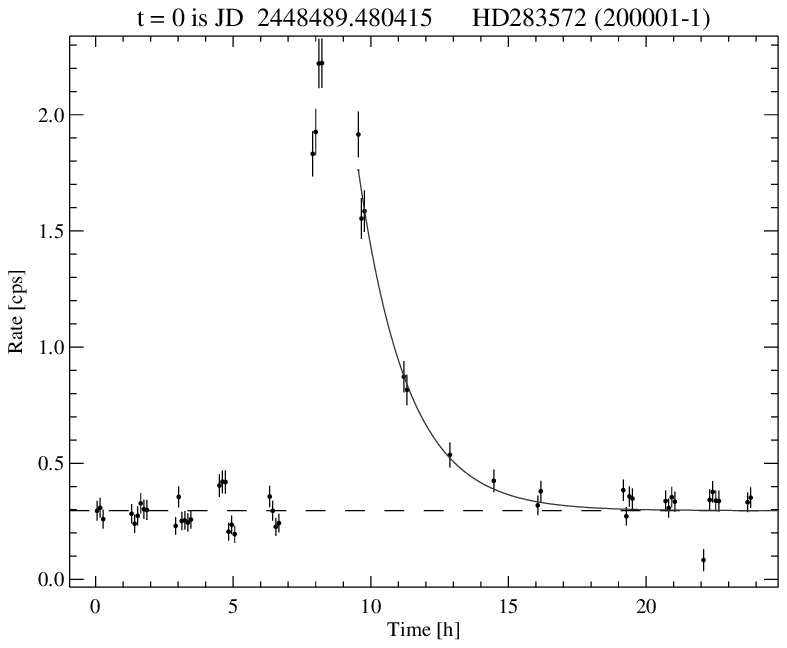}}}
\parbox{1cm}{\hspace*{1.cm}}
\parbox{4.5cm}{\resizebox{5.5cm}{!}{\includegraphics{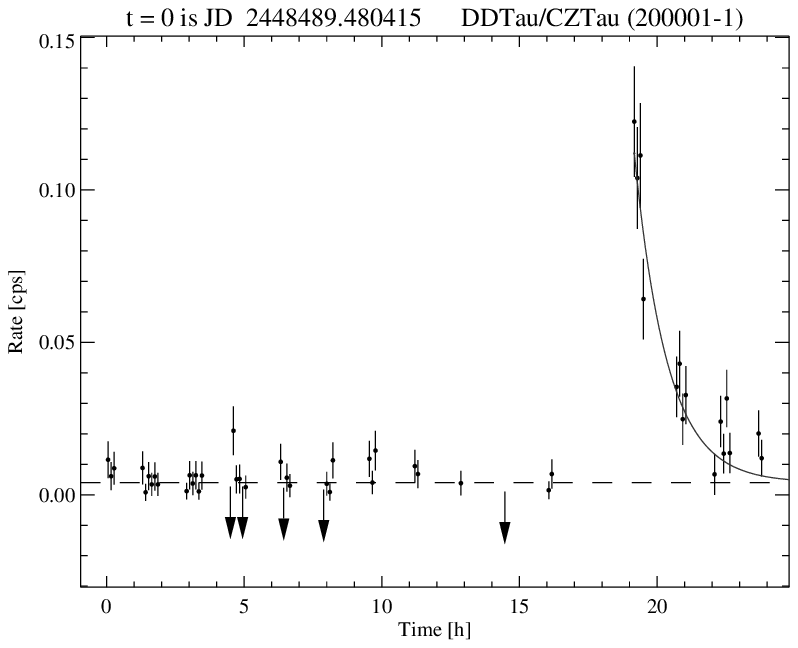}}}
\parbox{1cm}{\hspace*{1.cm}}
\parbox{4.5cm}{\resizebox{5.5cm}{!}{\includegraphics{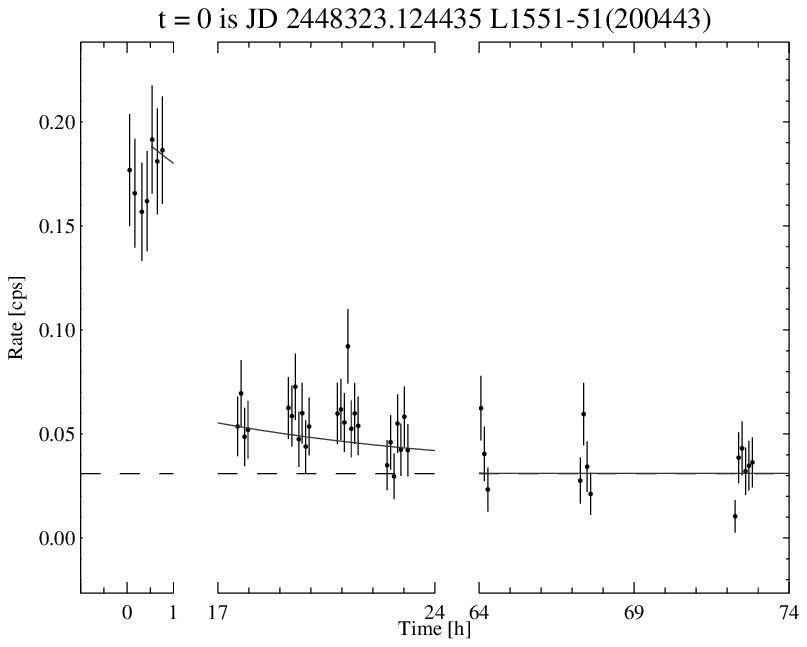}}}
}
\parbox{16cm}{
\parbox{4.5cm}{\resizebox{5.5cm}{!}{\includegraphics{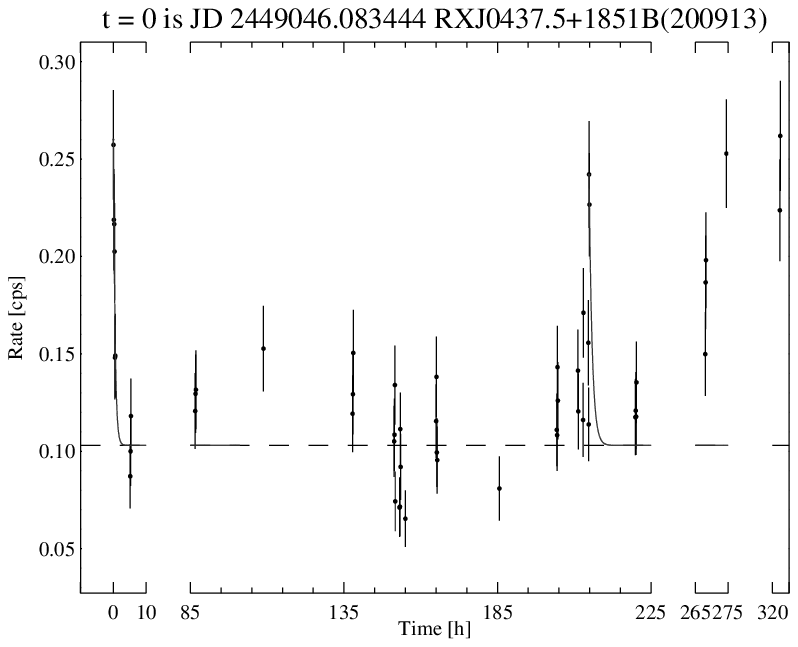}}}
\parbox{1cm}{\hspace*{1.cm}}
\parbox{4.5cm}{\resizebox{5.5cm}{!}{\includegraphics{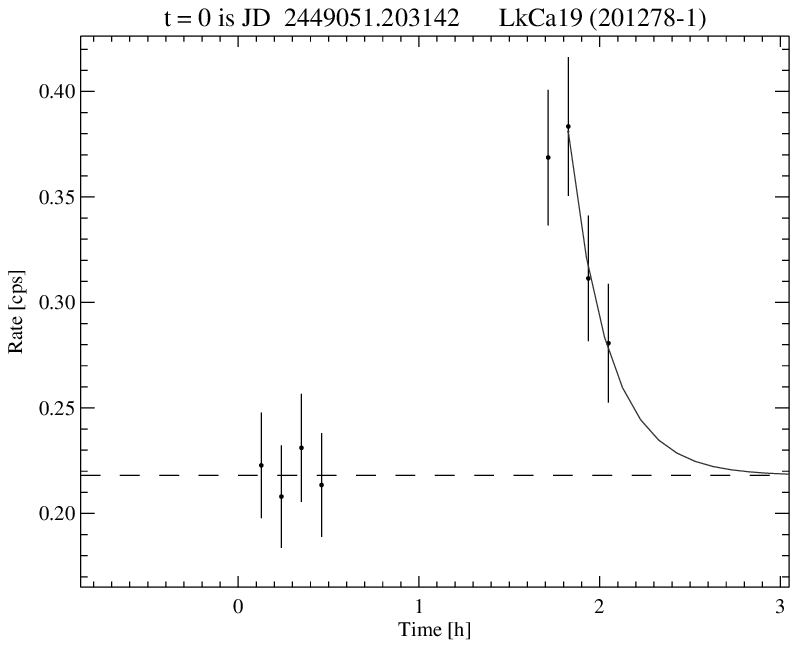}}}
\parbox{1cm}{\hspace*{1.cm}}
\parbox{4.5cm}{\resizebox{5.5cm}{!}{\includegraphics{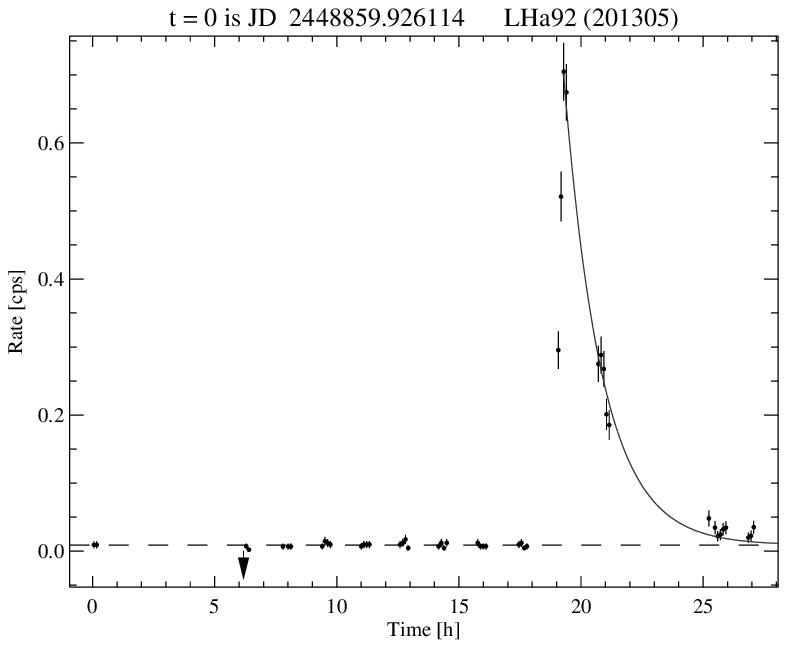}}}
}
\parbox{16cm}{
\parbox{4.5cm}{\resizebox{5.5cm}{!}{\includegraphics{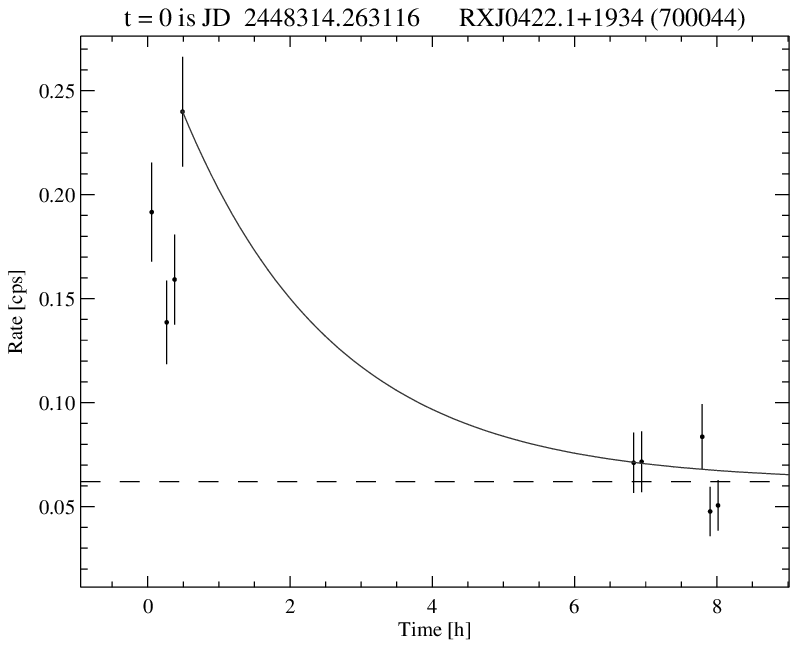}}}
\parbox{1cm}{\hspace*{1.cm}}
\parbox{4.5cm}{\resizebox{5.5cm}{!}{\includegraphics{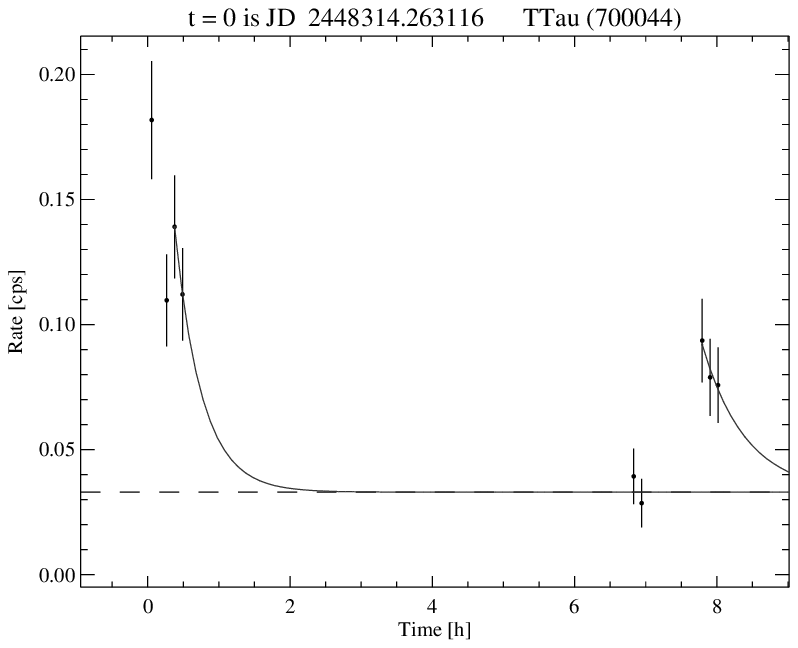}}}
\parbox{1cm}{\hspace*{1.cm}}
\parbox{4.5cm}{\resizebox{5.5cm}{!}{\includegraphics{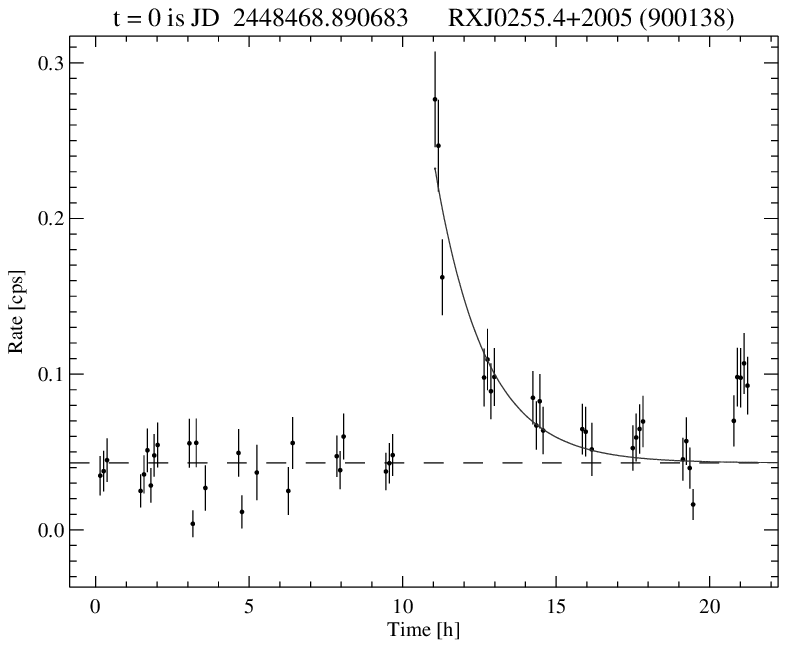}}}
}
\parbox{16cm}{
\parbox{4.5cm}{\resizebox{5.5cm}{!}{\includegraphics{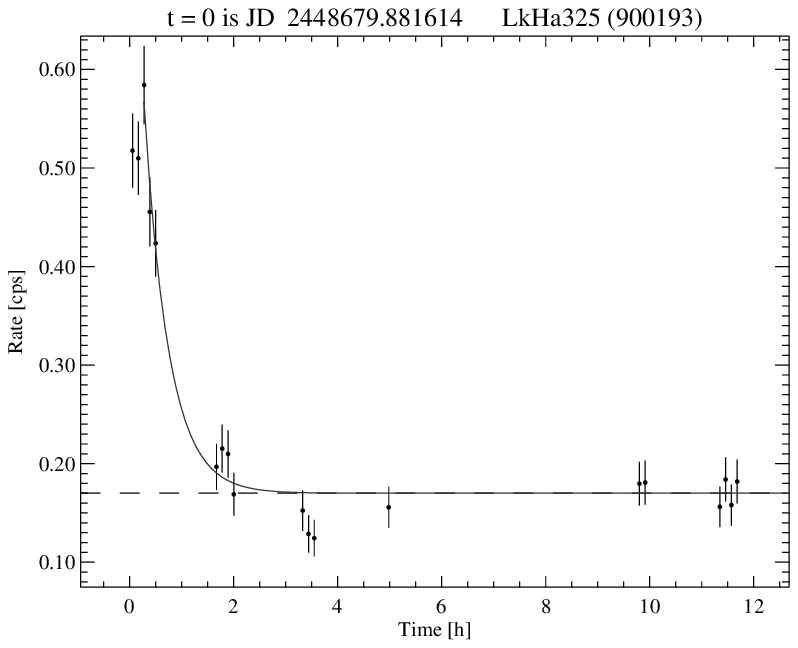}}}
\parbox{1cm}{\hspace*{1.cm}}
}
\caption{PSPC lightcurves for flares on TTSs in Taurus-Auriga and Perseus
detected by the procedure described in
Sect.~\ref{subsect:method_det}. Identification of the X-ray source and {\em
ROSAT} observation request number (in brackets) are given for each source. Binsize is 400\,s, 1\,$\sigma$ uncertainties. The dashed line represents the quiescent count rate. Solid lines are exponential fits to the data points belonging to the flare down to the quiescent emission. Background count rates are shown as upper limits when the background subtracted count rate is below zero.}
\label{fig:lcs_TTS}
\end{center}
\end{figure*}

\begin{figure*}
\begin{center}
\parbox{16cm}{
\parbox{4.5cm}{\resizebox{5.5cm}{!}{\includegraphics{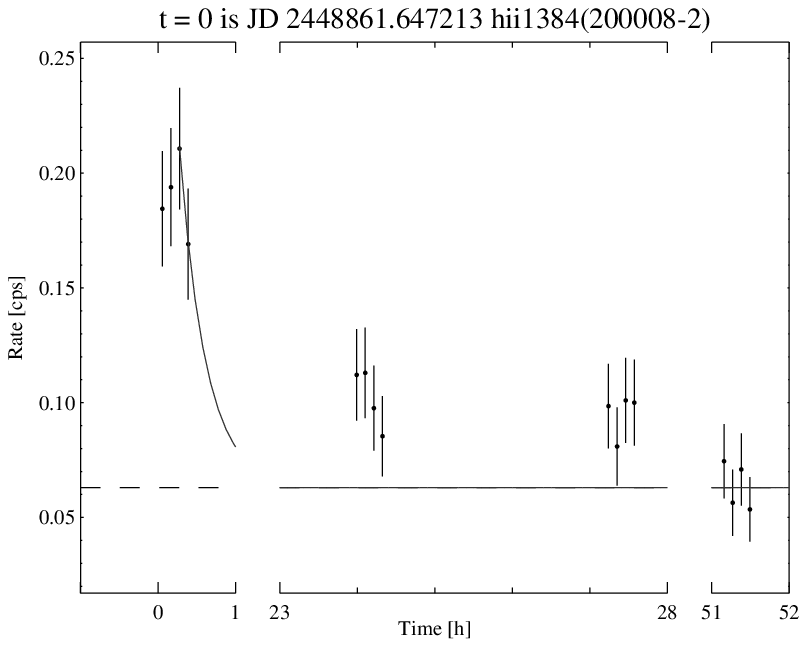}}}
\parbox{1cm}{\hspace*{1.cm}}
\parbox{4.5cm}{\resizebox{5.5cm}{!}{\includegraphics{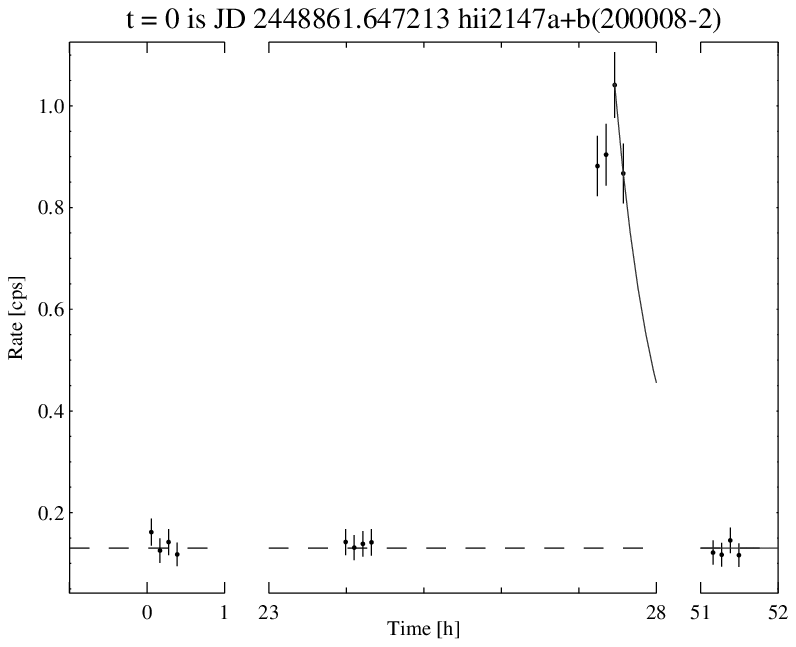}}}
\parbox{1cm}{\hspace*{1.cm}}
\parbox{4.5cm}{\resizebox{5.5cm}{!}{\includegraphics{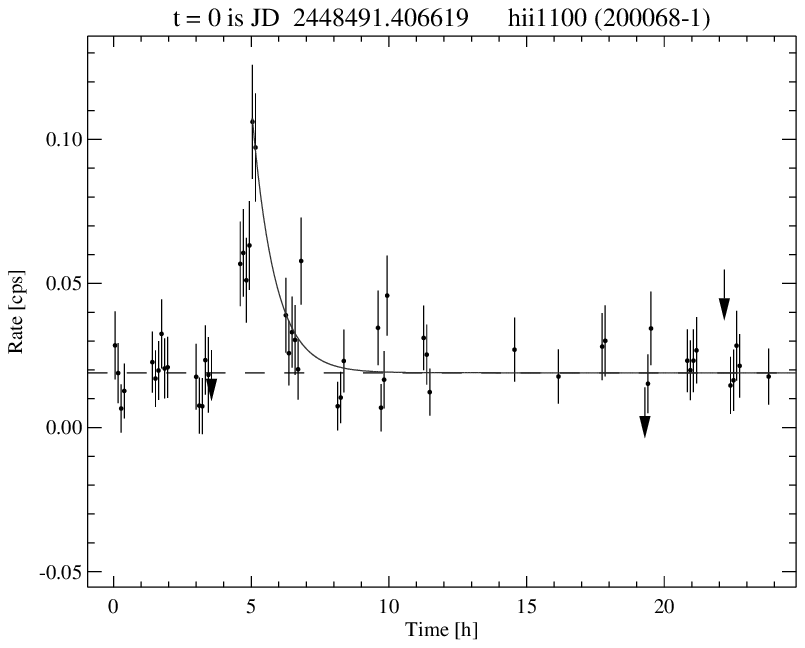}}}
}
\parbox{16cm}{
\parbox{4.5cm}{\resizebox{5.5cm}{!}{\includegraphics{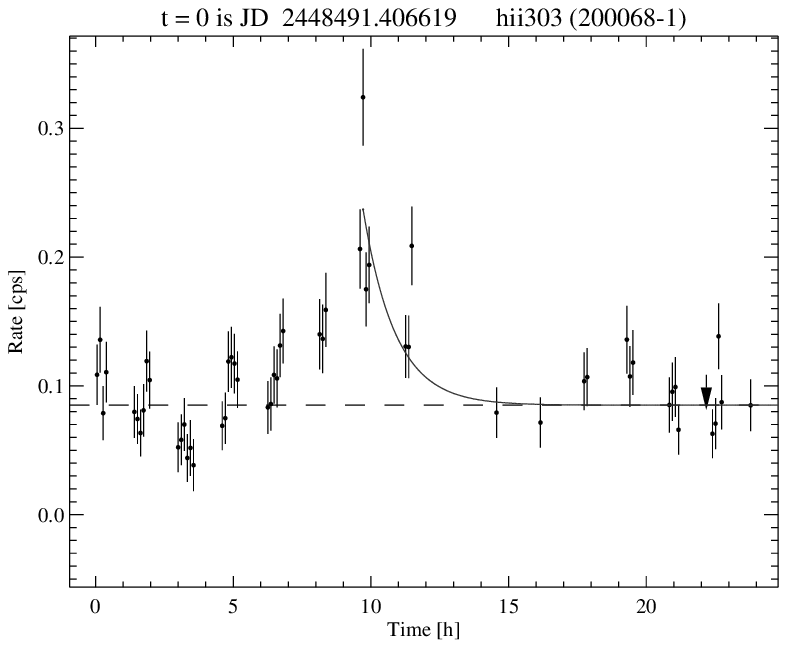}}}
\parbox{1cm}{\hspace*{1.cm}}
\parbox{4.5cm}{\resizebox{5.5cm}{!}{\includegraphics{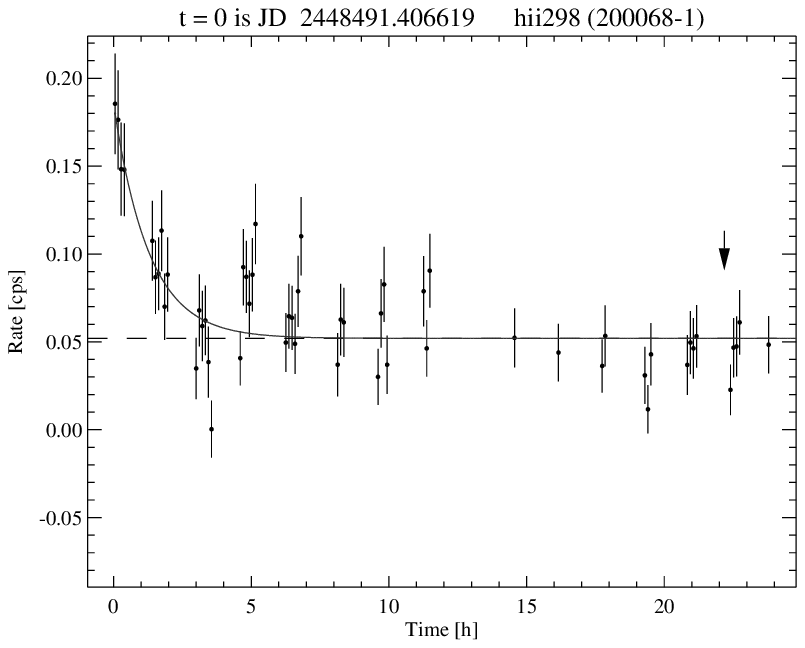}}}
\parbox{1cm}{\hspace*{1.cm}}
\parbox{4.5cm}{\resizebox{5.5cm}{!}{\includegraphics{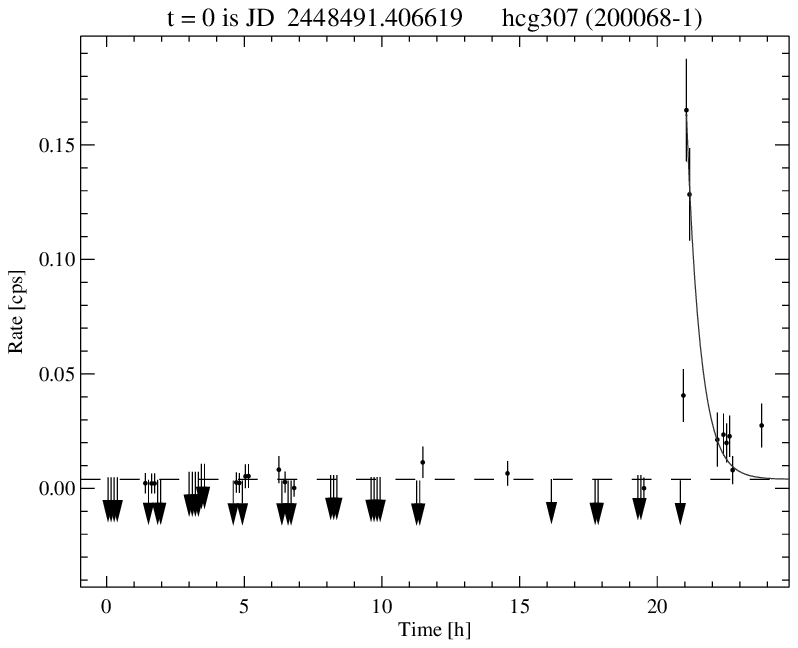}}}
}
\parbox{16cm}{
\parbox{4.5cm}{\resizebox{5.5cm}{!}{\includegraphics{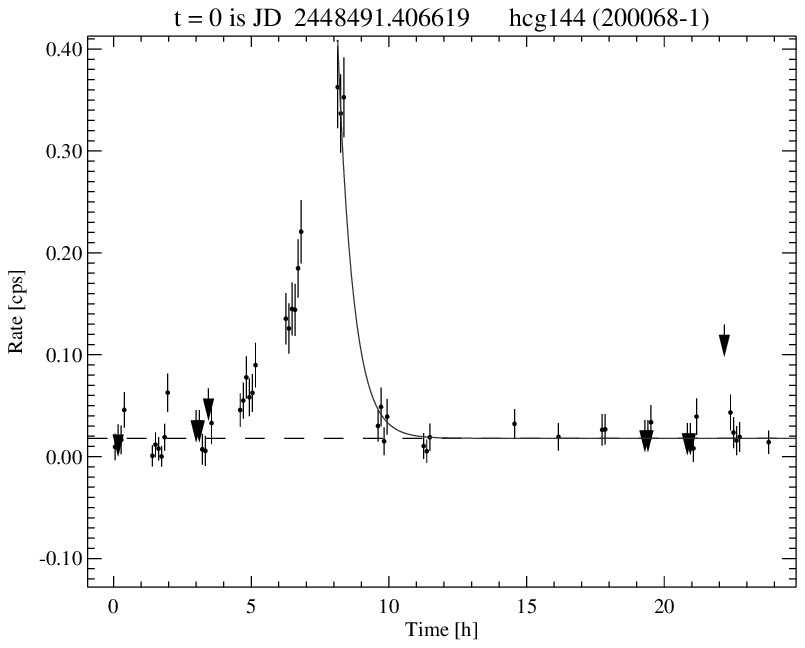}}}
\parbox{1cm}{\hspace*{1.cm}}
\parbox{4.5cm}{\resizebox{5.5cm}{!}{\includegraphics{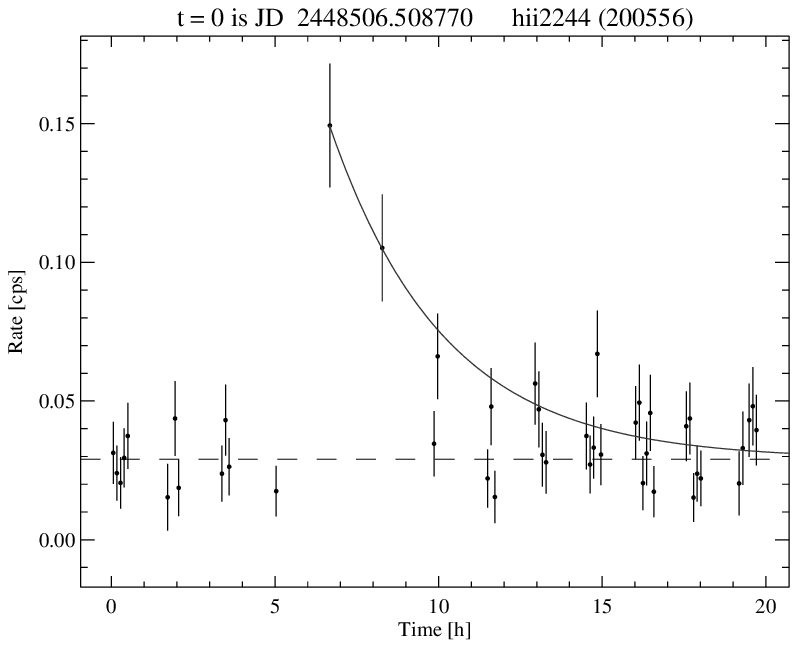}}}
\parbox{1cm}{\hspace*{1.cm}}
\parbox{4.5cm}{\resizebox{5.5cm}{!}{\includegraphics{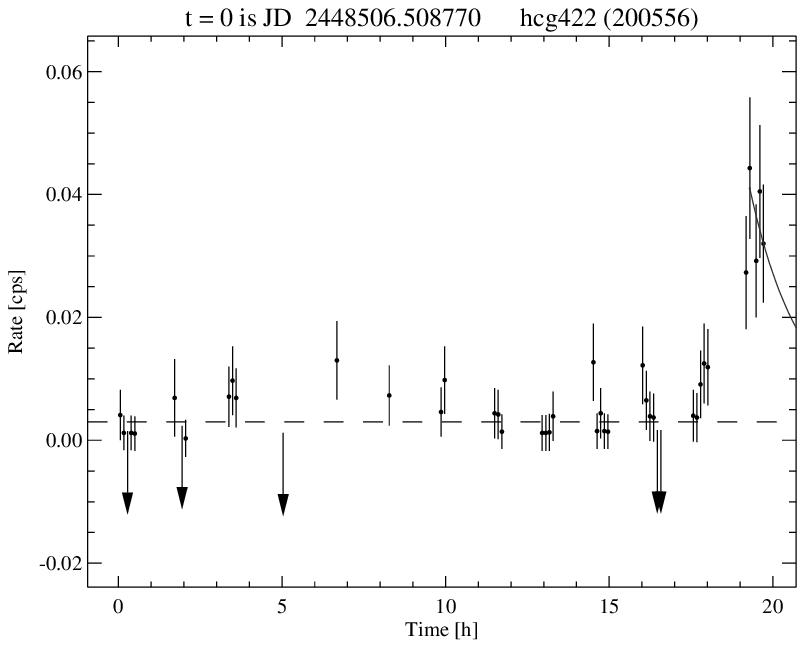}}}
}
\parbox{16cm}{
\parbox{4.5cm}{\resizebox{5.5cm}{!}{\includegraphics{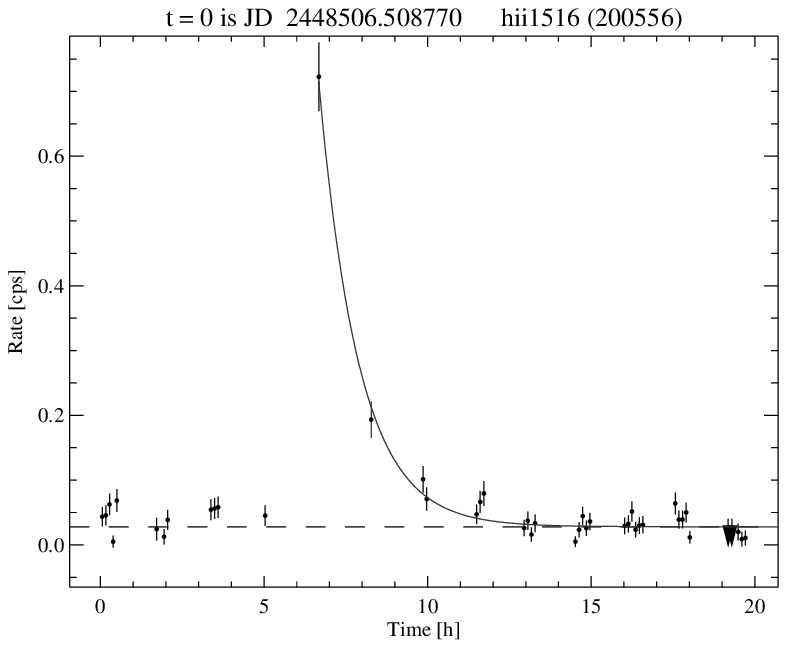}}}
\parbox{1cm}{\hspace*{1.cm}}
\parbox{4.5cm}{\resizebox{5.5cm}{!}{\includegraphics{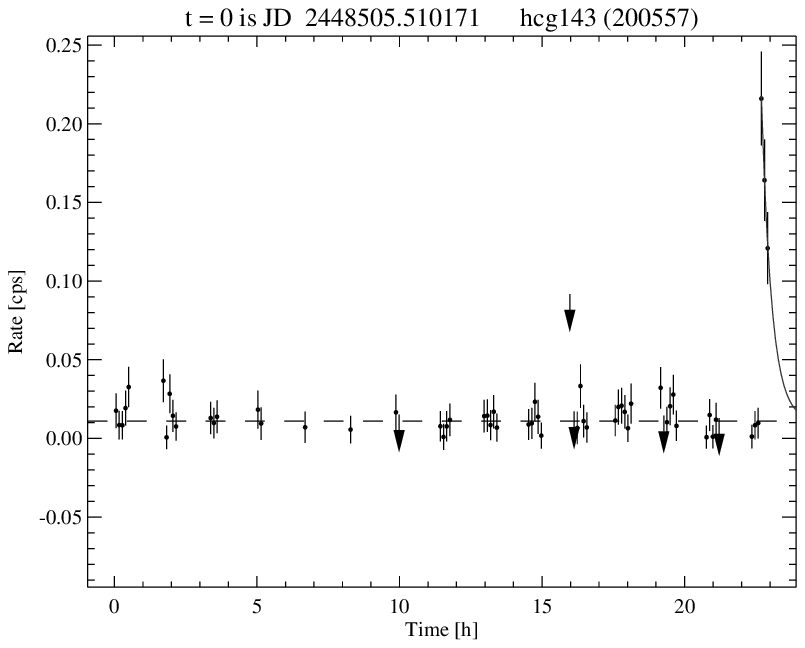}}}
\parbox{1cm}{\hspace*{1.cm}}
\parbox{4.5cm}{\resizebox{5.5cm}{!}{\includegraphics{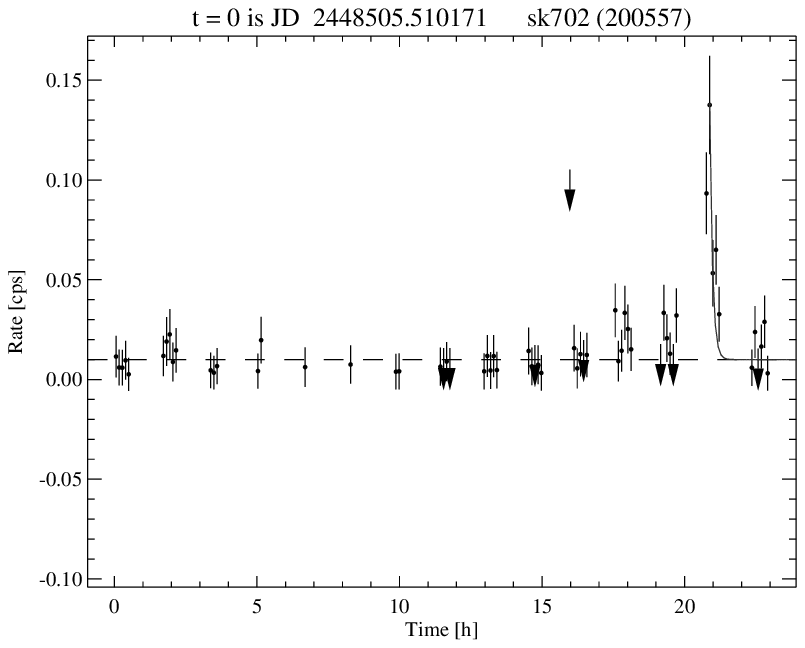}}}
}
\parbox{16cm}{
\parbox{4.5cm}{\resizebox{5.5cm}{!}{\includegraphics{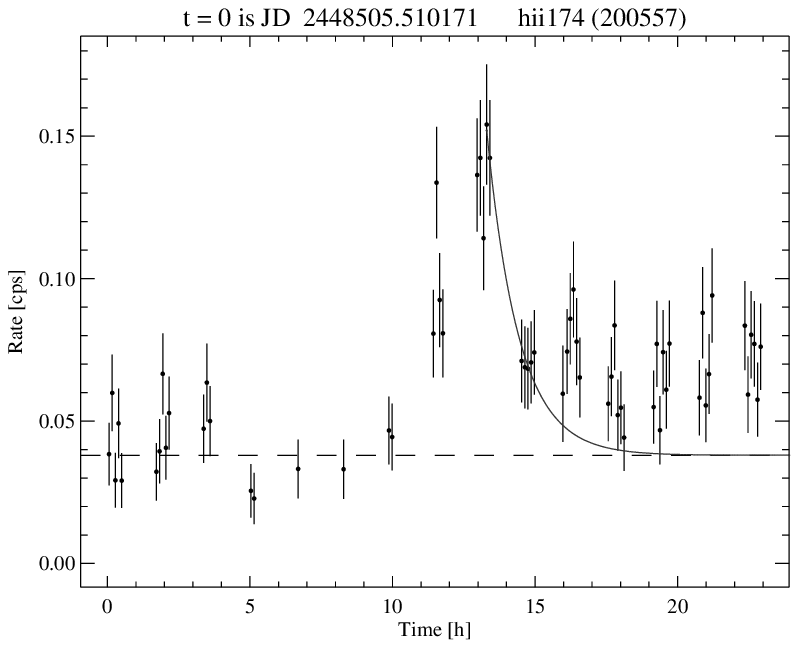}}}
\parbox{1cm}{\hspace*{1.cm}}
\parbox{4.5cm}{\resizebox{5.5cm}{!}{\includegraphics{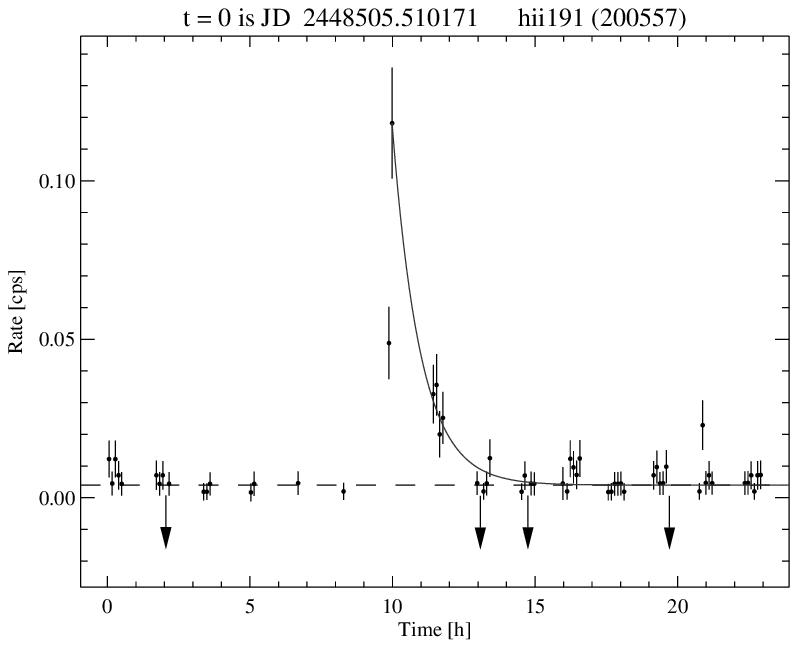}}}
\parbox{1cm}{\hspace*{1.cm}}
\parbox{4.5cm}{\resizebox{5.5cm}{!}{\includegraphics{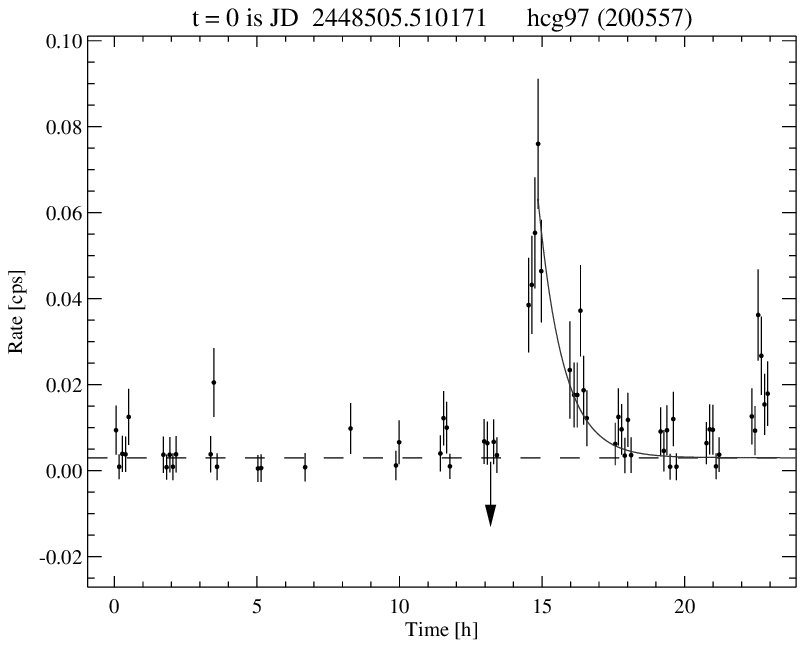}}}
}
\caption{PSPC lightcurves for flares on members of the Pleiades cluster
detected by the procedure described in
Sect~\ref{subsect:method_det}. Identification of the X-ray source and {\em
ROSAT} observation request number (in brackets) are given for each source. Binsize is 400\,s, 1\,$\sigma$ uncertainties. The dashed line represents the quiescent count rate. Solid lines are exponential fits to the data points belonging to the flare down to the quiescent emission. Background count rates are shown as upper limits when the background subtracted count rate is below zero.}
\label{fig:lcs_Ple}
\end{center}
\end{figure*}

\addtocounter{figure}{-1}

\begin{figure*}
\begin{center}
\parbox{16cm}{
\parbox{4.5cm}{\resizebox{5.5cm}{!}{\includegraphics{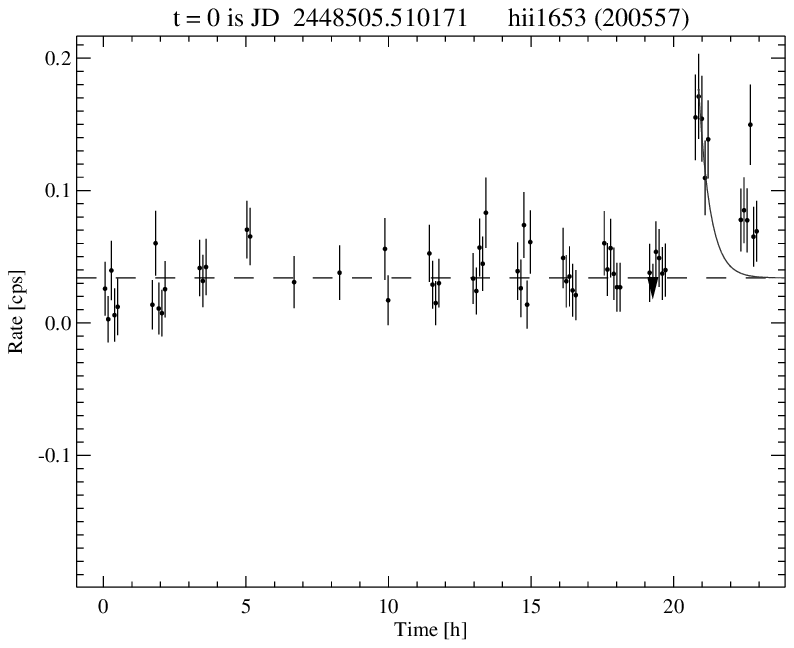}}}
\parbox{1cm}{\hspace*{1.cm}}
\parbox{4.5cm}{\resizebox{5.5cm}{!}{\includegraphics{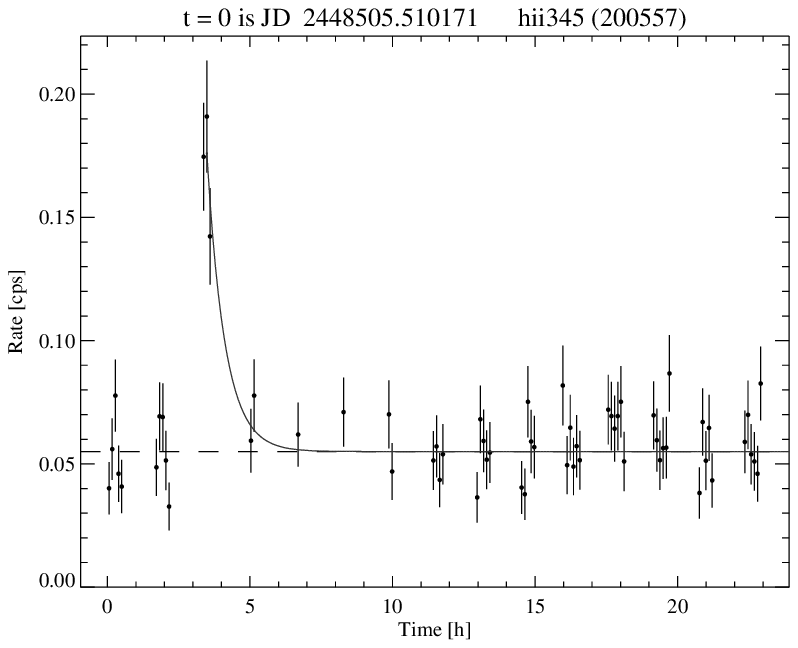}}}
\parbox{1cm}{\hspace*{1.cm}}
\parbox{4.5cm}{\resizebox{5.5cm}{!}{\includegraphics{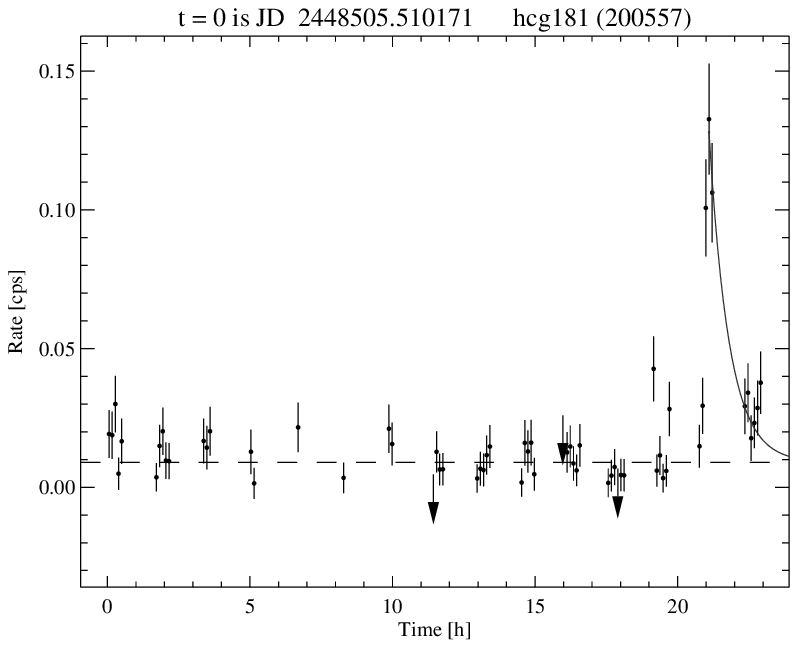}}}
}
\parbox{16cm}{
\parbox{4.5cm}{\resizebox{5.5cm}{!}{\includegraphics{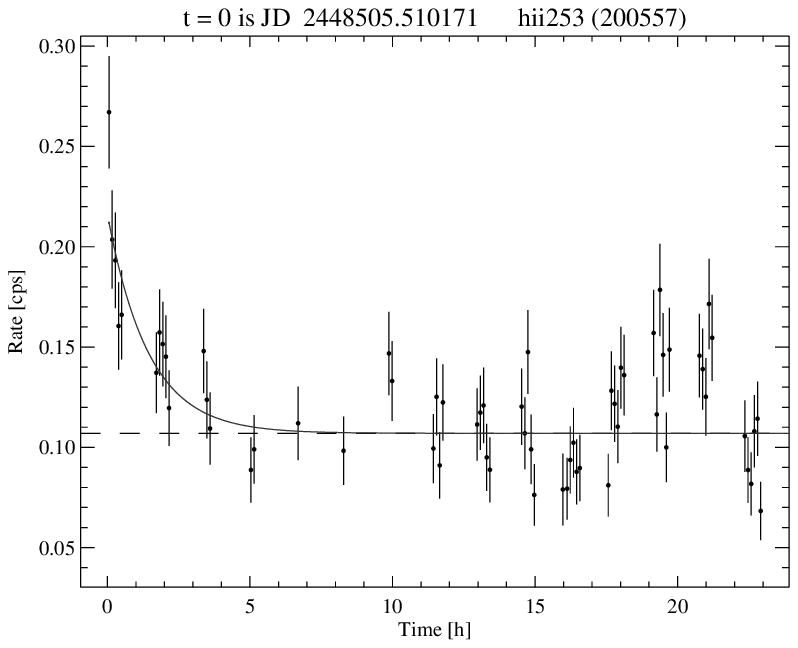}}}
\parbox{1cm}{\hspace*{1.cm}}
\parbox{4.5cm}{\resizebox{5.5cm}{!}{\includegraphics{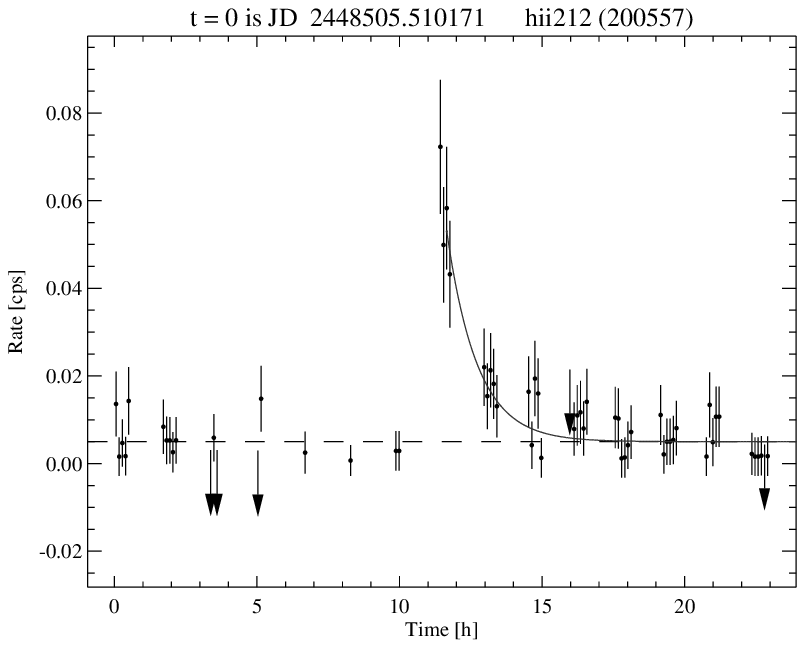}}}
}

\caption{{\em continued}}
\end{center}
\end{figure*}

\begin{figure*}
\begin{center}
\parbox{16cm}{
\parbox{4.5cm}{\resizebox{5.5cm}{!}{\includegraphics{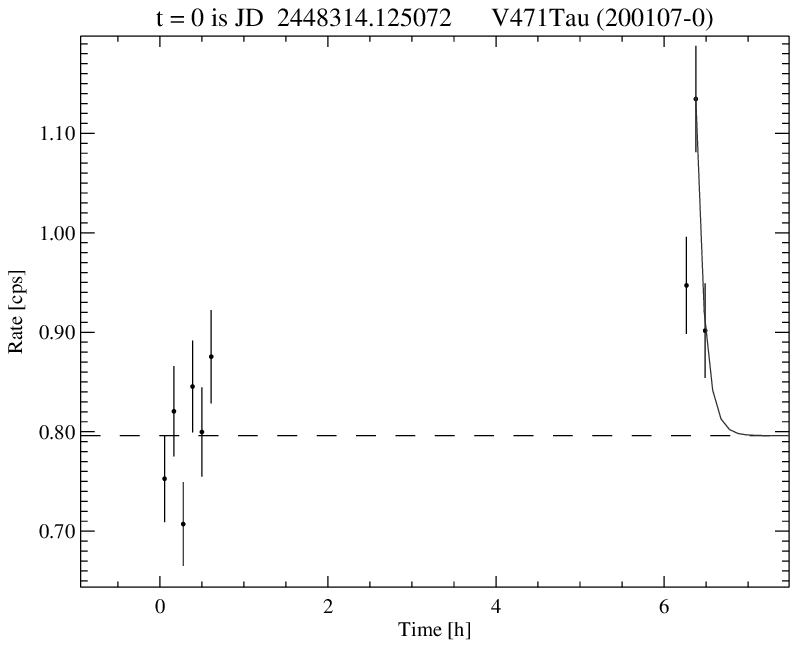}}}
\parbox{1cm}{\hspace*{1.cm}}
\parbox{4.5cm}{\resizebox{5.5cm}{!}{\includegraphics{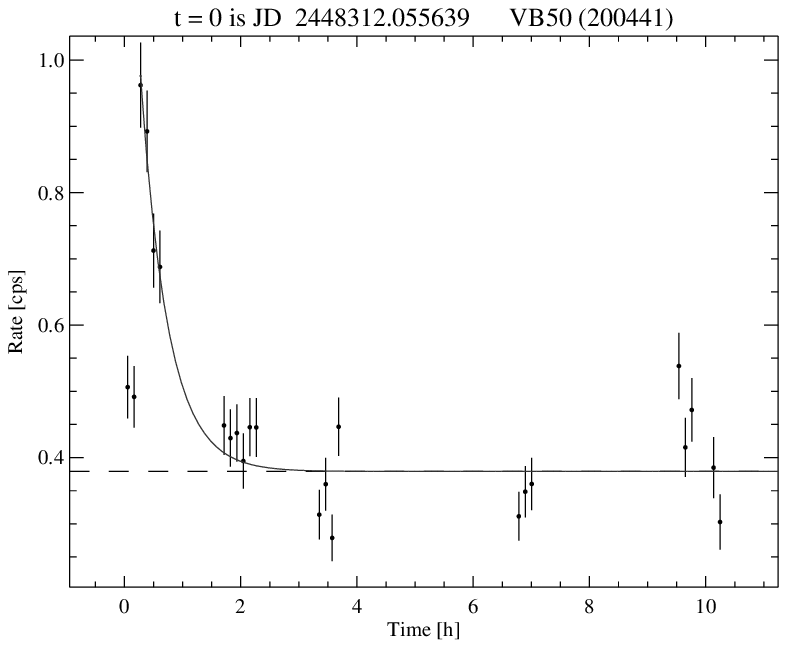}}}
\parbox{1cm}{\hspace*{1.cm}}
\parbox{4.5cm}{\resizebox{5.5cm}{!}{\includegraphics{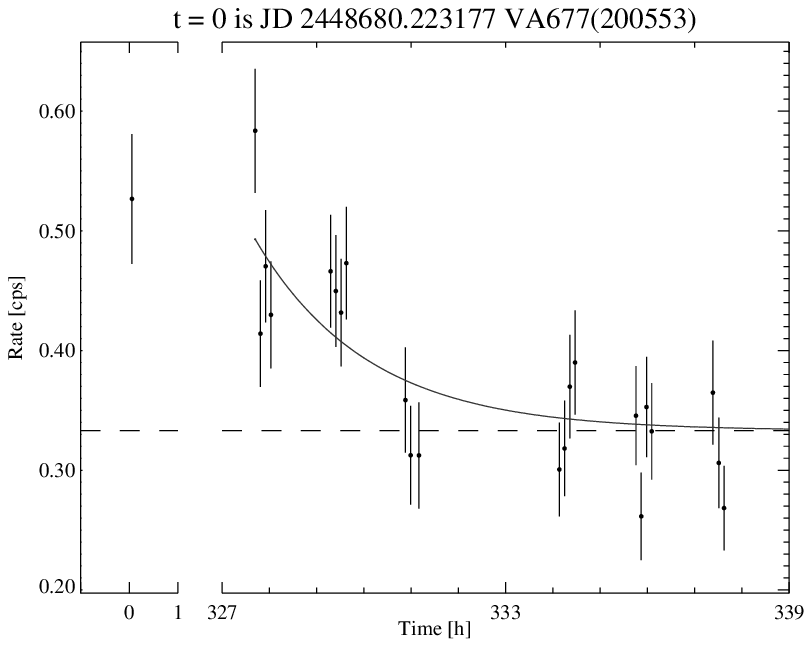}}}
}
\parbox{16cm}{
\parbox{4.5cm}{\resizebox{5.5cm}{!}{\includegraphics{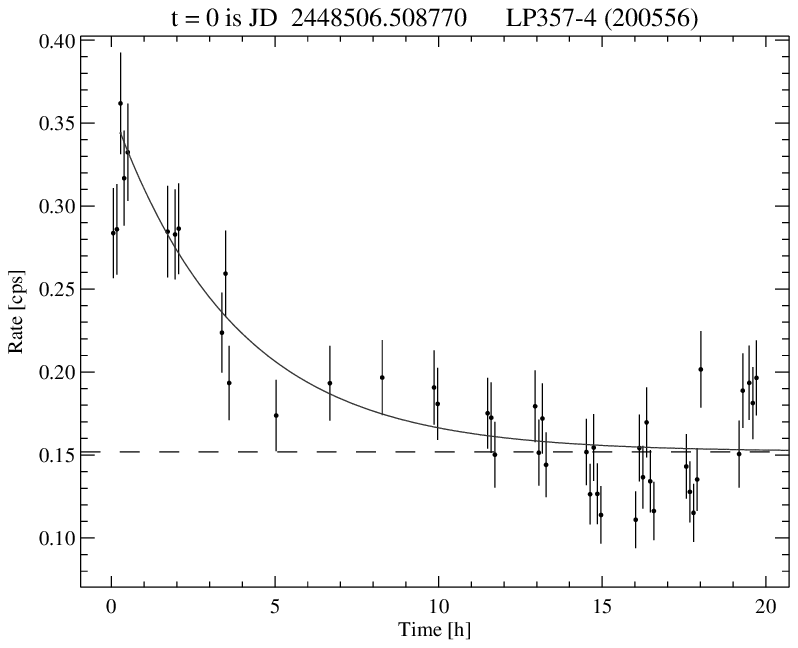}}}
\parbox{1cm}{\hspace*{1.cm}}
\parbox{4.5cm}{\resizebox{5.5cm}{!}{\includegraphics{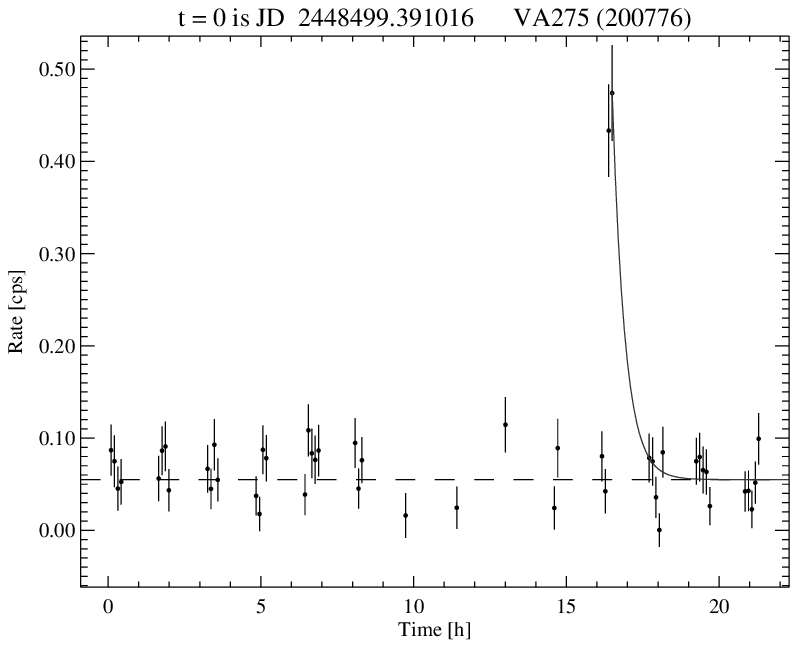}}}
\parbox{1cm}{\hspace*{1.cm}}
\parbox{4.5cm}{\resizebox{5.5cm}{!}{\includegraphics{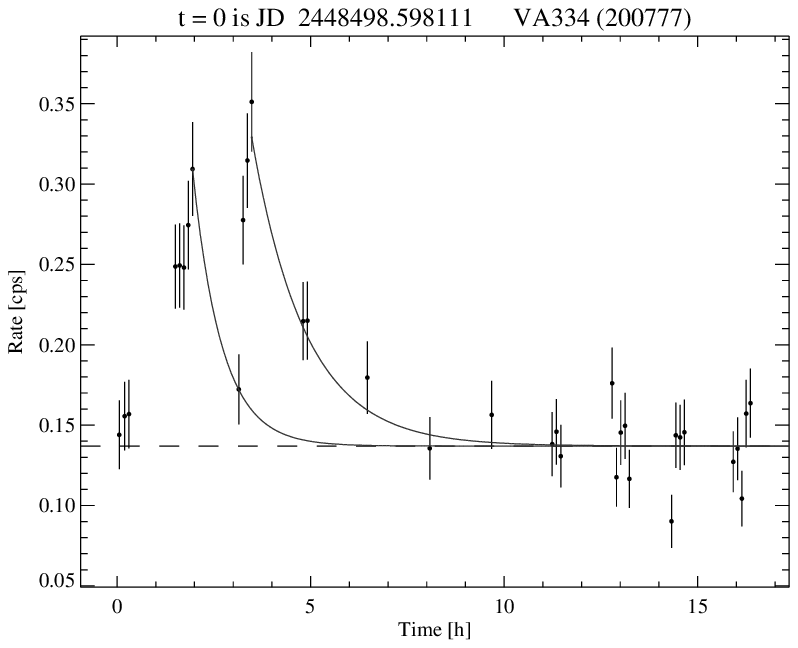}}}
}
\parbox{16cm}{
\parbox{4.5cm}{\resizebox{5.5cm}{!}{\includegraphics{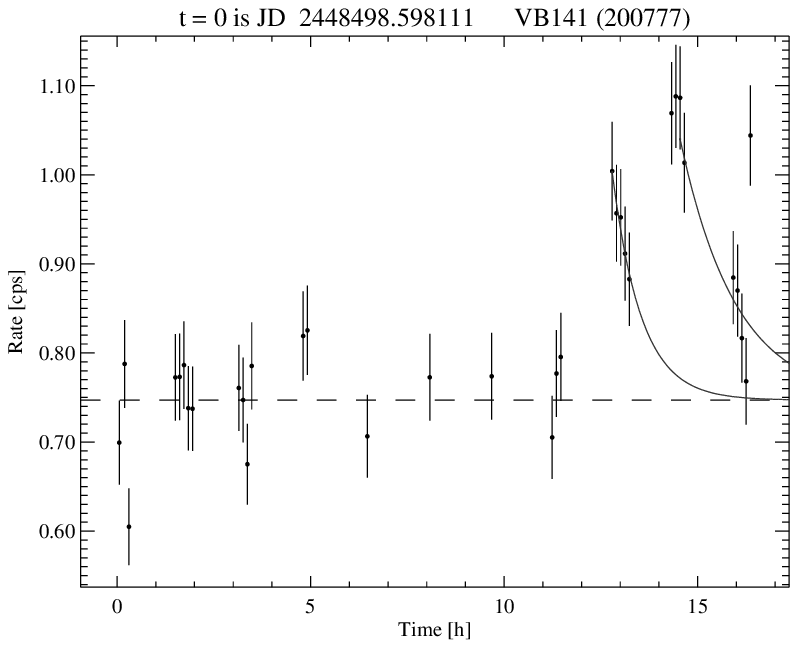}}}
\parbox{1cm}{\hspace*{1.cm}}
\parbox{4.5cm}{\resizebox{5.5cm}{!}{\includegraphics{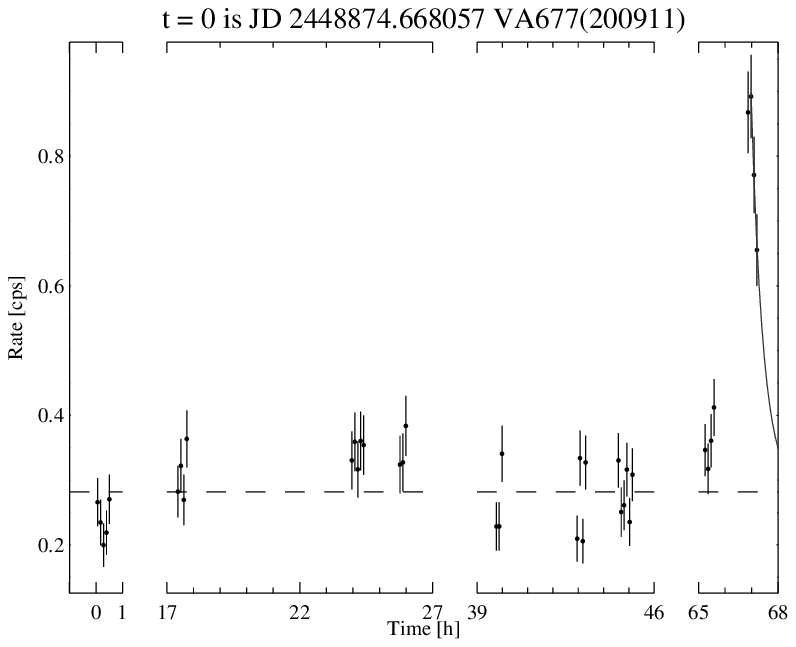}}}
\parbox{1cm}{\hspace*{1.cm}}
\parbox{4.5cm}{\resizebox{5.5cm}{!}{\includegraphics{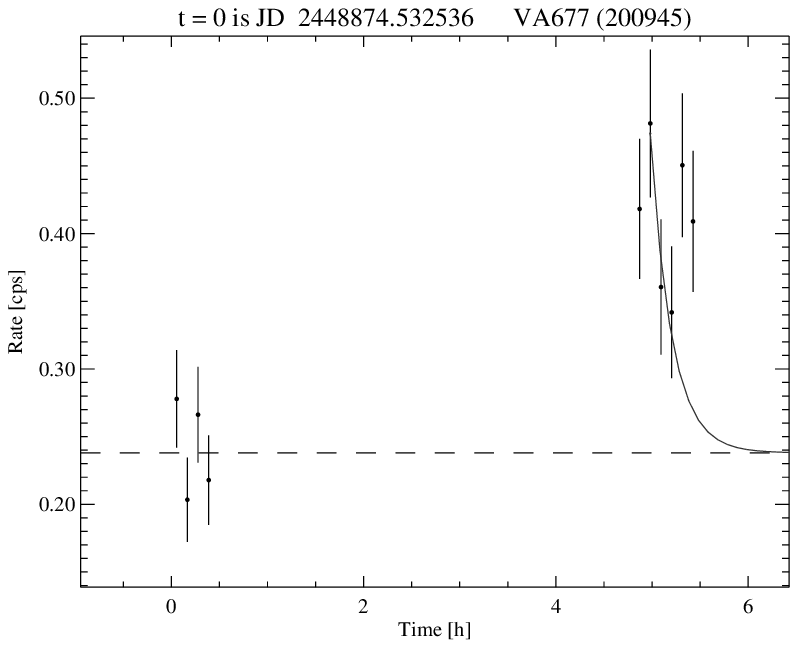}}}
}
\parbox{16cm}{
\parbox{4.5cm}{\resizebox{5.5cm}{!}{\includegraphics{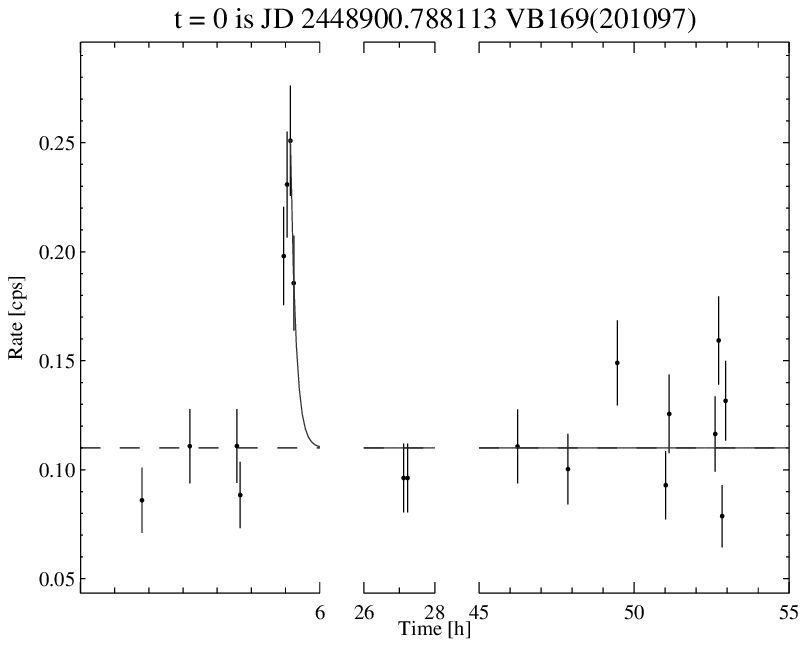}}}
\parbox{1cm}{\hspace*{1.cm}}
\parbox{4.5cm}{\resizebox{5.5cm}{!}{\includegraphics{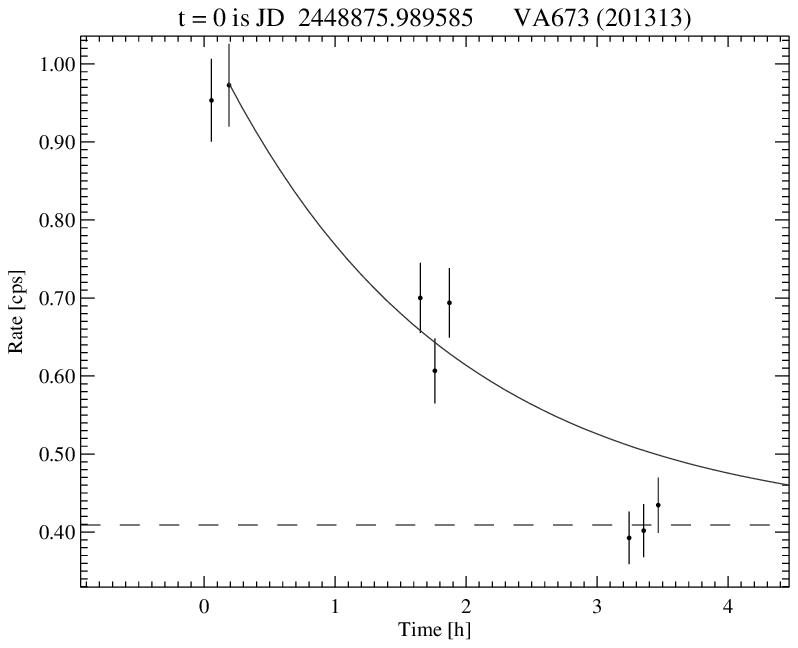}}}
\parbox{1cm}{\hspace*{1.cm}}
\parbox{4.5cm}{\resizebox{5.5cm}{!}{\includegraphics{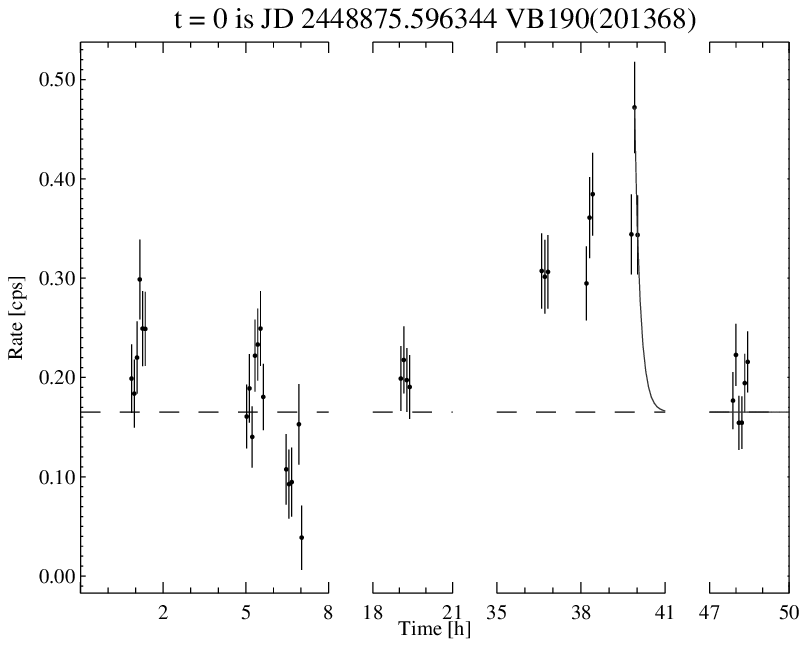}}}
}
\parbox{16cm}{
\parbox{4.5cm}{\resizebox{5.5cm}{!}{\includegraphics{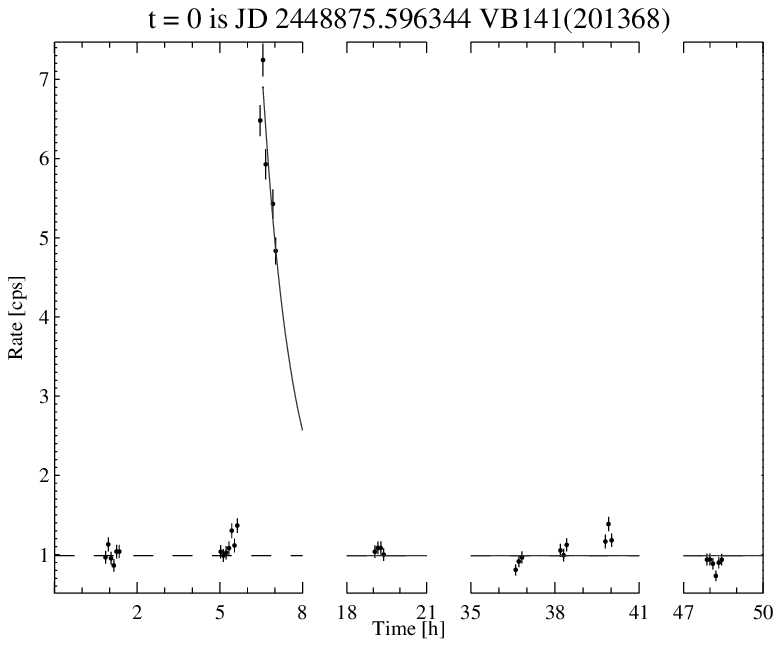}}}
\parbox{1cm}{\hspace*{1.cm}}
\parbox{4.5cm}{\resizebox{5.5cm}{!}{\includegraphics{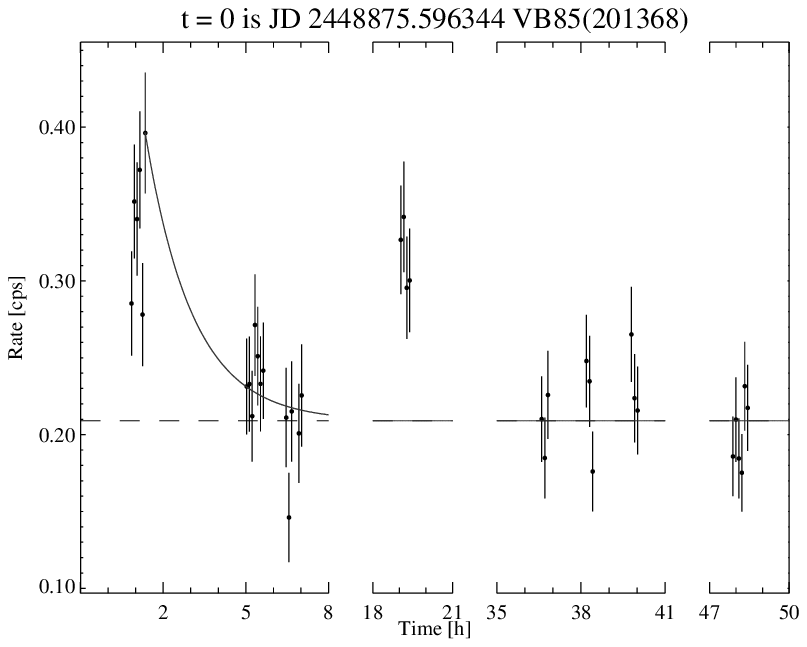}}}
\parbox{1cm}{\hspace*{1.cm}}
\parbox{4.5cm}{\resizebox{5.5cm}{!}{\includegraphics{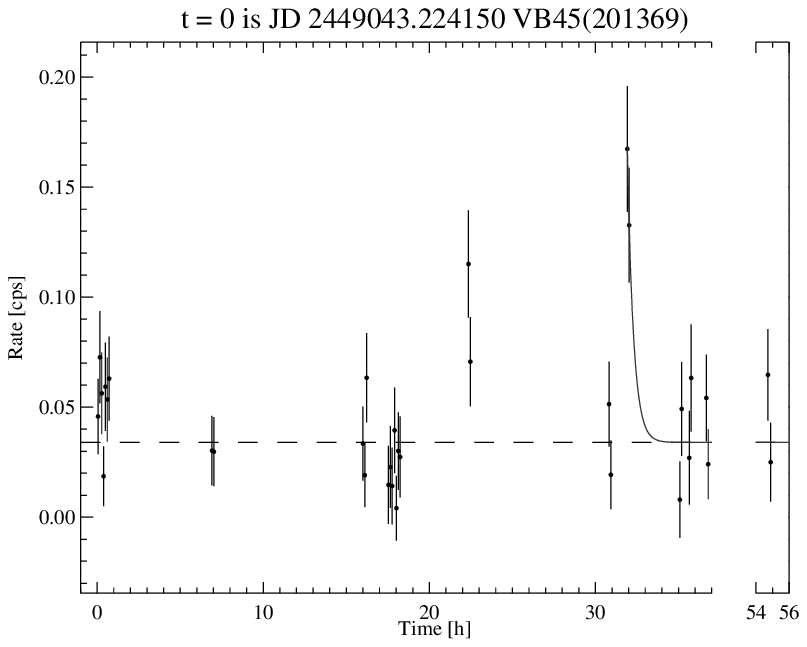}}}
}
\caption{PSPC lightcurves for flares on members of the Hyades cluster
detected by the procedure described in
Sect~\ref{subsect:method_det}. Identification of the X-ray source and {\em
ROSAT} observation request number (in brackets) are given for each source. Binsize is 400\,s, 1\,$\sigma$ uncertainties. The dashed line represents the quiescent count rate. Solid lines are exponential fits to the data points belonging to the flare down to the quiescent emission. Background count rates are shown as upper limits when the background subtracted count rate is below zero.}
\label{fig:lcs_Hya}
\end{center}
\end{figure*}

The lightcurves of all flares have been 
analyzed by fitting an exponential to the decay of each flare. 
In Figs.~\ref{fig:lcs_TTS},~\ref{fig:lcs_Ple},~and~\ref{fig:lcs_Hya} the fit 
function and measured mean quiescent count rate are displayed together with the data. 
In Tables~\ref{tab:det_flares_TTS},~\ref{tab:det_flares_Ple}~and~\ref{tab:det_flares_Hya} we give the result
of the modeling. Column~1 and column~2 contain the stellar identification 
of the X-ray source and the {\em ROSAT} observation request number (ROR). 
The mean quiescent 
count rate is given in column~3, the maximum count rate inferred from the 
exponential fit to the lightcurve in column~4, and the decay 
timescale $\tau_{\rm dec}$ from the fit in column~5. 
For flares with poor data sampling
we did not determine the errors of $\tau_{\rm dec}$.
Column~6 is the estimated rise time of the flare. Due to data gaps 
in most cases no reasonable estimate can be given.
Luminosities are listed in columns~7~and~8:
quiescent luminosity $L_{\rm qui}$, and 
maximum luminosity during the flare $L_{\rm max}$.
We assume that all stars in the system contribute the same level of
 X-ray emission during quiescence, but that only one component flares at
any one time. Therefore, 
for all multiple stars the observed quiescent count rate from 
column~3 has been divided by the number of components 
before the conversion to luminosity and energy.

For the conversion from count rates to luminosities we have used 
the mean {\em ROSAT} PSPC energy-conversion-factor (ECF) 
from \citey{Neuhaeuser95.1}, 
i.e. $ECF=1.1 \cdot 10^{11}\,{\rm cts\,cm^2/erg}$ and the distances given 
in Table~\ref{tab:opt_par}.
In order to eliminate uncertainties in the distance estimate we have 
computed ratios of luminosity (given in column~9). 
Here, $L_{\rm F}$ denotes the luminosity emitted during the 
flare, i.e. $L_{\rm F} = L_{\rm max} - L_{\rm qui}$. 
The total emitted energy during quiescence $E_{\rm qui}$ (column~10) 
and during the flare alone $E_{\rm F}$ (column~11) 
are inferred from the integration of the lightcurve 
between $t_{\rm max}$ and $t_{\rm max} + \tau$. 
The last column gives the reference for flares that have been 
published previously.
In the last two rows of
Tables~\ref{tab:det_flares_TTS},~\ref{tab:det_flares_Ple}~and~\ref{tab:det_flares_Hya}
we have listed the mean and median for each of the given parameters, except
$\tau_{\rm ris}$ which is not well constrained. The means and medians 
have been 
computed with the ASURV Kaplan-Meier estimator (see \cite{Feigelson85.1}), 
taking account of upper/lower limits. Lower limits of $L_{\rm max}$ 
occur when there is doubt
about whether the maximum emission of the flare has been observed (due to
a data gap near the observed maximum). Upper limits for $\tau_{\rm dec}$
occur when the decay is not observed because of a data gap between
maximum and post-flare quiescent count rate. Both luminosity and decay
timescale determine the flare energy, but the limits of
these two parameters carry opposite signs. We therefore consider all values
of $E_{\rm F}$ uncertain where $\tau_{\rm dec}$ or 
$L_{\rm max}$ are a limit (indicated by colons in Tables~\ref{tab:det_flares_TTS},~\ref{tab:det_flares_Ple}~and~\ref{tab:det_flares_Hya}) and have not included them in the 
computation of the mean and median.

\begin{table*}
\caption{Properties of stars in Tau-Aur-Per which flared during at
least one {\em ROSAT} PSPC observation. Spectral types marked with
an asterisk are determined from $B-V$ or $R-I$ given in the Open Cluster
database using the conversion given by
\protect\citey{Schmidt-Kaler82.1}. The meaning of the flags in column
`Multiplicity' is: B for binary, SB1 for single-lined spectroscopic binary,
and SB2 for double-lined spectroscopic binary. 
For one flare on TTSs it is not
clear whether it belongs to DD\,Tau, a cTTS, or CZ\,Tau, a wTTS, because
the spatial resolution of the PSPC is too low to resolve these stars.
DD\,Tau and CZ\,Tau both are binaries. Only flares on stars of spectral type G and later are
analysed in the remainder of this paper.}\label{tab:opt_par}
\begin{tabular}{lrlrrrlrrcr}\hline
Designation   & Distance & \multicolumn{2}{c}{Sp.Type (Ref)} &
               \multicolumn{2}{c}{vsini (Ref)} & Mult & \multicolumn{2}{c}{Binary sep} & \multicolumn{2}{c}{TTS} \\ 
               & \multicolumn{1}{c}{[pc]} &    &   &               \multicolumn{2}{c}{[km/s]} & & [$^{\prime\prime}$] & &               \multicolumn{2}{c}{Type} \\ \hline
\multicolumn{11}{c}{\bf T Tauri Stars} \\ \hline
 LkH$\alpha$270       &  350.0 & K7M0 &  (1)   &	       &     &      &	      &     & C & (1) \\ 
 BPTau                &  140.0 & K7   &  (1)   &	  10.0 & (17) &     &	      &     & C & (1) \\ 
 V410x-ray7           &  140.0 & M1   &  (2)   &	       &     &     &	      &     & W & (2) \\ 
 HD283572             &  140.0 & G5 I &  (1)   &	  75.6 & (18) &     &	      &     & W & (1) \\ 
 DDTau/CZTau          &  140.0 & M1/M1.5 &  (1) &	       &     &  B/B &	 0.57/0.33 & (27) & C/W & (1) \\ 
 L1551-51             &  140.0 & K7   &  (1)   &	  27.0 & (1) &    &        &      & W & (1) \\ 
 RXJ0437.5+1851\,B    &  140.0 & M0.5 &  (3)   &	  10.5 & (3) &  B   &	 4.30 & (28) & W & (3) \\ 
 LkCa19               &  140.0 & K0   &  (1)   &	  18.6 & (17) &     &	      &     & W & (1) \\ 
 LH$\alpha$92         &  300.0 & K0   &  (5)   &	       &     &      &	      &     & C & (1) \\ 
 RXJ0422.1+1934       &  140.0 & M4.5 &  (4)   &	       &     &  B   &	 12.0 & (28) & C & (4) \\ 
 TTau                 &  140.0 & K0   &  (1)   &	  20.7 & (17) &  B  &	 0.71 & (29) & C & (1) \\ 
 RXJ0255.4+2005       &   65.0 & K6   &  (6)   &	  10.0 & (6) &     &	      &  & W & (6) \\ 
 LkH$\alpha$325       &  350.0 & K7M0 &  (1)   &	       &     &  B  &	 11.0 & (30) & W & (1) \\ \hline
\multicolumn{11}{c}{\bf Pleiads} \\ \hline
 hii1384              &  116.0 & A4V  & (7) &	 215.0 & (19) &     &     &     & - &  \\ 
 hii2147              &  116.0 & G9   & (8) &	   6.9 & (20) & SB2 &     & (31) & - &  \\ 
 hii1100              &  116.0 & K3   & (8) &	   5.4 & (20) & B  & 0.78 & (32)  & - &  \\ 
 hii303               &  116.0 & K1.4$^*$ & (9) & 17.4 & (20) & B  & 1.81 & (32)  & - &  \\ 
 hii298               &  116.0 & K1.4$^*$ & (9) &  6.6 & (20) & B  & 5.69 & (32)  & - &  \\ 
 hcg307               &  116.0 &      &     &	  13.0 & (21) &     &      &     & - &  \\ 
 hcg144               &  116.0 &      &     &	       &     &      &      &     & - &  \\ 
 hii2244              &  116.0 & K2.5 & (10) &	  50.0 & (22) &     &	   &     & - &  \\ 
 hcg422               &  116.0 &      &     &	       &     &      &	   &     & - &  \\ 
 hii1516              &  116.0 & K7$^*$   & (11) & 105.0 & (20) &   &	   &     & - &  \\ 
 hcg143               &  116.0 & M2$^*$   & (11) &	 &     &    &	   &     & - &  \\ 
 sk702                &  116.0 &      &     &	       &     &      &	   &     & - &  \\ 
 hii174               &  116.0 & K0.8$^*$ & (9) & 28.0 & (23) &     &      &     & - &  \\ 
 hii191               &  116.0 & K7.7 & (12) &	   9.1 & (20) &     &	   &     & - &  \\ 
 hcg97                &  116.0 &      &     &	       &     &      &	   &     & - &  \\ 
 hii1653              &  116.0 & K6   & (8) &	  21.0 & (8) &      &	   &     & - &  \\ 
 hii345               &  116.0 & G8   & (10) &	  18.9 & (20) &     &	   &     & - &  \\ 
 hcg181               &  116.0 & M1.5$^*$ & (11) &	       &    &      &	      &     & - &  \\ 
 hii253               &  116.0 & G1   & (10) &	  37.0 & (23) &     &	   &     & - &  \\ 
 hii212               &  116.0 & M0.0 & (12) &	  10.0 & (8) &      &	   &     & - &  \\ \hline
\multicolumn{11}{c}{\bf Hyads} \\ \hline
 V471Tau              &   46.8 & K1.2$^*$ & (13) &	       &    &      &	      &     & - &  \\ 
 VB50                 &   44.9 & G1V  & (14) &	       &     & SB?  &      & (33)  & - &  \\ 
 VA677                &   57.0 & K    & (15) &	  24.0 & (24) & SB2 &      & (24)  & - &  \\ 
 LP357-4              &   46.3 & M3   & (15) &	       &     &      &	   &     & - &  \\ 
 VA275                &   46.3 & M2-3 & (15) &	  10.0 & (25) &     &      &     & - &  \\ 
 VB141                &   47.9 & A8$^*$   & (16) &     &      &     &	   &    & - &  \\ 
 VA334                &   41.0 & M0   & (15) &	   6.0 & (24) &     &      &     & - &  \\ 
 VB169                &   46.5 & A6.6$^*$ & (16) &	       &    &      &	      &     & - &  \\ 
 VA673                &   46.3 & M1   & (15) &	       &     & SB   &      & (34)  & - &  \\ 
 VB190                &   46.3 & K    & (15) &	   8.5 & (24) & SB  &	   & (32)  & - &  \\ 
 VB85                 &   41.2 & F5V  & (14) &	  55.0 & (26) &     &      & (35)  & - &  \\ 
 VB45                 &   47.2 & Am   & (14) &	  12.0 & (26) & SB1 &      & (26)  & - &  \\ \hline
\end{tabular}

{\bf Catalogues: } VB - \cite{Bueren52.1}, VA - \cite{Altena69.1}, LP - \cite{Luyten81.1}, hii - \cite{Hertzsprung47.1}, hcg - \cite{Haro82.2}, sk - \cite{Stauffer91.1}

{\bf References: } (1) - Herbig \& Bell 1988, (2) - Strom \& Strom 1994,
(3) - Wichmann et al. (in preparation), (4) - Mart\'\i n \& Magazz\`u 1999, (5) - Herbig
1998, (6) - Hearty et al. 2000, (7) - Mendoza 1956, (8) - Stauffer \& Hartmann 1987,
(9) - Johnson \& Mitchell 1958, (10) - Soderblom et al. 1993a, (11) -
Stauffer 1982, Stauffer 1984, (12) - Prosser et al. 1991, (13) -
priv. comm. between J. Stauffer and E. Weis according to the Open Cluster
data base, (14) - Morgan \& Hiltner 1965, (15) - Pesch 1968, (16) - Morel \& Magnenat
1978, (17) - Hartmann \& Stauffer 1989, (18) - Walter et al. 1987, (19) - Anderson
et al. 1966, (20) - Queloz et al. 1998, (21) - Jones et al. 1996, (22) -
Stauffer et al. 1984, (23) - Soderblom et al. 1993b, (24) - Stauffer et
al. 1997, (25) - Stauffer et al. 1987, (26) - Kraft 1965, (27) - Leinert et
al. 1993, (28) - K\"ohler \& Leinert 1998, (29) - Ghez et al. 1993, (30) -
Cohen \& Kuhi 1979, (31) - Raboud \& Mermilliod 1998, (32) - Bouvier et al
1997, (33) - Griffin et al. 1985, (34) - Stefanik \& Latham 1992, (35) -
Ziskin 1993

\end{table*}

\begin{sidewaystable*}
\caption{Parameters derived from the lightcurves of flares on T Tauri
Stars detected in the {\em ROSAT} PSPC observations from Table~1 (see
Sect.~\ref{subsect:flarepar} for an explanation). The last column gives
the reference for flares that have been presented elsewhere in the
literature: [a] \protect\citey{Strom94.1},
[b] \protect\citey{Preibisch93.1}, but newly reduced here. In the last two
rows we give the mean (determined by taking account of lower/upper limits)
and the median for each parameter.}
\label{tab:det_flares_TTS}
\newcolumntype{d}[1]{D{.}{.}{#1}}
\begin{tabular}{llccr@{}d{2}@{}crrrr@{}d{3}@{}rcrc}\hline
Desig. & \newrule ROR & $I_{\rm qui}$ & $I_{\rm max}$ &
       \multicolumn{3}{c}{$\tau_{\rm dec}$} & \multicolumn{1}{c}{$\tau_{\rm
       ris}$} & \multicolumn{1}{c}{$L_{\rm qui}$} &
       \multicolumn{1}{c}{$L_{\rm max}$} & \multicolumn{3}{c}{$L_{\rm
       F}/L_{\rm qui}$} &       $E_{\rm qui}$ & \multicolumn{1}{c}{$E_{\rm
       F}$} & Notes \\
       &  \newrule   & [cps]           & [cps]           &       \multicolumn{3}{c}{[h]}    & \multicolumn{1}{c}{[h]} & \multicolumn{1}{c}{[erg/s]}       & \multicolumn{1}{c}{[erg/s]}     &    & &                     &  [erg]      &       \multicolumn{1}{c}{[erg]}   &   \\ \hline
 LkH$\alpha$270         	 &	  \newrule 	   180185 	 &	 0.061	 $\pm$	 0.008	 &	 0.256	 $\pm$	 0.008	 &	$<$ &	  8.85	 &	 $	 	  	 	  	   	 $	 &	 $	 <	  7.84	 $	 &	  8.13e+30	 &	$>$  3.41e+31	 &	$>$ &	  3.197	 &	 	  	 &	  2.59e+35	 &	  5.23e+35:	 &	  \\ 
 BPTau           	 &	  \newrule 	 200001-0 	 &	 0.037	 $\pm$	 0.004	 &	 0.247	 $\pm$	 0.034	 &	 &	  2.67	 &	 $	 ^{+	  0.96	 }_{-	  0.92	  } 	 $	 &	 $	 >	  0.22	 $	 &	  7.89e+29	 &	  5.27e+30	 &	 &	  5.676	 &	 $\pm$	  1.110	 &	  7.58e+33	 &	  2.73e+34	 &	  \\ 
 V410x-ray7      	 &	  \newrule 	 200001-0 	 &	 0.050	 $\pm$	 0.011	 &	 0.114	 $\pm$	 0.018	 &	 &	  0.80	 &	 $	 ^{+	  7.22	 }_{-	  0.53	  } 	 $	 &	 $	 >	  0.00	 $	 &	  1.07e+30	 &	 $>$ 2.43e+30	 &	$>$ &	  1.280	 &	 	  	 &	  3.07e+33	 &	 $>$ 2.52e+33:	 &	  \\ 
 HD283572        	 &	  \newrule 	 200001-1 	 &	 0.296	 $\pm$	 0.069	 &	 1.762	 $\pm$	 0.101	 &	 &	  1.78	 &	 $	 ^{+	  0.18	 }_{-	  0.19	  } 	 $	 &	 $	 <	  1.56	 $	 &	  6.31e+30	 &	  3.76e+31	 &	 &	  4.953	 &	 $\pm$	  1.226	 &	  4.04e+34	 &	  1.27e+35	 &	 [a] \\ 
 DDTau/CZTau           	 &	  \newrule 	 200001-1 	 &	 0.004	 $\pm$	 0.005	 &	 0.113	 $\pm$	 0.013	 &	 &	  1.19	 &	 $	 ^{+	  0.28	 }_{-	  0.27	  } 	 $	 &	 $	 <	  2.99	 $	 &	  2.13e+28	 &	 $>$ 2.41e+30	 &	$>$ &	 112.000	 &	 	 	 &	  9.14e+31	 &	$>$  6.24e+33:	 &	 [a] \\ 
 L1551-51        	 &	  \newrule 	   200443 	 &	 0.031	 $\pm$	 0.015	 &	 0.188	 $\pm$	 0.027	 &	$<$ &	  8.83	 &	 $	 	  	 	  	   	 $	 &	 $	 >	  0.48	 $	 &	  6.61e+29	 &	  4.01e+30	 &	 &	  5.065	 &	 $\pm$	  2.645	 &	  2.10e+34	 &	 $>$ 6.74e+34:	 &	  \\ 
 RXJ0437.5+1851 	 &	  \newrule 	   200913 	 &	 0.103	 $\pm$	 0.027	 &	 0.259	 $\pm$	 0.037	 &	 &	  0.49	 &	 $	 ^{+	  0.27	 }_{-	  0.18	  } 	 $	 &	 $	 >	  0.00	 $	 &	  1.10e+30	 &	  5.52e+30	 &	 &	  4.029	 &	 $\pm$	  1.304	 &	  1.94e+33	 &	  3.75e+33	 &	  \\ 
 RXJ0437.5+1851 	 &	  \newrule 	   200913 	 &	 0.103	 $\pm$	 0.027	 &	 0.242	 $\pm$	 0.027	 &	 &	  0.94	 &	 $	 	  	 	  	   	 $	 &	 $	 >	  0.11	 $	 &	  1.10e+30	 &	  5.16e+30	 &	 &	  3.699	 &	 $\pm$	  1.133	 &	  3.72e+33	 &	  6.35e+33	 &	  \\ 
 LkCa19          	 &	  \newrule 	 201278-1 	 &	 0.218	 $\pm$	 0.010	 &	 0.382	 $\pm$	 0.040	 &	 &	  0.22	 &	 $	 ^{+	  0.22	 }_{-	  0.09	  } 	 $	 &	 $	 <	  1.36	 $	 &	  4.65e+30	 &	  8.15e+30	 &	 &	  0.752	 &	 $\pm$	  0.192	 &	  3.68e+33	 &	  1.72e+33	 &	  \\ 
 LH$\alpha$92           	 &	  \newrule 	   201305 	 &	 0.009	 $\pm$	 0.009	 &	 0.699	 $\pm$	 0.029	 &	 &	  1.55	 &	 $	 ^{+	  0.11	 }_{-	  0.12	  } 	 $	 &	 $	 <	  1.50	 $	 &	  8.81e+29	 &	  6.84e+31	 &	 &	 76.667	 &	 $\pm$	 76.741	 &	  4.92e+33	 &	  2.38e+35	 &	[b]  \\ 
 RXJ0422.1+1934  	 &	  \newrule 	   700044 	 &	 0.062	 $\pm$	 0.016	 &	 0.240	 $\pm$	 0.016	 &	$<$ &	  2.15	 &	 $	 	  	 	  	   	 $	 &	 $	 >	  0.43	 $	 &	  6.61e+29	 &	 $>$ 5.12e+30	 &	$>$ &	  6.742	 &	 	  	 &	  5.12e+33	 &	  1.86e+34:	 &	  \\ 
 TTau           	 &	  \newrule 	   700044 	 &	 0.033	 $\pm$	 0.008	 &	 0.139	 $\pm$	 0.008	 &	 &	  0.38	 &	 $	 	  	 	  	   	 $	 &	 $	 >	  0.00	 $	 &	  3.52e+29	 &	 $>$ 2.96e+30	 &	$>$ &	  7.424	 &	 	  	 &	  4.81e+32	 &	 $>$ 1.94e+33:	 &	  \\ 
 TTau           	 &	  \newrule 	   700044 	 &	 0.033	 $\pm$	 0.008	 &	 0.092	 $\pm$	 0.018	 &	 &	  0.61	 &	 $	 ^{+	  5.46	 }_{-	  0.41	  } 	 $	 &	 $	 >	  0.85	 $	 &	  3.52e+29	 &	 $>$ 1.96e+30	 &	$>$ &	  4.576	 &	 	  	 &	  7.73e+32	 &	 $>$ 1.74e+33:	 &	  \\ 
 RXJ0255.4+2005  	 &	  \newrule 	   900138 	 &	 0.043	 $\pm$	 0.025	 &	 0.232	 $\pm$	 0.032	 &	 &	  1.62	 &	 $	 ^{+	  0.29	 }_{-	  0.30	  } 	 $	 &	 $	 <	  1.37	 $	 &	  1.98e+29	 &	  1.07e+30	 &	 &	  4.395	 &	 $\pm$	  2.724	 &	  1.15e+33	 &	  3.19e+33	 &	  \\ 
 LkH$\alpha$325         	 &	  \newrule 	   900193 	 &       0.170	 $\pm$	 0.013	 &	 0.566	 $\pm$	 0.038	 &	 &       0.47	 &	 $	 ^{+	  0.19	 }_{-	  0.16	  } 	 $       &	 $	 >	  0.22	 $	 &	  1.13e+31 &       7.54e+31	 &	 &	  5.659	 &	 $\pm$	  0.627	 &       1.92e+34	 &	  5.59e+34	 & \\ \hline
MEAN   & \newrule & $0.084 \pm 0.021$ & $0.369 \pm 0.105$ &
       \multicolumn{3}{c}{$1.049 \pm 0.194$} & & 2.51e+30 & 2.84e+31 & &
       \multicolumn{2}{c}{$35.501 \pm 13.161$} & 2.48e+34 & 5.79e+34 & \\
MEDIAN & \newrule & 0.047 & 0.241 & \multicolumn{3}{c}{0.792} & & 8.35e+29
       & 6.15e+30 & & \multicolumn{2}{c}{$5.192$} & 3.70e+33 & 6.35e+33 & \\ \hline 
\end{tabular}
\end{sidewaystable*}

\begin{sidewaystable*}
\caption{Parameters derived from the lightcurves of flares detected on
members of the Pleiades cluster in the {\em ROSAT} PSPC observations from
Table~1. The last column gives the reference for flares that
have been presented elsewhere in the literature: [a]
\protect\citey{Gagne95.1}, but newly reduced here. In the last two 
rows we give the mean (determined by taking account
of lower/upper limits) and the median for each parameter.}
\label{tab:det_flares_Ple}
\newcolumntype{d}[1]{D{.}{.}{#1}}
\begin{tabular}{llccr@{}d{2}@{}crrrr@{}d{3}@{}rcrc}\hline
Desig. & \newrule ROR & $I_{\rm qui}$ & $I_{\rm max}$ &
       \multicolumn{3}{c}{$\tau_{\rm dec}$} & \multicolumn{1}{c}{$\tau_{\rm ris}$} & \multicolumn{1}{c}{$L_{\rm qui}$} & \multicolumn{1}{c}{$L_{\rm max}$} & \multicolumn{3}{c}{$L_{\rm F}/L_{\rm qui}$} & $E_{\rm qui}$ & \multicolumn{1}{c}{$E_{\rm F}$} & Notes \\
       & \newrule    & [cps]           & [cps]           &       \multicolumn{3}{c}{[h]}    & \multicolumn{1}{c}{[h]} & \multicolumn{1}{c}{[erg/s]}       & \multicolumn{1}{c}{[erg/s]}     & & &       &  [erg]      & [erg]         &  \\ \hline
 hii2147        	 &	  \newrule 	 200008-2 	 &	 0.130	 $\pm$	 0.017	 &	 1.041	 $\pm$	 0.017	 &	 &	  0.52	 &	 $	 	  	 	  	   	 $	 &	 $	 <	  3.13	 $	 &	  9.52e+29	 &	  1.52e+31	 &	 &	 15.015	 &	 $\pm$	  1.985	 &	  1.78e+33	 &	  1.59e+34	 &	  \\ 
 hii1384         	 &	  \newrule 	 200008-2 	 &	 0.063	 $\pm$	 0.010	 &	 0.211	 $\pm$	 0.027	 &	 &	  0.34	 &	 $	 ^{+	  3.03	 }_{-	  0.19	  } 	 $	 &	 $	 >	  0.22	 $	 &	  9.22e+29	 &	  3.09e+30	 &	 &	  2.349	 &	 $\pm$	  0.590	 &	  1.13e+33	 &	  1.66e+33	 &	  \\ 
 hii298          	 &	  \newrule 	 200068-1 	 &	 0.052	 $\pm$	 0.024	 &	 0.180	 $\pm$	 0.032	 &	 &	  1.28	 &	 $	 ^{+	  0.82	 }_{-	  0.54	  } 	 $	 &	 $	 >	  0.00	 $	 &	  3.81e+29	 &	 $>$ 2.64e+30	 &	$>$ &	  5.923	 &	 	  	 &	  1.75e+33	 &	 $>$ 5.43e+33:	 &	  \\ 
 hii303          	 &	  \newrule 	 200068-1 	 &	 0.085	 $\pm$	 0.027	 &	 0.236	 $\pm$	 0.039	 &	 &	  1.17	 &	 $	 ^{+	  0.81	 }_{-	  1.01	  } 	 $	 &	 $	 	  3.12	 $	 &	  6.22e+29	 &	  3.45e+30	 &	 &	  4.553	 &	 $\pm$	  1.742	 &	  2.62e+33	 &	  5.92e+33	 &	  \\ 
 hcg307          	 &	  \newrule 	 200068-1 	 &	 0.004	 $\pm$	 0.007	 &	 0.163	 $\pm$	 0.025	 &	 &	  0.50	 &	 $	 ^{+	  0.23	 }_{-	  0.24	  } 	 $	 &	 $	  	  0.22	 $	 &	  5.86e+28	 &	  2.39e+30	 &	 &	 39.750	 &	 $\pm$	 69.865	 &	  1.05e+32	 &	  2.64e+33	 &	  \\ 
 hii1100         	 &	  \newrule 	 200068-1 	 &	 0.019	 $\pm$	 0.009	 &	 0.107	 $\pm$	 0.023	 &	 &	  0.83	 &	 $	 ^{+	  0.69	 }_{-	  0.49	  } 	 $	 &	 $	  	  1.48	 $	 &	  1.39e+29	 &	  1.57e+30	 &	 &	 10.263	 &	 $\pm$	  5.452	 &	  4.16e+32	 &	  2.43e+33	 &	 [a] \\ 
 hcg144          	 &	  \newrule 	 200068-1 	 &	 0.018	 $\pm$	 0.016	 &	 0.407	 $\pm$	 0.046	 &	 &	  0.56	 &	 $	 ^{+	  0.14	 }_{-	  0.13	  } 	 $	 &	 $	  	  4.58	 $	 &	  2.64e+29	 &	 $>$ 5.96e+30	 &	$>$ &	 21.611	 &	 	 	 &	  5.31e+32	 &	 $>$ 7.24e+33:	 &	  \\ 
 hii2244         	 &	  \newrule 	   200556 	 &	 0.029	 $\pm$	 0.013	 &	 0.149	 $\pm$	 0.013	 &	 &	  3.49	 &	 $	 	  	 	  	   	 $	 &	 $	 <	  1.64	 $	 &	  4.25e+29	 &	 $>$ 2.18e+30	 &	$>$ &	  4.138	 &	 	  	 &	  5.33e+33	 &	 $>$ 1.39e+34:	 &	  \\ 
 hcg422          	 &	  \newrule 	   200556 	 &	 0.003	 $\pm$	 0.004	 &	 0.041	 $\pm$	 0.009	 &	$<$ &	  1.56	 &	 $	 	 	 	  	   	 $	 &	 $	  	  1.61	 $	 &	  4.39e+28	 &	  6.00e+29	 &	 &	 12.667	 &	 $\pm$	 17.205	 &	  2.47e+32	 &	 $<$ 1.97e+33:	 &	  \\ 
 hii1516         	 &	  \newrule 	   200556 	 &	 0.028	 $\pm$	 0.019	 &	 0.714	 $\pm$	 0.071	 &	 &	  1.22	 &	 $	 ^{+	  0.15	 }_{-	  0.16	  } 	 $	 &	 $	 <	  1.64	 $	 &	  4.10e+29	 &	 $>$ 1.05e+31	 &	$>$ &	 24.500	 &	 	 	 &	  1.80e+33	 &	 $>$ 2.78e+34:	 &	 [a] \\ 
 hcg97           	 &	  \newrule 	   200557 	 &	 0.003	 $\pm$	 0.005	 &	 0.063	 $\pm$	 0.015	 &	 &	  0.88	 &	 $	 ^{+	  0.50	 }_{-	  0.39	  } 	 $	 &	 $	  	  1.44	 $	 &	  4.39e+28	 &	  9.22e+29	 &	 &	 20.000	 &	 $\pm$	 33.747	 &	  1.39e+32	 &	  1.77e+33	 &	 [a] \\ 
 hii1653         	 &	  \newrule 	   200557 	 &	 0.034	 $\pm$	 0.019	 &	 0.176	 $\pm$	 0.039	 &	 &	  0.40	 &	 $	 ^{+	  1.21	 }_{-	  0.22	  } 	 $	 &	 $	 <	  1.16	 $	 &	  4.98e+29	 &	  2.58e+30	 &	 &	  4.176	 &	 $\pm$	  2.660	 &	  7.17e+32	 &	  1.91e+33	 &	  \\ 
 hii174          	 &	  \newrule 	   200557 	 &	 0.038	 $\pm$	 0.014	 &	 0.153	 $\pm$	 0.023	 &	 &	  1.12	 &	 $	 ^{+	  0.30	 }_{-	  0.28	  } 	 $	 &	 $	 <	  3.31	 $	 &	  5.56e+29	 &	  2.24e+30	 &	 &	  3.026	 &	 $\pm$	  1.321	 &	  2.24e+33	 &	  4.28e+33	 &	 [a] \\ 
 hii191          	 &	  \newrule 	   200557 	 &	 0.004	 $\pm$	 0.004	 &	 0.119	 $\pm$	 0.021	 &	 &	  0.99	 &	 $	 ^{+	  0.21	 }_{-	  0.19	  } 	 $	 &	 $	 <	  1.70	 $	 &	  5.86e+28	 &	  1.74e+30	 &	 &	 28.750	 &	 $\pm$	 29.243	 &	  2.09e+32	 &	  3.79e+33	 &	 [a] \\ 
 hii253          	 &	  \newrule 	   200557 	 &	 0.107	 $\pm$	 0.025	 &	 0.213	 $\pm$	 0.030	 &	 &	  1.44	 &	 $	 ^{+	  0.70	 }_{-	  0.56	  } 	 $	 &	 $	 >	  0.00	 $	 &	  1.57e+30	 &	 $>$ 3.12e+30	 &	$>$ &	  0.991	 &	 	  	 &	  8.12e+33	 &	 $>$ 5.07e+33:	 &	  \\ 
 hii212          	 &	  \newrule 	   200557 	 &	 0.005	 $\pm$	 0.005	 &	 0.053	 $\pm$	 0.014	 &	 &	  1.07	 &	 $	 ^{+	  0.68	 }_{-	  0.50	  } 	 $	 &	 $	 <	  1.44	 $	 &	  7.32e+28	 &	$>$  7.76e+29	 &	$>$ &	  9.600	 &	 	 	 &	  2.82e+32	 &	 $>$ 1.72e+33:	 &	 [a] \\ 
 hii345          	 &	  \newrule 	   200557 	 &	 0.055	 $\pm$	 0.013	 &	 0.176	 $\pm$	 0.027	 &	 &	  0.63	 &	 $	 ^{+	  0.39	 }_{-	  0.48	  } 	 $	 &	 $	 <	  1.32	 $	 &	  8.05e+29	 &	  2.58e+30	 &	 &	  2.200	 &	 $\pm$	  0.753	 &	  1.83e+33	 &	  2.53e+33	 &	 [a] \\ 
 hcg181          	 &	  \newrule 	   200557 	 &	 0.009	 $\pm$	 0.007	 &	 0.128	 $\pm$	 0.021	 &	 &	  0.70	 &	 $	 ^{+	  0.32	 }_{-	  0.30	  } 	 $	 &	 $	  	  0.33	 $	 &	  1.32e+29	 &	  1.87e+30	 &	 &	 13.222	 &	 $\pm$	 10.574	 &	  3.32e+32	 &	  2.77e+33	 &	 [a] \\ 
 sk702           	 &	  \newrule 	   200557 	 &	 0.010	 $\pm$	 0.009	 &	 0.138	 $\pm$	 0.009	 &	 &	  0.10	 &	 $	 	  	 	  	   	 $	 &	 $	  	  1.27	 $	 &	  1.46e+29	 &	  2.02e+30	 &	 &	 12.800	 &	 $\pm$	 11.590	 &	  5.27e+31	 &	  4.39e+32	 &	  \\ 
 hcg143          	 &	  \newrule 	   200557 	 &       0.011	 $\pm$	 0.009	 &	 0.217	 $\pm$	 0.030	 &	 &       0.36	 &	 $	 ^{+	  0.23	 }_{-	  0.14	  } 	 $       &	 $	  	  0.11	 $	 &	  1.61e+29	 &       3.18e+30	 &	 &	 18.727	 &	 $\pm$	 15.585	 &       2.09e+32	 &	  2.47e+33	 &	 [a] \\ \hline
MEAN   & \newrule & $0.035 \pm 0.008$ & $0.234 \pm 0.052$ &
       \multicolumn{3}{c}{$0.919 \pm 0.156$} & & 4.13 e+29 & 5.33e+30 & &
       \multicolumn{2}{c}{$16.792 \pm 3.059$} & 1.49e+33 & 3.73e+33 & \\
MEDIAN & \newrule & 0.019 & 0.163 & \multicolumn{3}{c}{0.762} & & 2.64e+29
       & 2.51e+30 & & \multicolumn{2}{c}{$12.996$} & 5.31e+32 & 2.50e+33 & \\ \hline
\end{tabular}
\end{sidewaystable*}

\begin{sidewaystable*}
\caption{Parameters derived from the lightcurves of flares detected on
members of the Hyades cluster in the {\em ROSAT} PSPC observations from
Table~1. No analysis of flares on Hyads are reported in the literature. In 
the last two rows we give the mean (determined by taking account of
lower/upper limits) and the median for each parameter.}
\label{tab:det_flares_Hya}
\newcolumntype{d}[1]{D{.}{.}{#1}}
\begin{tabular}{llccr@{}d{2}@{}rrrrr@{}d{3}@{}rcrc}\hline
Desig. & \newrule ROR & $I_{\rm qui}$ & $I_{\rm max}$ &
       \multicolumn{3}{c}{$\tau_{\rm dec}$} & \multicolumn{1}{c}{$\tau_{\rm ris}$} & \multicolumn{1}{c}{$L_{\rm qui}$} & \multicolumn{1}{c}{$L_{\rm max}$} & \multicolumn{3}{c}{$L_{\rm F}/L_{\rm qui}$} & $E_{\rm qui}$ & \multicolumn{1}{c}{$E_{\rm F}$} & Notes \\
       & \newrule    & [cps]           & [cps]       &       \multicolumn{3}{c}{[h]}    & \multicolumn{1}{c}{[h]} & \multicolumn{1}{c}{[erg/s]}       &       \multicolumn{1}{c}{[erg/s]}     & & &       &  [erg]      & \multicolumn{1}{c}{[erg]}         &  \\ \hline
 V471Tau       	 &	  \newrule 	 200107-0 	 &	 0.796	 $\pm$	 0.062	 &	 1.135	 $\pm$	 0.062	 &	 &	  0.10	 &	 $	 	  	 	  	   	 $	 &	 $	 <	  5.87	 $	 &	  1.90e+30	 &	  2.70e+30	 &	 &	  0.426	 &	 $\pm$	  0.115	 &	  6.83e+32	 &	  1.75e+32	 &	  \\ 
 VB50            	 &	  \newrule 	   200441 	 &	 0.379	 $\pm$	 0.082	 &	 0.975	 $\pm$	 0.101	 &	 &	  0.47	 &	 $	 ^{+	  0.17	 }_{-	  0.14	  } 	 $	 &	 $	  	  0.22	 $	 &	  4.16e+29	 &	  2.14e+30	 &	 &	  4.145	 &	 $\pm$	  1.065	 &	  7.03e+32	 &	  1.41e+33	 &	  \\ 
 VA677           	 &	  \newrule 	   200553 	 &	 0.333	 $\pm$	 0.071	 &	 0.493	 $\pm$	 0.080	 &	 &	  2.38	 &	 $	 ^{+	  1.68	 }_{-	  0.87	  } 	 $	 &	 $	 <	 327.6	 $	 &	  5.89e+29	 &	 $>$ 1.74e+30	 &	$>$ &	  1.961	 &	 	  	 &	  5.04e+33	 &	 $>$ 3.07e+33:	 &	  \\ 
 LP357-4         	 &	  \newrule 	   200556 	 &	 0.152	 $\pm$	 0.029	 &	 0.345	 $\pm$	 0.034	 &	 &	  3.75	 &	 $	 ^{+	  1.75	 }_{-	  1.15	  } 	 $	 &	 $	 >	  0.22	 $	 &	  3.54e+29	 &	  8.05e+29	 &	 &	  1.270	 &	 $\pm$	  0.381	 &	  4.79e+33	 &	  3.83e+33	 &	  \\ 
 VA275           	 &	  \newrule 	   200776 	 &	 0.055	 $\pm$	 0.028	 &	 0.474	 $\pm$	 0.028	 &	$<$ &	  0.42	 &	 $	 	  	 	  	   	 $	 &	 $	 <	  0.22	 $	 &	  1.28e+29	 &	  1.11e+30	 &	 &	  7.618	 &	 $\pm$	  3.945	 &	  1.94e+32	 &	 $<$ 9.30e+32:	 &	  \\ 
 VB141           	 &	  \newrule 	   200777 	 &	 0.747	 $\pm$	 0.055	 &	 1.040	 $\pm$	 0.078	 &	 &	  1.44	 &	 $	 ^{+	  0.76	 }_{-	  0.53	  } 	 $	 &	 $	 <	  1.20	 $	 &	  1.86e+30	 &	 $>$ 2.60e+30	 &	$>$ &	  0.392	 &	 	  	 &	  9.67e+33	 &	 $>$ 2.39e+33:	 &	  \\ 
 VA334           	 &	  \newrule 	   200777 	 &	 0.137	 $\pm$	 0.021	 &	 0.329	 $\pm$	 0.021	 &	$<$ &	  1.39	 &	 $	 	  	 	  	   	 $	 &	 $	  	  0.33	 $	 &	  2.51e+29	 &	 $>$ 6.02e+29	 &	$>$ &	  1.401	 &	 	  	 &	  1.25e+33	 &	  1.11e+33:	 &	  \\ 
 VB141           	 &	  \newrule 	   200777 	 &	 0.747	 $\pm$	 0.055	 &	 1.003	 $\pm$	 0.075	 &	 &	  0.75	 &	 $	 ^{+	  0.95	 }_{-	  0.37	  } 	 $	 &	 $	 <	  1.32	 $	 &	  1.86e+30	 &	  2.50e+30	 &	 &	  0.343	 &	 $\pm$	  0.127	 &	  5.03e+33	 &	  1.09e+33	 &	  \\ 
 VA334           	 &	  \newrule 	   200777 	 &	 0.137	 $\pm$	 0.021	 &	 0.309	 $\pm$	 0.021	 &	 &	  0.75	 &	 $	 	  	 	  	   	 $	 &	 $	 <	  1.64	 $	 &	  2.51e+29	 &	  5.65e+29	 &	 &	  1.255	 &	 $\pm$	  0.290	 &	  6.76e+32	 &	  5.41e+32	 &	  \\ 
 VA677           	 &	  \newrule 	   200911 	 &	 0.282	 $\pm$	 0.056	 &	 0.896	 $\pm$	 0.083	 &	 &	  0.46	 &	 $	 ^{+	  0.26	 }_{-	  0.16	  } 	 $	 &	 $	 <	  1.61	 $	 &	  4.98e+29	 &	  3.17e+30	 &	 &	  5.355	 &	 $\pm$	  1.232	 &	  8.25e+32	 &	  2.26e+33	 &	  \\ 
 VA677           	 &	  \newrule 	   200945 	 &	 0.238	 $\pm$	 0.036	 &	 0.474	 $\pm$	 0.073	 &	 &	  0.22	 &	 $	 ^{+	  0.34	 }_{-	  0.11	  } 	 $	 &	 $	 <	  4.59	 $	 &	  4.21e+29	 &	 $>$ 1.68e+30	 &	$>$ &	  2.983	 &	 	  	 &	  3.33e+32	 &	 $>$ 4.26e+32:	 &	  \\ 
 VB169           	 &	  \newrule 	   201097 	 &	 0.110	 $\pm$	 0.025	 &	 0.251	 $\pm$	 0.025	 &	 &	  0.18	 &	 $	 	  	 	  	   	 $	 &	 $	 <	  1.67	 $	 &	  2.59e+29	 &	  5.90e+29	 &	 &	  1.282	 &	 $\pm$	  0.434	 &	  1.68e+32	 &	  1.35e+32	 &	  \\ 
 VA673           	 &	  \newrule 	   201313 	 &	 0.409	 $\pm$	 0.022	 &	 0.975	 $\pm$	 0.066	 &	 &	  1.78	 &	 $	 ^{+	  0.48	 }_{-	  0.35	  } 	 $	 &	 $	 >	  0.13	 $	 &	  4.77e+29	 &	 $>$ 2.27e+30	 &	$>$ &	  3.768	 &	 	  	 &	  3.06e+33	 &	 $>$ 5.35e+33:	 &	  \\ 
 VB190           	 &	  \newrule 	   201368 	 &	 0.165	 $\pm$	 0.061	 &	 0.472	 $\pm$	 0.061	 &	 &	  0.20	 &	 $	 	  	 	  	   	 $	 &	 $	 <	 20.52	 $	 &	  1.92e+29	 &	  1.10e+30	 &	 &	  4.721	 &	 $\pm$	  1.931	 &	  1.39e+32	 &	  3.33e+32	 &	  \\ 
 VB141           	 &	  \newrule 	   201368 	 &	 0.985	 $\pm$	 0.136	 &	 6.897	 $\pm$	 0.183	 &	 &	  1.21	 &	 $	 ^{+	  0.16	 }_{-	  0.18	  } 	 $	 &	 $	 <	  1.14	 $	 &	  2.46e+30	 &	  1.72e+31	 &	 &	  6.002	 &	 $\pm$	  0.860	 &	  1.07e+34	 &	  4.08e+34	 &	  \\ 
 VB85            	 &	  \newrule 	   201368 	 &	 0.209	 $\pm$	 0.027	 &	 0.396	 $\pm$	 0.027	 &	$<$ &	  1.91	 &	 $	 	  	 	  	   	 $	 &	 $	 >	  0.55	 $	 &	  3.86e+29	 &	 $>$ 7.31e+29	 &	$>$ &	  0.895	 &	 	  	 &	  2.65e+33	 &	  1.50e+33:	 &	  \\ 
 VB45            	 &	  \newrule 	   201369 	 &       0.034	 $\pm$	 0.020	 &	 0.167	 $\pm$	 0.020	 &       &	  0.37	 &	 $	 	  	 	  	         $	 &	 $	 <	  0.98	 $	 &	  4.12e+28       &	 $>$ 4.05e+29	 &	$>$ &	  8.824	 &	        	 &	  5.49e+31	 &	 $>$ 2.72e+32:	 &       \\ \hline
MEAN   & \newrule & $0.348 \pm 0.069$ & $0.978 \pm 0.367$ &
       \multicolumn{3}{c}{$0.902 \pm 0.234$} & & 7.26e+29 & 4.24e+30 & &
       \multicolumn{2}{c}{$4.433 \pm 0.767$} & 2.71e+33 & 5.62e+33 \\
MEDIAN & \newrule & 0.223 & 0.473 & \multicolumn{3}{c}{0.463} & & 4.01e+29
       & 2.34e+30 & & \multicolumn{2}{c}{$4.468$} & 7.64e+32 & 8.17e+32 \\ \hline
\end{tabular}
\end{sidewaystable*}

\section{Observational selection effects}\label{sect:bias}

It is the purpose of this paper to compare the flare activity of
different stars, and thus some attention has to be drawn to observational 
selection effects. In this section, 
we will discuss how observational restrictions
influence the search for flares. At several points during the data
analysis, we are confronted with the problem of finding a
representation of the data which is free from these biases.

\begin{figure}
\begin{center}
\resizebox{9cm}{!}{\includegraphics{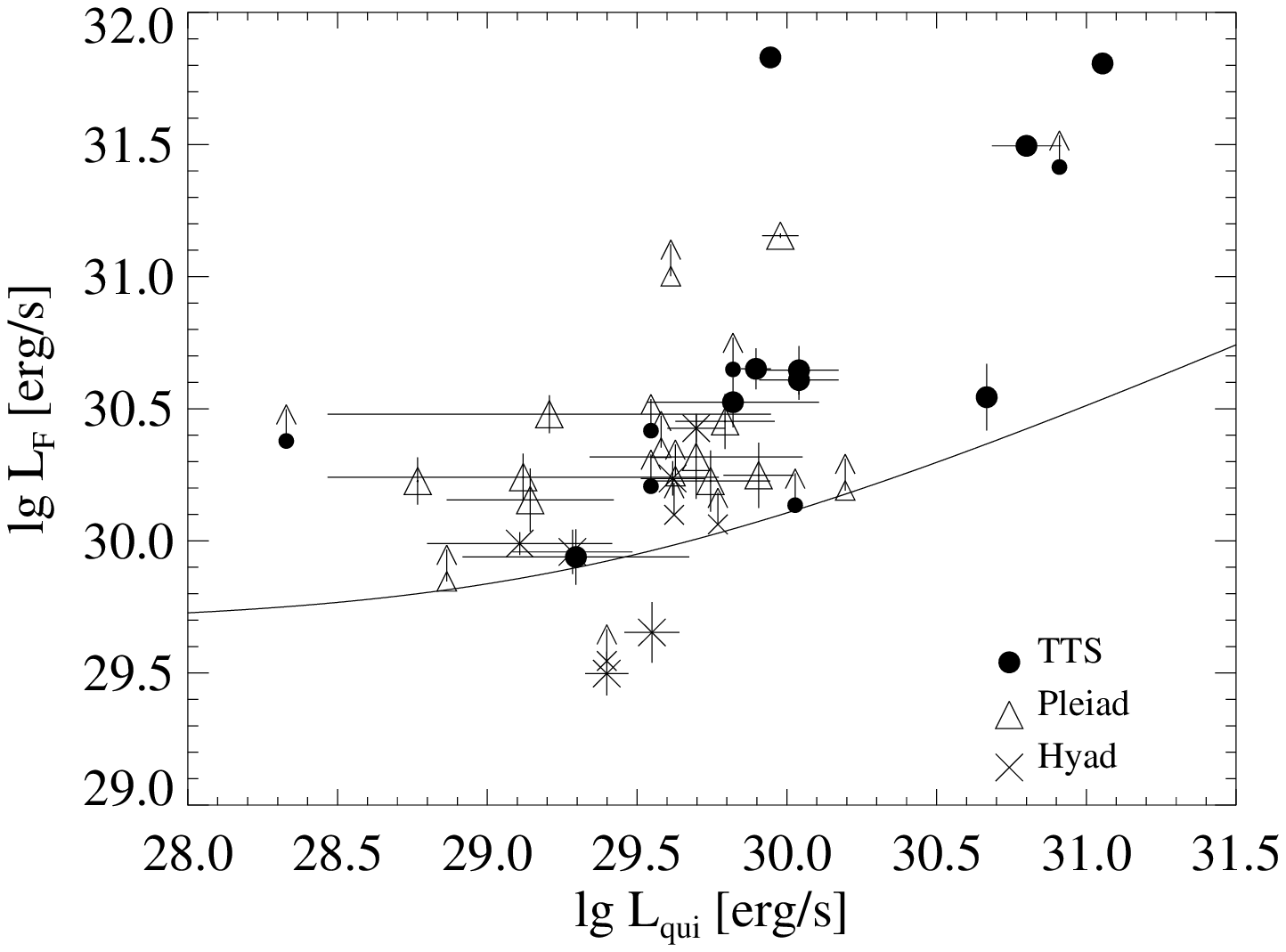}}
\caption{Correlation between quiescent and flare luminosity. 
TTSs, Pleiads, and Hyads are represented by different plotting symbols: TTS
-- circles, Pleiads -- triangles, Hyads -- crosses. The solid
curve is the minimum flare strength needed for detection of a flare with
given $L_{\rm qui}$ if the star is at a distance of 140\,pc. The data
points below that curve represent no contradiction to the calculated
threshold because the detection threshold increases with stellar distance,
i.e. apparent brightness of the stars, and
therefore Pleiads and Hyads may show flares with smaller $L_{\rm F}$ for
given $L_{\rm qui}$.}
\label{fig:Lf_Lq}
\end{center}
\end{figure}

The major difficulty with the statistical evaluation of flares on different
stars is that
the sensitivity of the flare detection process depends on the measured
(quiescent) count rate, which determines the signal-to-noise (S/N), 
and hence on the distance to the star. The 
observational bias consists in the fact 
that for bright stars ($L_{\rm qui}$ large) the minimum luminosity 
$L_{\rm F}$ of a detectable flare is higher than for a faint star. 
The result is, among others, that at first hand it can not be decided
whether any observed correlation between $L_{\rm qui}$ and $L_{\rm F}$
is real or produced by this effect.
In Fig.~\ref{fig:Lf_Lq} we have plotted the flare luminosity 
$L_{\rm F}$ against the quiescent luminosity $L_{\rm qui}$.

\begin{figure}
\begin{center}
\resizebox{8.5cm}{!}{\includegraphics{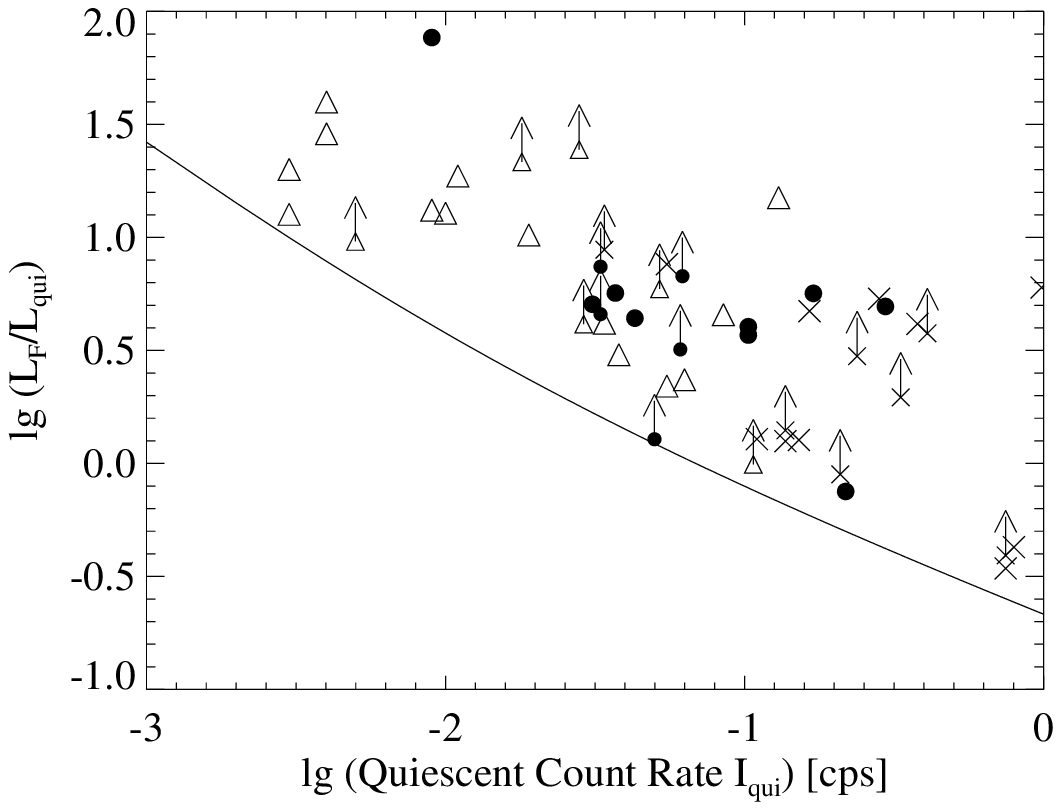}}
\caption{Relative flare strength, $L_{\rm F}/L_{\rm qui}$,
as a function of the quiescent count rate
$I_{\rm qui}$ in double logarithmic scale. 
Assuming that the flare is characterized by a rise
in count rate of 3\,$\sigma$ above the mean during a single 400\,s time
bin, the solid curve gives the minimum values of $L_{\rm F}/L_{\rm qui}$
for which a flare will be detected on stars with given $I_{\rm qui}$. The
observed values consistently lie above this line. The meaning of the
plotting symbols is the same as in Fig.~\ref{fig:Lf_Lq}. For clarity we
have omitted error bars.}
\label{fig:det_accuracy}
\end{center}
\end{figure}

The contribution of the observational bias to this correlation 
can be estimated as follows:
For each quiescent count rate $I_{\rm qui}$ we can determine the minimum
strength $L_{\rm F}/L_{\rm qui}$ needed for a flare to be detected, if we
assume that a flare is found whenever there is a rise in count rate of at 
least 3\,$\sigma$ within one 400\,s time bin. (In our actual flare search we
were even more conservative; see Sect.~\ref{sect:detect}.) 
Hypothetical events of that kind obey
a detection threshold curve for $L_{\rm F}/L_{\rm qui}$ as shown in 
Fig.~\ref{fig:det_accuracy}. 
As mentioned above, the minimum flare luminosity needed for detection of 
a flare becomes
larger with increasing quiescent brightness. In contrast, 
the required luminosity ratio, i.e. the relative strength of the events,
decreases when $I_{\rm qui}$ increases. 
Note also, that the curve in Fig.~\ref{fig:det_accuracy}
is distance independent. But the relation between $L_{\rm qui}$ and
the corresponding minimum $L_{\rm F}$ of a detectable flare differs for 
stars at different distances. In Fig.~\ref{fig:Lf_Lq} we have overplotted
the theoretical threshold for detection of a flare on a star
at 140\,pc distance. Note, that the slope of the data 
in Fig.~\ref{fig:Lf_Lq} is somewhat steeper than the increase of the threshold 
imposed by the S/N. This seems to indicate an intrinsic correlation between
quiescent and flare luminosity. We have subtracted the theoretical threshold 
value for $L_{\rm F}$ from the observed flare luminosity for each of the
stars from Fig.~\ref{fig:Lf_Lq}. Correlation tests for the difference
between
threshold and observed value for $L_{\rm F}$,
$(L_{\rm F,theo} - L_{\rm F,obs})$, with 
$L_{\rm qui}$ show that the correlation is of low
significance, $\alpha$=0.05.
The data points below the theoretical curve are all Pleiads or Hyads. They 
do not contradict the threshold curve,
since Pleiades and Hyades stars are closer than
140\,pc and therefore have a lower flare detection threshold.


\section{Statistical comparison of the flaring stars}\label{sect:statcomp}

We present now a statistical analysis of the X-ray flares from
Tables~\ref{tab:det_flares_TTS},~\ref{tab:det_flares_Ple}~and~\ref{tab:det_flares_Hya}. 
A detailed discussion of the (quiescent) X-ray 
properties of all detected and undetected 
stars is postponed to a later paper (Stelzer et al., in preparation).

In this section, different
flare parameters will be checked for dependence on age, circumstellar
environment, and rotation rate 
(Sect.~\ref{subsect:lumfunct},~\ref{subsect:cw},~\ref{subsect:corrvsini}.)
to see whether any of these properties has an effect on the
characteristic luminosity and time scales of coronal activity.
For the statistical comparison of the flaring stars the 
ASURV package version 1.2 (\cite{Feigelson85.1}) was used. 

First we compare the flaring populations of TTSs, Pleiads, and Hyads
concerning their effective temperatures.
We have converted spectral types to effective temperatures
using the conversion given in \citey{Kenyon95.1}
for PMS stars earlier than M0, and \citey{Luhman99.1} for PMS M-type stars
intermediate between dwarfs and giants. For Pleiades and Hyades stars
we have used the conversion of 
\citey{Schmidt-Kaler82.1}.
We have applied two-sample tests to each pair of $T_{\rm
eff}$-distributions
to reveal possible differences between flaring stars of the three
groups. Henceforth, we denote the probability that the distributions
are similar by $\alpha$. In all but one of the comparisons we found
$\alpha > 0.2$, and therefore no significant differences in $T_{\rm eff}$.
The exception is the logrank test between TTSs and Hyads where
$\alpha = 0.03$.

Most flares occurred on G, K and M stars. However, some events were observed 
on A  and F stars. Stars of intermediate spectral type, 
lacking both a convection-driven dynamo and a strong stellar wind, 
seem to have no efficient mechanism to
generate X-ray flares. Therefore, it is often assumed that X-ray emission 
apparently seen on A or B stars, can be attributed to an (unknown) late-type 
companion (see e.g. \cite{Stauffer94.1}, \cite{Gagne94.1}, 
\cite{Panzera99.1}). 
The same arguments can be applied to explain X-ray
flares on these stars. 
In any case, the emission mechanism of early-type stars is
different from that of late-type stars. 
From the sharp onset of rotation-activity relations in dwarf stars
\citey{Walter83.1} has argued that the onset of solar-like dynamo
activity occurs abruptly at about spectral type F5.
To ensure that no stars with X-ray generation mechanisms other than
stellar dynamos are included, 
we have excluded the stars of spectral type F and earlier 
from the statistical analysis presented in this paper,
i.e. we have restricted the flare sample to events on G, K, and M stars.
This limitation provides samples which have similar
$T_{\rm eff}$ distributions, i.e. $\alpha > 0.2$ also for the
two-sample test between TTSs and Hyads (see Table~\ref{tab:teff} for the
detailed results). This justifies to 
combine all flaring late-type stars for the statistical analysis.
In the following the stellar sample is restricted to G, K, and M stars.
If not explicitly mentioned the two flares on known white-dwarf systems 
(on V471\,Tau and VA\,673) are excluded from the sample, 
since the white dwarf could be responsible for the X-ray event instead of
its late-type companion. 
\begin{table}
\caption{Results of two-sample tests between each pair of stellar samples 
(TTSs,
Pleiads, and Hyads) with respect to the effective temperature $T_{\rm eff}$
of the flaring star. Only flares on G, K, or M stars have been admitted. The analysis was performed with the ASURV package. Next
to the mean and median of $T_{\rm eff}$ we give the probability that the 
null hypothesis of two distributions being the same is true derived
from  Gehan's generalized Wilcoxon test and the logrank test. The values suggest that there is no
significant difference between the spectral types of the flaring TTSs,
Pleiads, and Hyads and justify to combine G, K, and M stars for the
comparison of flares from these different stellar groups.}
\label{tab:teff}
\begin{center}
\begin{tabular}{lccccc} \hline
\multicolumn{6}{c}{$\lg{T_{\rm eff}}$} \\ \hline
Sample & Sample & MEAN & MED. & Prob. & Prob. \\
Name   & Size   &      &     & GW (HV) & logrank \\ \hline
& & & & 0.395 & 0.525 \\
TTSs & 15 & 3.63 $\pm$ 0.08 & 3.600 & & \\
Pleiads & 14 & 3.66 $\pm$ 0.07 & 3.675 & & \\ \hline
& & & & 0.897 & 0.972 \\
TTSs & 15 & (see above) & & & \\
Hyads & 11 & 3.63 $\pm$ 0.07 & 3.595 & & \\ \hline
& & & & 0.212 & 0.401 \\
Pleiads & 14 & (see above) & & & \\
Hyads & 11 & (see above) & & & \\ \hline
\end{tabular}
\end{center}
\end{table}

\subsection{Flare frequency of MS stars and spectral type}\label{subsect:sptypes}

It is interesting to ask whether the depth of the
convection zone has any influence on the occurrence of surface
flares. Since the relative size of the convection zone increases for later
spectral types, the
distribution of flares onto stars of different spectral types may help to 
solve this question. 
What we really want to check
is whether the flare frequency depends on stellar mass, which corresponds
to spectral type on the MS. Because PMS stars still evolve through the 
Hertzsprung-Russell diagram (HRD),
i.e. change their spectral type, we exclude the TTSs from this part of the
analysis. Flaring Pleiades and Hyades stars 
are combined to increase the sample size.

\begin{figure}
\begin{center}
\resizebox{9cm}{!}{\includegraphics{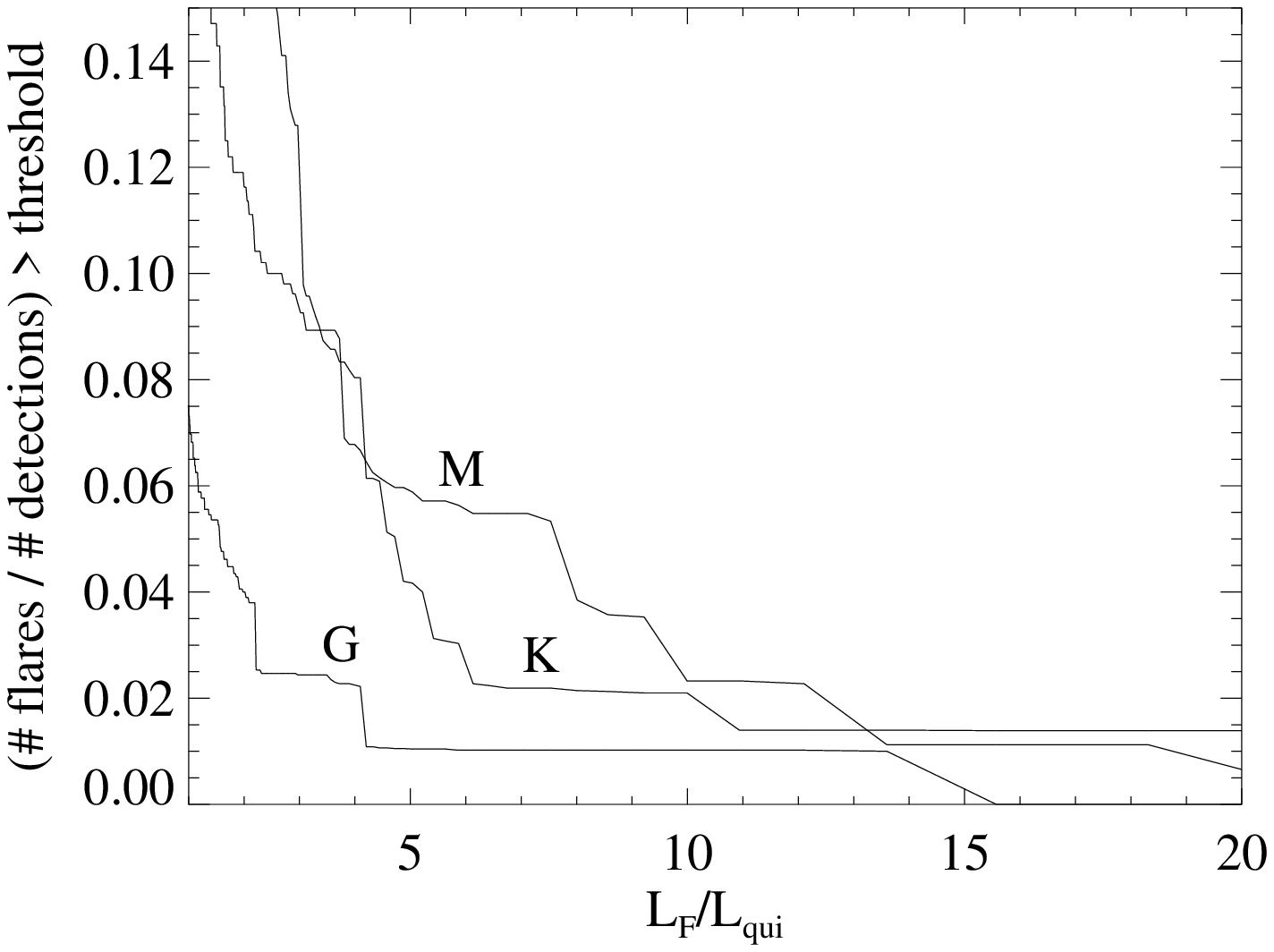}}
\caption{Fraction of flares among detected stars above a given threshold
$L_{\rm F}/L_{\rm qui}$ for different spectral types. Shown are G
stars (dashed line), K stars (solid line) and M stars (dotted
line). Pleiades and Hyades stars have been combined. TTSs are not
considered since they are on the PMS where the relation between the stellar
interior structure and the X-ray emission might be different. 
Throughout the examined range of $L_{\rm F}/L_{\rm qui}$ the number
of flares per detected stars is lowest for spectral type G.}
\label{fig:detsens_spt}
\end{center}
\end{figure}

We have studied the spectral type distribution of flares 
by comparing the number of flares on stars of a certain spectral type
to the total number of detected stars of that spectral type. 
The detection sensitivity for
flares of a given strength $L_{\rm F}/L_{\rm qui}$ is different for each
star because it depends on the level of quiescent emission $I_{\rm qui}$ 
(see Sect.~\ref{sect:bias}). $I_{\rm qui}$,
depends on the spectral type of the star. For this reason, a 
simple comparison between numbers of flares and numbers of detected stars
of each spectral type would be misleading.
The observational bias can, however, be eliminated if the 
numbers (of flares and detections) are evaluated above a certain 
threshold $L_{\rm F}/L_{\rm qui}$. 
We compare the number of flares with measured luminosity ratio above 
a critical value $({L_{\rm F}/L_{\rm qui}})_{\rm crit}$ 
to the number of detected stars 
for which $I_{\rm qui}$ exceeds the minimum value needed for detection
of a flare of that critical strength. We have compiled these numbers 
for a reasonable range of values 
$L_{\rm F}/L_{\rm qui}$, and show the result 
in Fig.~\ref{fig:detsens_spt}. Plotted are the number of flares exceeding
$L_{\rm F}/L_{\rm qui}$ divided by the number of detected stars that are
bright enough for detection of flares with that value of 
$L_{\rm F}/L_{\rm qui}$.
G stars clearly show the smallest rate of events throughout all of
the observed range of flare strengths. 

\subsection{Age of flaring stars (Luminosity functions)}\label{subsect:lumfunct}

To study how the flare activity of young late-type stars 
evolves with stellar age 
we have computed luminosity distribution functions (LDF) and performed 
two-sample tests for three subsamples of stars: TTSs, Pleiads, and Hyads. 

\begin{figure}
\begin{center}
\rotatebox{270}{\resizebox{7cm}{!}{\includegraphics{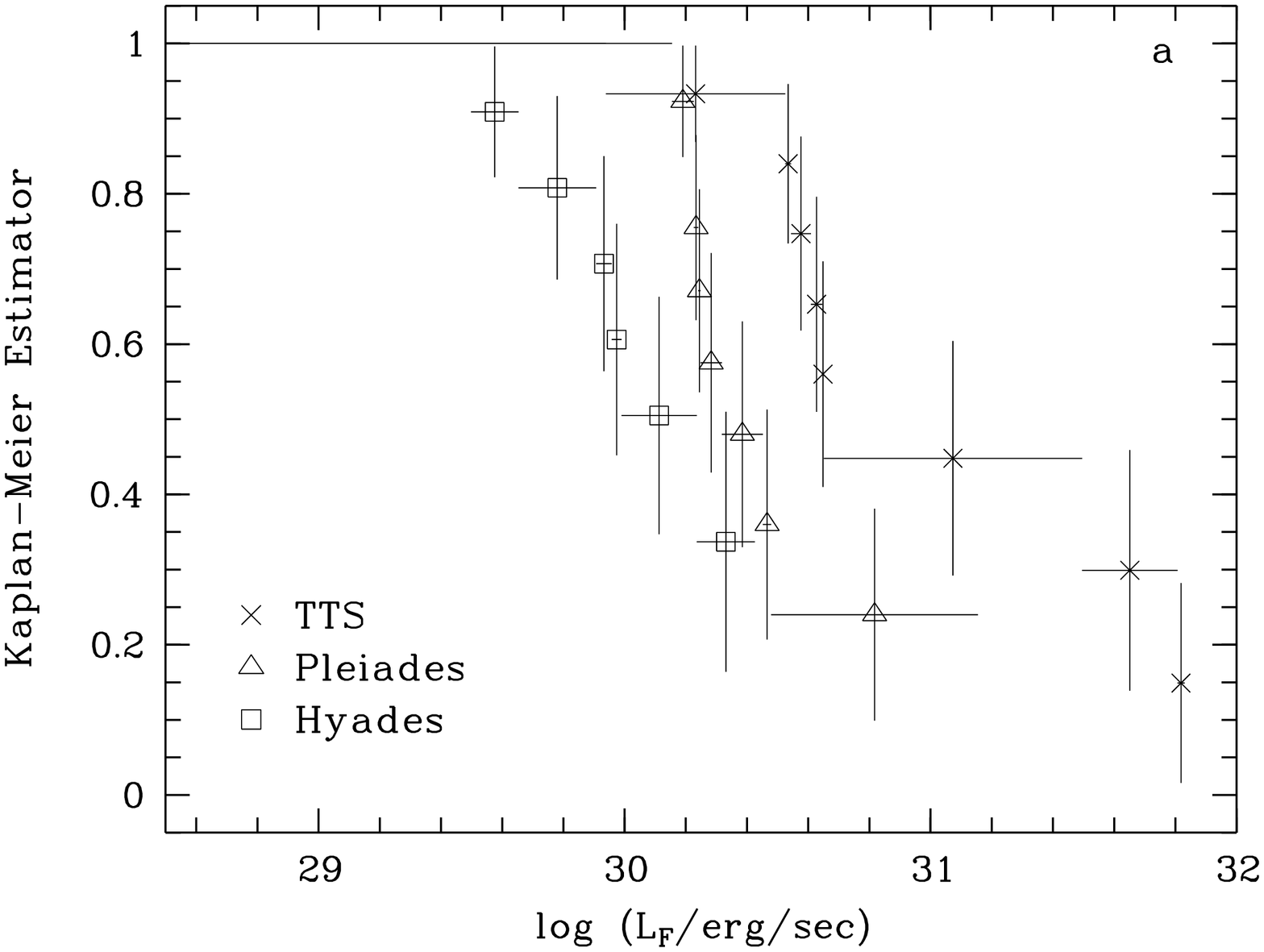}}}
\rotatebox{270}{\resizebox{7cm}{!}{\includegraphics{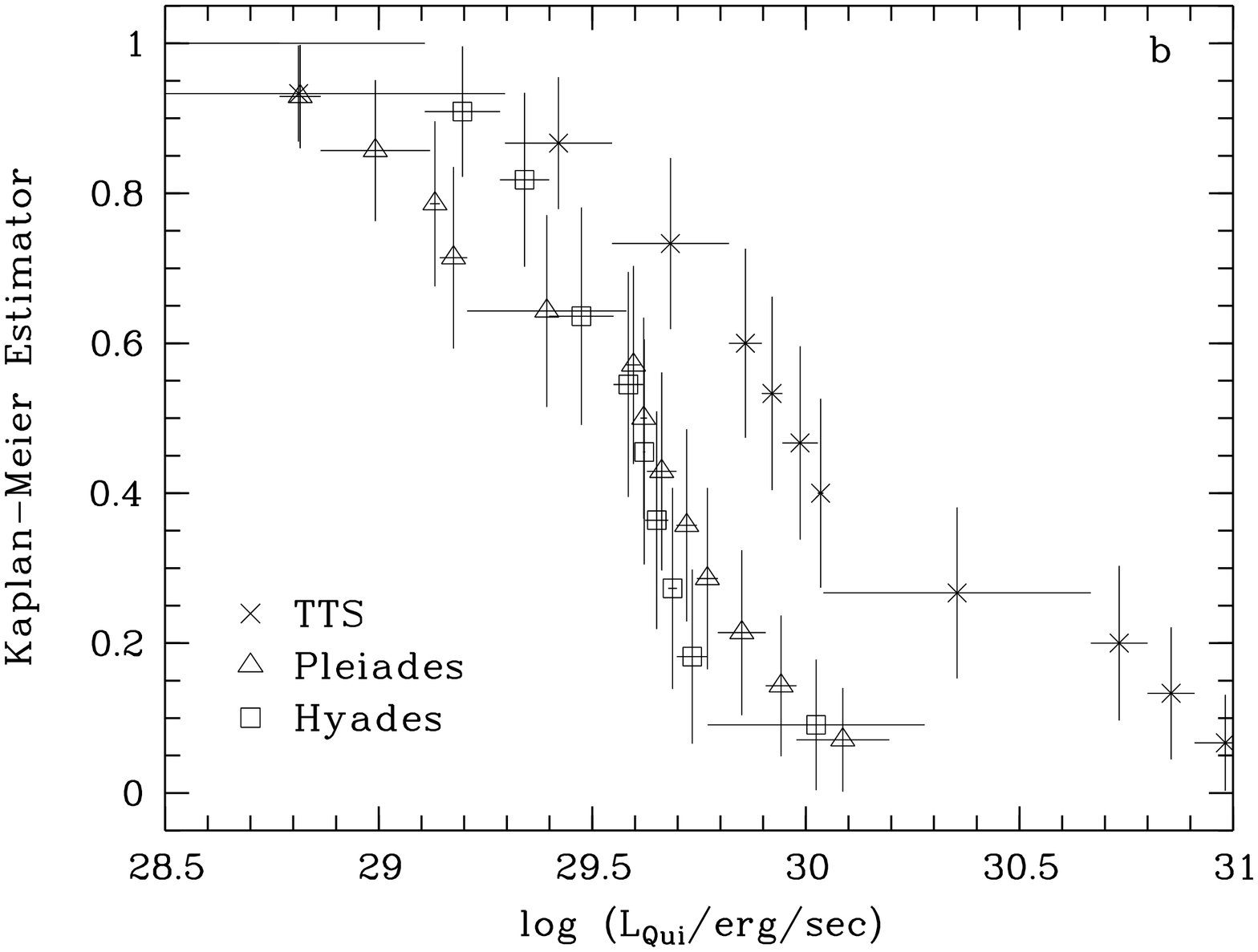}}}
\caption{Luminosity functions for the flaring TTSs, Pleiads, and Hyads
during the X-ray flare {\bf (a)} and during quiescence {\bf (b)}. 
Stars of spectral type earlier than G or unknown spectral type, and known
white-dwarf systems are excluded. Note that the quiescent distribution of the Hyades stars is consistent with that of the Pleiads, due to the selection of flaring stars only (see discussion in the text).}
\label{fig:lumfunct}
\end{center}
\end{figure}

Maximum likelihood distributions for TTSs, Pleiads, and Hyads are presented 
in Fig.~\ref{fig:lumfunct} for both flare luminosity $L_{\rm F}$ 
and mean luminosity during the {\em quiescent} part of flare observations 
$L_{\rm qui}$. Note, that Fig.~\ref{fig:lumfunct} (b) contains no
upper limits because only stars which have shown a flare are included, and 
$L_{\rm qui}$ during flare observations can be extracted from the 
lightcurves.
The flare luminosity in Fig.~\ref{fig:lumfunct}\,(a) includes
upper limits. Since $L_{\rm F} = L_{\rm max} - L_{\rm qui}$, upper
limits for $L_{\rm max}$ (see
Table~\ref{tab:det_flares_TTS},~\ref{tab:det_flares_Ple},
and~\ref{tab:det_flares_Hya}) translate to upper limits for $L_{\rm F}$.
LDFs for all non-flaring stars (detections and non-detections) will be
shown elsewhere (Stelzer et al., in preparation).

\begin{table}
\caption{Results of two-sample tests for the flare luminosity $L_{\rm F}$
and quiescent luminosity of flare observations $L_{\rm qui}$. 
The number of
upper limits to the luminosity are listed in brackets in column `Sample
Size'. See Table~\ref{tab:teff} for the meaning of the remaining columns.
All distributions of $L_{\rm F}$ are significantly different from
each other. Flaring Pleiads and Hyads have similar $L_{\rm qui}$ distributions.}
\label{tab:L_gkm_2s}
\begin{tabular}{lrrrr}\hline
Sample Names & \multicolumn{2}{c}{Sample sizes} & Prob. GW & Prob logrank \\ \hline
\multicolumn{5}{c}{Flare luminosity $L_{\rm F}$} \\ \hline
TTSs - Pleiads  & 15(6) & 14(5) & $0.021$ & $0.017$ \\
TTSs - Hyads    & 15(6) &  9(3) & $0.001$ & $0.000$ \\
Pleiads - Hyads & 14(5) &  9(3) & $0.007$ & $0.010$ \\ \hline
\multicolumn{5}{c}{Quiescent luminosity of flaring stars $L_{\rm qui}$} \\ \hline
TTSs - Pleiads  & 15(0) & 14(0) & $0.025$ & $0.010$ \\
TTSs - Hyads    & 15(0) & 9(0)  & $0.007$ & $0.001$ \\
Pleiads - Hyads & 14(0) & 9(0)  & $0.615$ & $0.243$ \\ \hline
\end{tabular}
\end{table}

\begin{table}
\caption{Mean and median flare and quiescent luminosity of flaring TTSs,
Pleiads, and Hyads. The sample of cTTSs was to small to compute the median
of $L_{\rm F}$.}
\label{tab:L_gkm_mean}
\begin{tabular}{lrlcc}\hline
Sample Name & \multicolumn{2}{c}{Sample Size} & Mean & Median \\ \hline
\multicolumn{5}{c}{Flare Luminosity $\lg{L_{\rm F}}$} \\ \hline
TTSs        & 15 & (6) & $31.05 \pm 0.19$ & $30.65$ \\
Pleiads     & 14 & (5) & $30.51 \pm 0.12$ & $30.30$ \\
Hyads       & 9  & (3) & $30.06 \pm 0.11$ & $29.99$ \\ \hline
cTTSs       & 6  & (4) & $31.44 \pm 0.32$ & $-$ \\
wTTSs       & 8  & (1) & $30.81 \pm 0.22$ & $30.58$ \\ \hline
\multicolumn{5}{c}{Quiescent Luminosity of Flaring Stars $\lg{L_{\rm qui}}$} \\ \hline
TTSs        & 15 & (0) & $29.98 \pm 0.17$ & $29.92$ \\
Pleiads     & 14 & (0) & $29.52 \pm 0.11$ & $29.61$ \\
Hyads       & 9  & (0) & $29.50 \pm 0.07$ & $29.47$ \\ \hline
cTTSs       & 6  & (0) & $29.94 \pm 0.19$ & $29.82$ \\
wTTSs       & 8  & (0) & $30.22 \pm 0.19$ & $30.04$ \\ \hline
\end{tabular}
\end{table}

Two-sample tests were applied to each pair of LDFs to search for 
differences.
The results are given in Table~\ref{tab:L_gkm_2s} (for $L_{\rm F}$
and $L_{\rm qui}$). 
The null hypothesis of two samples being the same 
is rejected for all pairs of flare luminosity distributions 
at significance levels $\alpha < 0.05$. The quiescent luminosity of
{\em flaring} TTSs
is different from both the quiescent luminosity of {\em flaring} Pleiads and {\em flaring} Hyads. Usually the quiescent 
luminosity functions of Pleiades and Hyades stars are also 
found to be distinct (see e.g. \cite{Caillault96.1}). However, 
we find no difference ($\alpha > 0.61$) 
between the quiescence luminosities
of the flaring stars in these two clusters.
Using the relation between $I_{\rm qui}$ and the 
threshold for $L_{\rm F}/L_{\rm qui}$ (see Fig.~\ref{fig:det_accuracy}) we
have determined that 
more than 90\% of the detected Hyades stars are bright enough for
detection of a flare whose strength $L_{\rm F}/L_{\rm qui}$ is equal to 
the mean observed for flares on late-type 
Hyads, i.e. $L_{\rm F}/L_{\rm qui}=4.527$. The fact that mostly 
{\em X-ray bright} 
Hyades stars display flaring activity is therefore not a selection effect. 
Instead, flaring Hyads indeed are overluminous compared to the
non-flaring Hyades stars detected by the {\em ROSAT} PSPC.

The mean luminosities of TTSs, Pleiads, and Hyads and their standard deviations
derived with inclusion of upper limits are given in Table~\ref{tab:L_gkm_mean}.

\subsection{Flaring cTTSs and wTTSs}\label{subsect:cw}

So far we have not distinguished between cTTSs and wTTSs, because
it is a matter of debate whether all cTTSs are younger than wTTSs.
However, they are clearly distinguished by their circumstellar environment.
The disks of cTTSs may influence flare activity.
We have, therefore, compared cTTSs and wTTSs with respect to several flare
parameters (see Table~\ref{tab:cw_2s} for the results). 
Significant differences are found in the flare luminosity 
$L_{\rm F}$ and relative strength of the flare $L_{\rm F}/L_{\rm qui}$.
The decay timescale does not depend on the type of TTS.
The values given in 
Table~\ref{tab:cw_2s} have been derived from all flares on TTSs except
the one on DD\,Tau\,/\,CZ\,Tau. 
DD\,Tau is a cTTS binary and CZ\,Tau a wTTS binary.
The four stars are not resolved in the PSPC
image. It is therefore 
impossible to classify this flare concerning the type of TTS.
We have performed two further series of two-sample tests 
in which the event is included. In one of these series of tests the flare
is attributed to DD\,Tau, and the other time to 
CZ\,Tau. The significance of the results did not change.

The mean flare and quiescent luminosities of cTTSs and wTTSs are given
in Table~\ref{tab:L_gkm_mean}. cTTSs, although characterized by lower
quiescent emission, show stronger flares than wTTSs.

\begin{table}
\caption{Results of two-sample tests for differences between cTTSs and 
wTTSs with respect to the flare parameters 
$L_{\rm qui}$, $L_{\rm F}$, $\tau_{\rm dec}$, and $L_{\rm F} / L_{\rm qui}$. 
The flare on the unresolved stars DD\,Tau/CZ\,Tau is not
considered here, since it is not clear whether the event should be
attributed to the wTTS (CZ\,Tau) or to the cTTS (DD\,Tau).
The meaning of columns~2--5 is the same as in Table~\ref{tab:L_gkm_2s}.}
\label{tab:cw_2s}
\begin{tabular}{lrrcc} \hline
Parameter & \multicolumn{2}{c}{Sample Size} & Prob. GW & Prob logrank \\  
          & cTTSs & wTTSs & & \\ \hline
$\lg{L_{\rm F}}$        & 6(4) & 8(1) & 0.053 & 0.046 \\
$\lg{L_{\rm qui}}$      & 6(0) & 8(0) & 0.206 & 0.264 \\
$\tau_{\rm dec}$        & 6(2) & 8(1) & 0.714 & 0.752 \\
$L_{\rm F}/L_{\rm qui}$ & 6(4) & 8(1) & 0.007 & 0.002 \\ \hline
\end{tabular}
\end{table}

\subsection{$v\,\sin{i}$ of the flaring stars}\label{subsect:corrvsini}

\begin{figure}
\begin{center}
\resizebox{9cm}{!}{\includegraphics{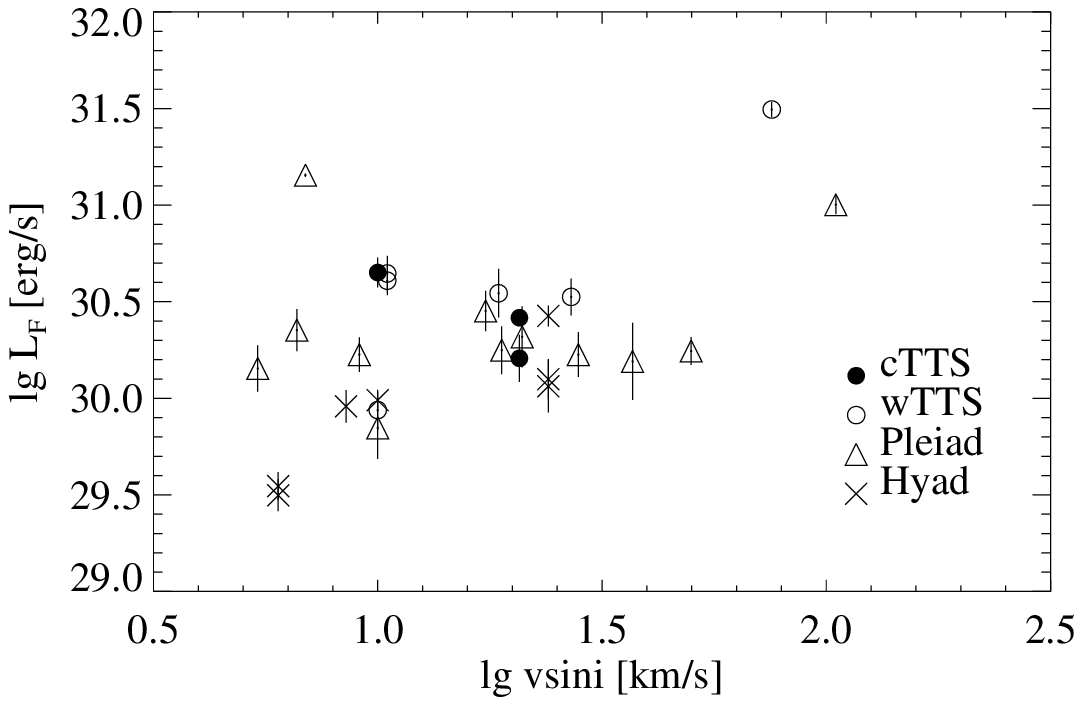}}
\resizebox{9cm}{!}{\includegraphics{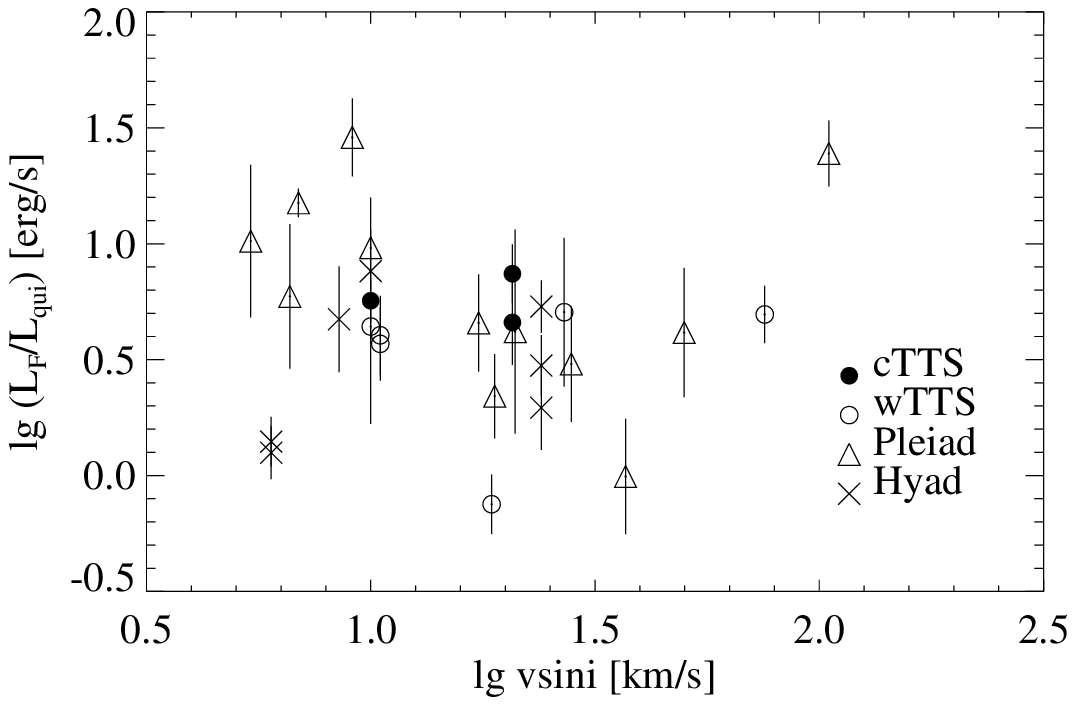}}
\caption{X-ray luminosities versus projected rotational
velocity $v\,\sin{i}$ in double logarithmic scale: {\em top} - flare luminosity $L_{\rm F}$, {\em bottom} -
luminosity ratio $L_{\rm F}/L_{\rm qui}$. While $L_{\rm F}$ and
$v\,\sin{i}$ may show a weak positive correlation, $L_{\rm F}/L_{\rm qui}$ does not depend on the rotation.}
\label{fig:lx_vsini}
\end{center}
\end{figure}

Stellar rotation is one of the necessary conditions for magnetic
activity. We have, therefore, examined the influence of the stellar rotation
rate on the characteristics of X-ray flares. 
The relation between flare parameters and  projected rotational 
velocity, $v\,\sin{i}$, is shown in Fig.~\ref{fig:lx_vsini}.
The statistical significance for
correlations between some flare parameters and $v\,\sin{i}$ is given in 
Table~\ref{tab:vsini_2s} (columns~2~and~3). The weak correlation between
luminosity, both $L_{\rm F}$ and $L_{\rm qui}$, and $v\,\sin{i}$ is significant.
The decay time $\tau_{\rm dec}$ and the relative flare strength 
$L_{\rm F}/L_{\rm qui}$, on the other hand, are not related to
$v\,\sin{i}$.

We have studied the flaring population in terms of differences in
flare characteristics between
slow and fast rotators. The boundary was set to 20\,km/s because this 
choice gives two samples of about equal size: 16 slow and 
12 fast rotators showed an X-ray flare. Stars from
Tables~\ref{tab:det_flares_TTS},~\ref{tab:det_flares_Ple},~and~\ref{tab:det_flares_Hya}
for which no measurement of $v\,\sin{i}$ is available are ignored. The result
of two sample tests for the parameters $L_{\rm F}$, $L_{\rm qui}$,
$\tau_{\rm dec}$,
and $L_{\rm F}/L_{\rm qui}$ are presented in the remaining columns of 
Table~\ref{tab:vsini_2s}. In no case the null 
hypothesis that slow and fast rotators are drawn from the
same distribution was rejected at significance level $\alpha < 0.05$. 

\begin{table}
\caption{Relation between flare parameters $L_{\rm qui}$, $L_{\rm F}$,
$\tau_{\rm dec}$, and $L_{\rm F} / L_{\rm qui}$ and the projected
rotational velocity $v\,\sin{i}$: Columns~2~and~3 are results of
correlation tests between each of the listed parameters and
$v\,\sin{i}$. The remaining columns give results of two-sample tests for
differences between slow and fast rotators with boundary at $20\,{\rm
km/s}$. The samples consist of 16 slow versus 12 fast rotators.}
\label{tab:vsini_2s}
\begin{tabular}{lccrr} \hline
Parameter & Pearson r & Spearman r & Prob. & Prob \\ 
          & (Signif.) & (Signif.)  & GW    & logrank \\ \hline
$\lg{L_{\rm F}}$        & 0.479 (0.005) & 0.372 (0.026) & 0.075 & 0.091 \\
$\lg{L_{\rm qui}}$      & 0.468 (0.006) & 0.496 (0.004) & 0.269 & 0.528 \\
$\tau_{\rm dec}$        & 0.247 (0.103) & 0.221 (0.129) & 0.487 & 0.584 \\
$L_{\rm F}/L_{\rm qui}$ & -0.017 (0.466) & -0.287 (0.069) & 0.399 & 0.547
\\ \hline
\end{tabular}
\end{table}

\section{Flare rates}\label{sect:rate}

In this section, we will derive flare rates as a means to 
determine the activity level for a stellar sample with distinct properties. 
The characteristic properties which will be
examined are (a) stellar age (comparing TTSs, Pleiads, and Hyads),
(b) stellar rotation (comparing slow and fast rotators), and
(c) stellar multiplicity (comparing close binaries to other stars).
Flare rates will be computed separately for each group of stars.

We assume that the duration of the active state is represented by
the decay timescale $\tau_{\rm dec}$, i.e. the generally poorly
restricted rise times $\tau_{\rm ris}$ are neglected. 
This is certainly wrong for the flare
on hcg\,144 which seems to have a somehow reversed character (slow rise and
rapid decay). However, hcg\,144 is a star of unknown spectral type and
therefore not part of the group to be examined here.
To compile the flare rates, $F = \sum{(\tau_i)} / T_{\rm obs}$, 
we have added up the decay timescales $\tau_{\rm dec}$ of the flares and 
divided this sum by the total observing time, $T_{\rm obs}$, 
of all detections (flaring and non-flaring stars). 
Only the nearest identification of each X-ray source has been
considered in the compilation of $T_{\rm obs}$. But for 
multiple systems we have multiplied the observing time by the number of 
components.
For the compilation of $T_{\rm obs}$ we have eliminated data gaps
larger than 1\,h, the typical flare duration.
This provides us the fraction of the total observing 
time during which the stars
are observed in the active state.

In practice $\sum{(\tau_i)}$ is computed from the sample mean $\bar{\tau}$ 
returned by ASURV's Kaplan-Meier estimator. This way we ensure that upper
limits to $\tau_{\rm dec}$ are taken into account. The Kaplan-Meier estimator
returns also the uncertainty of $\bar{\tau}$. To include the spread
of the data in the estimation of $F$ we have converted this uncertainty
of the mean to the sample variance $\sigma_\tau$. Consequently:
\begin{equation}
F = \frac{\bar{\tau} \cdot N}{T_{\rm obs}} \pm \frac{\sigma_\tau \cdot \sqrt{N}}{T_{\rm obs}}
\end{equation}

\subsection{Flare rate and stellar age}

The evolution of flare rates with stellar age is examined by 
comparing the flare frequency of TTSs to that of the Pleiades and the Hyades.
15 flares have occurred on TTSs, 14 on late-type
Pleiads, and 11 on late-type Hyads. 
We have derived the following values for the flare rate
$F$: $0.86 \pm 0.16$\% (TTSs), 
$0.67 \pm 0.13$\% (Pleiads), and $0.86 \pm 0.32$\% (Hyads). 
When the white-dwarf binaries are excluded the 
flare rate for Hyades declines to 
$F_{\rm H,no WD} = 0.71 \pm 0.30$\%. 

The flare rates are biased for several reasons which will be explained next.
First, 
the flare detection limit is determined by the S/N, which in turn depends
on the distance to the
star. Therefore, flare
rates of TTSs, Pleiads, and Hyads are only comparable
above a limiting minimum strength of the flare, expressed by a 
threshold $L_{\rm F} / L_{\rm qui}$. And, secondly,
incomplete data sampling might lead to wrong conclusions about the
decay timescale of individual events and thus contaminate the 
resulting $F$.

To solve the first problem we
have scaled the quiescent count rate of all flaring stars 
to a distance of 140\,pc, the distance of most of the TTSs. I.e.
we have multiplied all quiescent count rates with a factor 
$(\frac{d}{140\,pc})^2$.
These theoretical values of $I_{\rm qui}$ correspond to higher flare
detection thresholds $L_{\rm F} / L_{\rm qui}$ for all stars except 
the ones in Perseus. All flares from Perseus stars would be detected at
140\,pc, since they are further away than this distance. 
The observed luminosity ratios of all flaring stars have then been compared
to the theoretical threshold needed if the star were at 140\,pc. All
flares for which the observed value is below this requirement should
be neglected when the flare rates are computed. It turns out that
all flares on Pleiads remain above the 140\,pc threshold. But only
7 out of 11 flares on Hyads (one of the 7 is a white-dwarf system) 
have $L_{\rm F} / L_{\rm qui}$ high enough
to be detected at a distance of 140\,pc. Now the comparison of our
different samples is free from the sensitivity bias. And we derive
flare rates $F$ of 
$0.86 \pm 0.16$\% (TTSs), 
$0.67 \pm 0.13$\% (Pleiads), and $0.46 \pm 0.19$\% (Hyads).
Without the white-dwarf binary $F$ decreases for the Hyades
to $0.32 \pm 0.17$\%.

The uncertainties in the measurement of 
the flare duration are
less easy to overcome. The large flare rate of TTSs is partially due to
two extraordinary long events of duration $> 8\,{\rm h}$ 
(see Table~\ref{tab:det_flares_TTS}). The decay times of both of these
flares are considered to be an upper limit. 
If these two events are discarded from the sample of flares, 
$F_{\rm TTS} = 0.74 \pm 0.14$\%.

We have also compiled $F$ for cTTSs and wTTSs separately to see
whether the circumstellar environment has any influence on the frequency 
of the flare activity. Among the events on TTSs, 
6 are observed on cTTSs and 8 on wTTSs.
An additional flare was seen from the unresolved stars DD\,Tau\,/\,CZ\,Tau.
The classification of this event within the subgroups of TTSs remains
therefore unclear, and complicates the comparison of $F$ for the two
classes of TTSs. At first, the event on DD\,Tau\,/\,CZ\,Tau has been
eliminated from the sample, thus that 6 flares on cTTSs are opposed to
8 flares on wTTSs.
The respective flare rates are 
$F_{\rm c} = 1.09 \pm 0.39$\% and
$F_{\rm w} = 0.65 \pm 0.16$\%. 
When the ambiguous event is 
counted on the side of the cTTSs $F_{\rm c}$ rises to 
$1.28 \pm 0.37$\%.
When it is attributed to the wTTS CZ\,Tau instead, $F_{\rm w}$ becomes
$0.76 \pm 0.16$\%. 
Note, that even though the number of flares on wTTSs is higher than the 
number of flares on cTTSs,
the flare rate for wTTSs is lower than the flare rate for cTTSs.
This is possible because of differences in the total observing time.

\begin{figure}
\begin{center}
\resizebox{9cm}{!}{\includegraphics{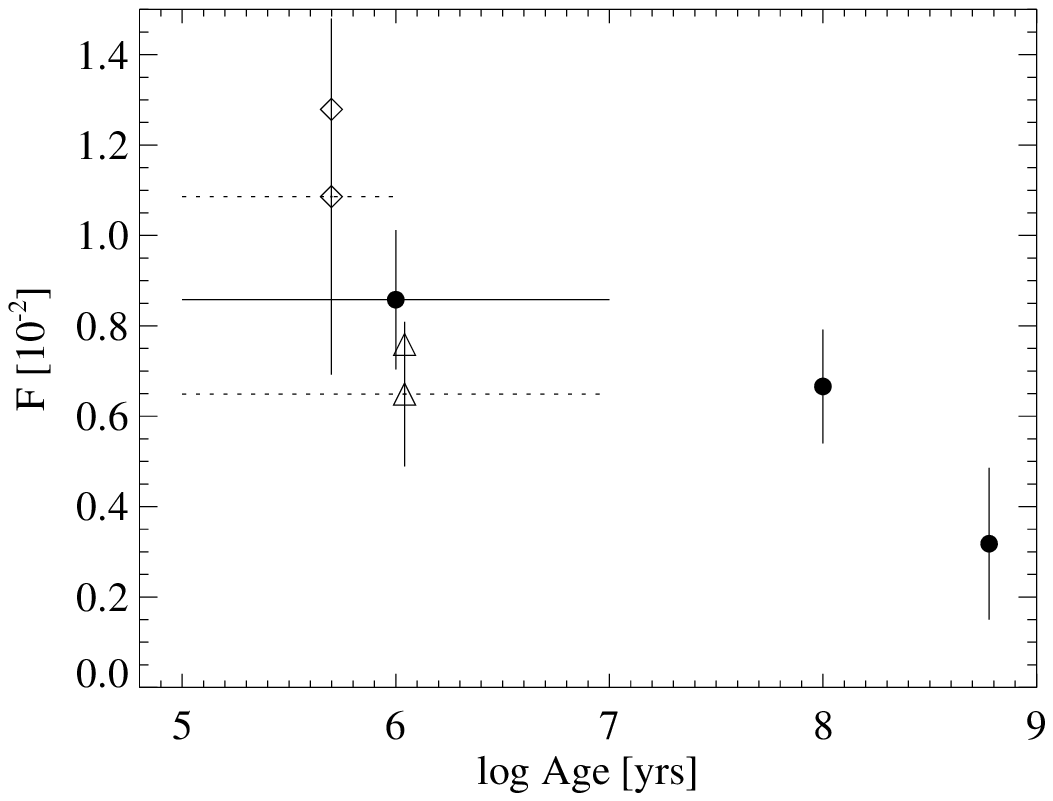}}
\caption{Flare rate $F = \sum{(\tau_i)} / T_{\rm obs}$ as a function of
stellar age for TTSs, Pleiads, and Hyads. To eliminate biases due to the
S/N-dependence of the detection sensitivity for flares, only events which
are bright enough for detection at a uniform distance of 140\,pc are
considered. Furthermore, flares which can not definitely be assigned to a
late-type star are excluded. The flare rates for cTTSs (diamonds) and wTTSs
(triangles) alone are overplotted. The upper of the two symbols applies if
the ambiguous event on DD\,Tau\,/\,CZ\,Tau is included in the sample (see
text). The horizontal bars represent the age spread of TTSs.}
\label{fig:frates}
\end{center}
\end{figure}

$F$ as a function of stellar age is displayed in Fig.~\ref{fig:frates}.
The decline of the flare rate with stellar age is obvious. Rates for cTTSs
and wTTSs are symbolized by diamonds and triangles, respectively. The
location of the lower diamond and triangle describes the flare rates 
without the event on DD\,Tau\,/\,CZ\,Tau. The upper diamond and
triangle are values for $F$ if this flare is included 
in the respective group of TTSs.

\subsection{Flare rate and rotational velocity}

Here we examine whether the flare
{\em frequency} depends on rotation. This is done by computing $F$ 
(defined as before) for fast rotators on the one hand 
($v\,\sin{i} > 20\,{\rm km/s}$) and slow
rotators on the other hand ($v\,\sin{i} < 20\,{\rm km/s}$).
Again, only late-type stars are considered.
The resulting rates are 
$F_{\rm slow} = 0.55 \pm 0.10$\% and 
$F_{\rm fast} = 1.55 \pm 0.38$\%.
Thus there is a clear trend towards an increase of
flare activity with increasing rotational velocity.

\subsection{Flare rate and multiplicity}

Another interesting question is whether the circumstellar
surroundings have any influence on the flare frequency.
The coronal activity may e.g. change if there are interactions 
between the magnetic fields of binaries.
Such interactions are expected to take place only in {\em close}
binaries.
To search for such a connection we, therefore, discriminate between 
spectroscopic binaries on the one hand and all others,
i.e. singles or visual multiples.
The flare rate $F$ is computed in the same way as before.
Since the observation time of each
stellar system has been multiplied by the number of components, the
flare rates should be about equal for both samples if the 
underlying physics are the same. 
However, we find that the flare rate of spectroscopic binaries
is enhanced by more than a factor of two:
$F_{\rm non-SB} = 0.64 \pm 0.12$\%
and $F_{\rm SB} = 1.43 \pm 0.25$\%.
Note, that the study of individual flare parameters (similar to 
the analysis of Sect.~\ref{sect:statcomp}) has 
shown no difference for these two samples.

\section{Hardness Ratios}\label{sect:hr}

For most of the flaring sources not enough counts are collected by the PSPC 
to compare the different levels of X-ray emission in a detailed spectral 
analysis. Therefore, we use hardness ratios to mark spectral changes. 
{\em ROSAT} PSPC hardness ratios are defined by:
\begin{equation}
HR\,1 = \frac{H-S}{H+S} \quad HR\,2 = \frac{H_2-H_1}{H_2-H_1}
\end{equation}
where $S$, $H$, $H_1$, and $H_2$ 
denote the count rates in the {\em ROSAT} PSPC
soft (0.1-0.4\,keV), hard (0.5-2.0\,keV), hard1 (0.5-0.9\,keV) and hard2 
(0.9-2.0\,keV) band respectively.
For each flare observation $HR\,1$ and $HR\,2$ are computed for three activity 
stages representing the quiescent state (pre- and post-flare), the 
rise and the decay, respectively. Sometimes no counts are measured in one
or more of the energy bands. Whenever this is the case we have derived
upper limits for the hardness ratio making use of the background counts
in that energy band at the source location.

\begin{figure*}
\begin{center}
\parbox{16cm}{
\parbox{4.5cm}{\resizebox{5.2cm}{!}{\includegraphics{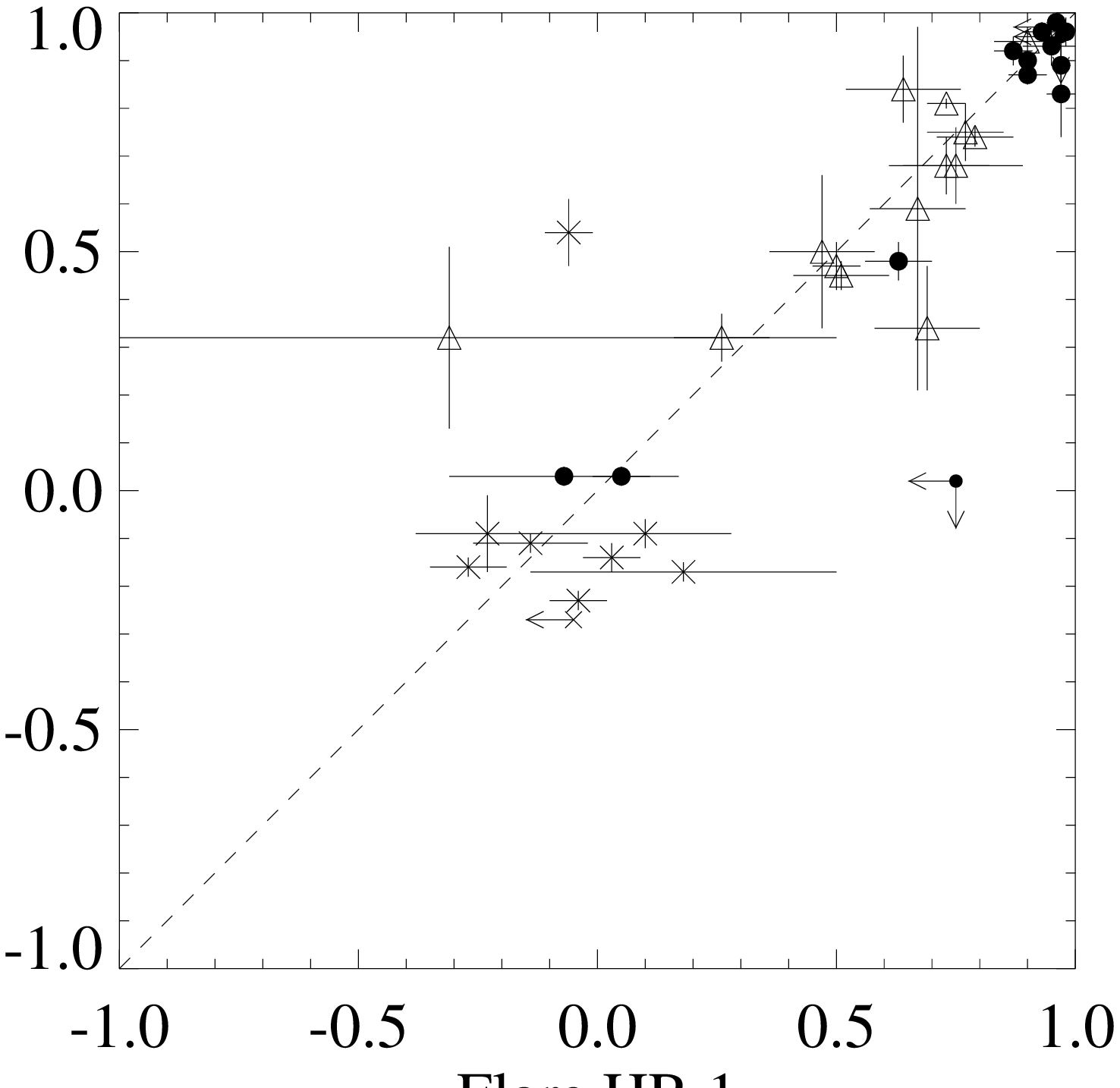}}}
\parbox{1cm}{\hspace*{1.cm}}
\parbox{4.5cm}{\resizebox{5.2cm}{!}{\includegraphics{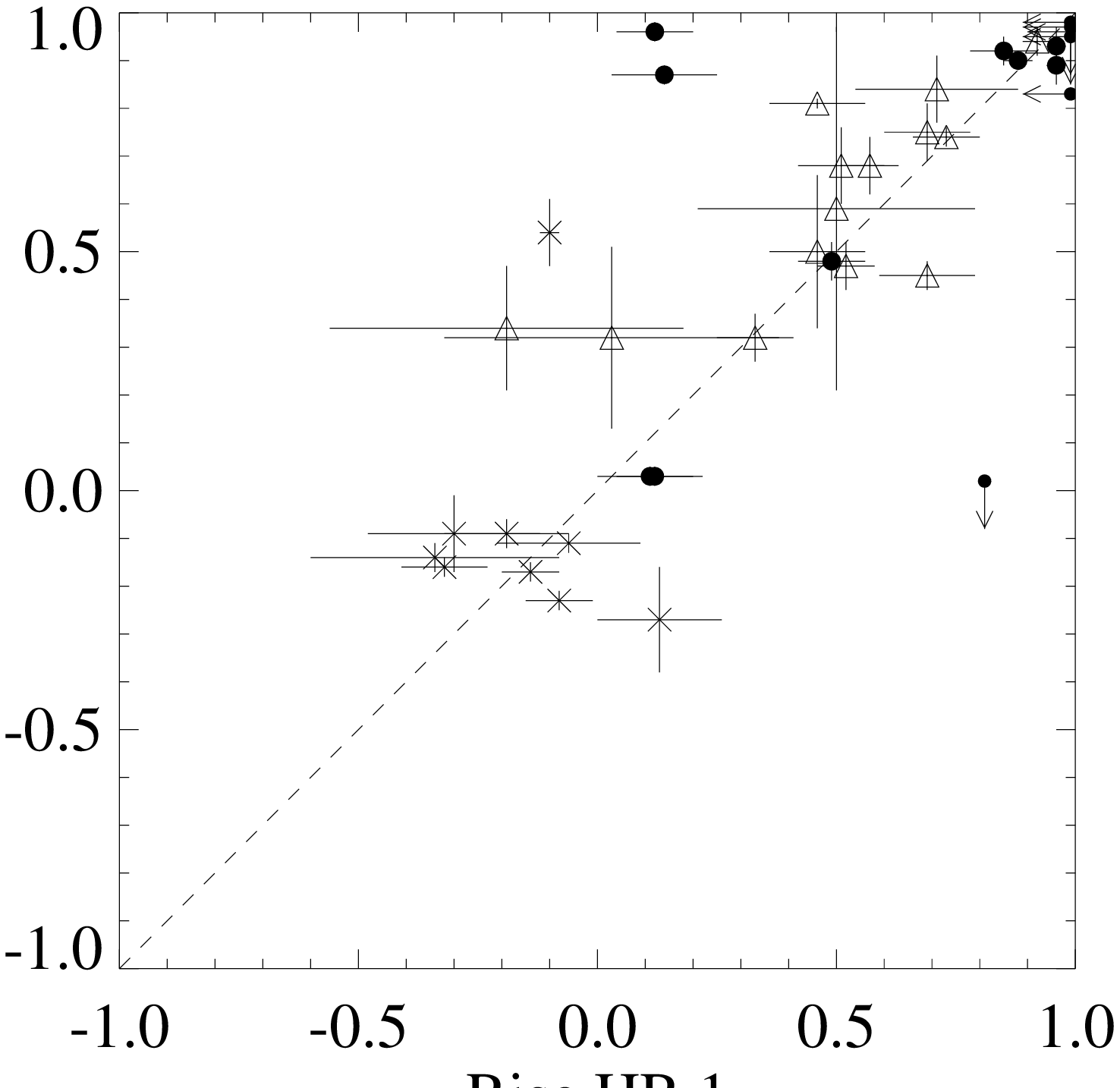}}}
\parbox{1cm}{\hspace*{1.cm}}
\parbox{4.5cm}{\resizebox{5.2cm}{!}{\includegraphics{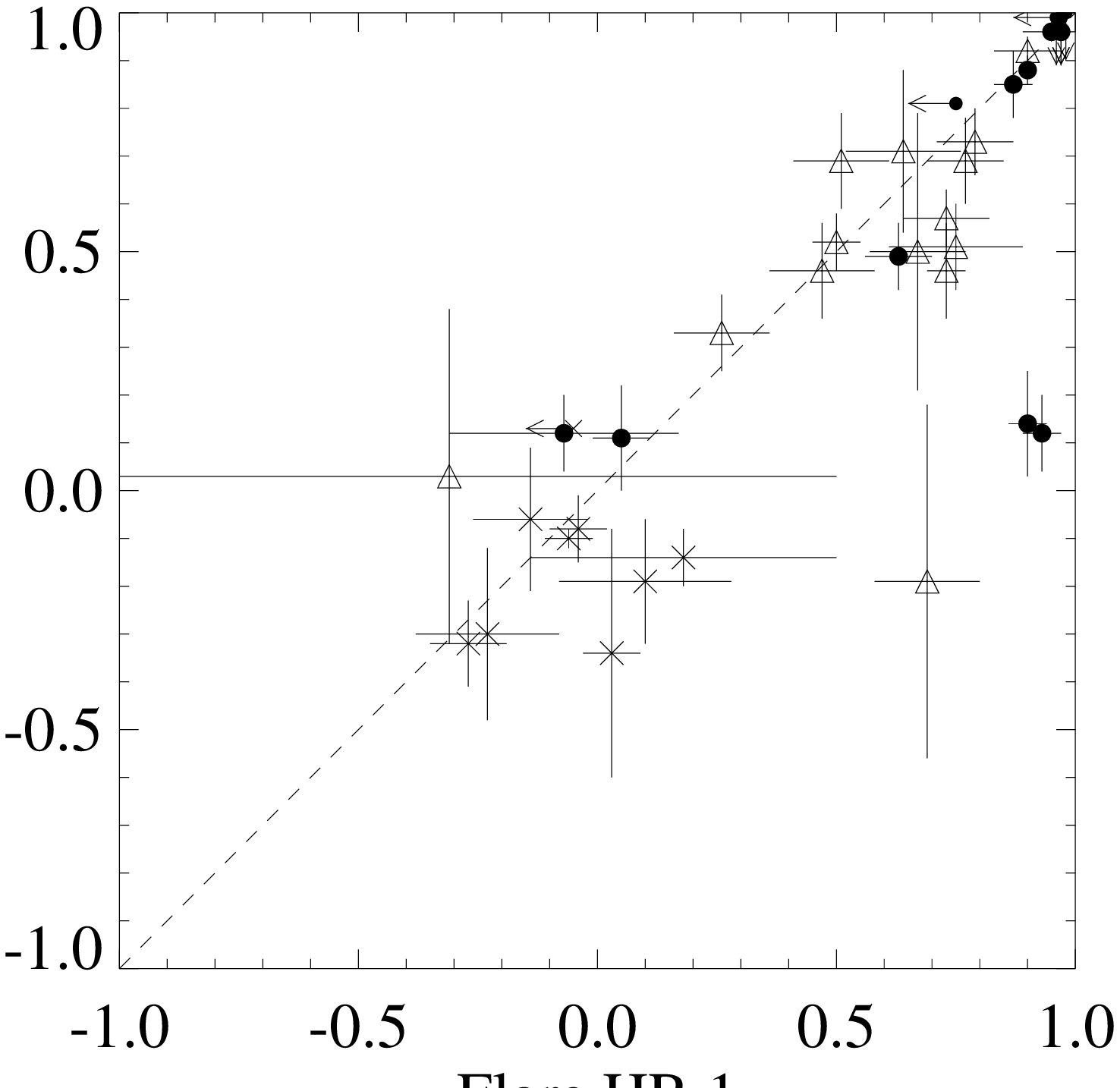}}}
}
\parbox{16cm}{
\parbox{4.5cm}{\resizebox{5.2cm}{!}{\includegraphics{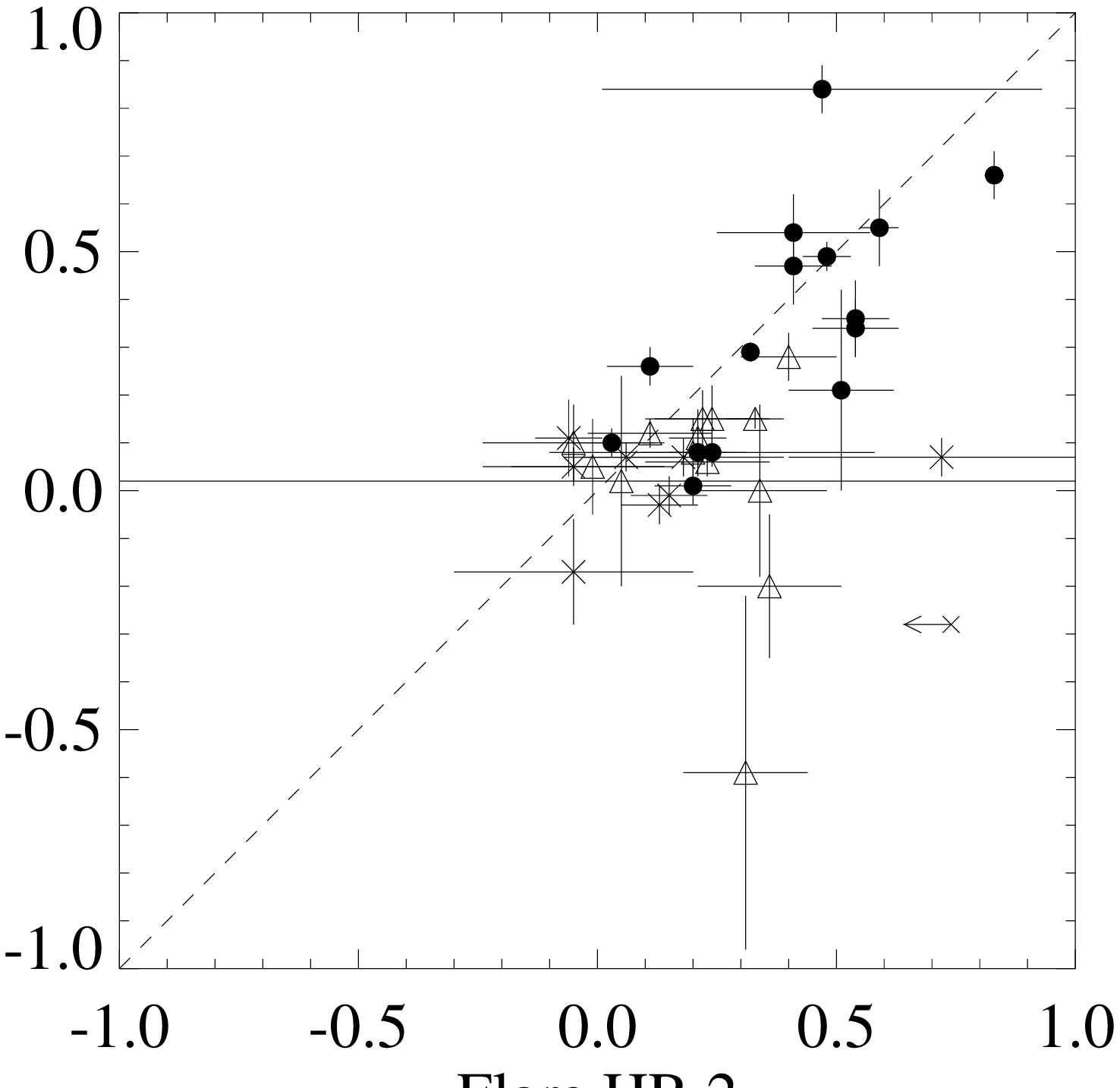}}}
\parbox{1cm}{\hspace*{1.cm}}
\parbox{4.5cm}{\resizebox{5.2cm}{!}{\includegraphics{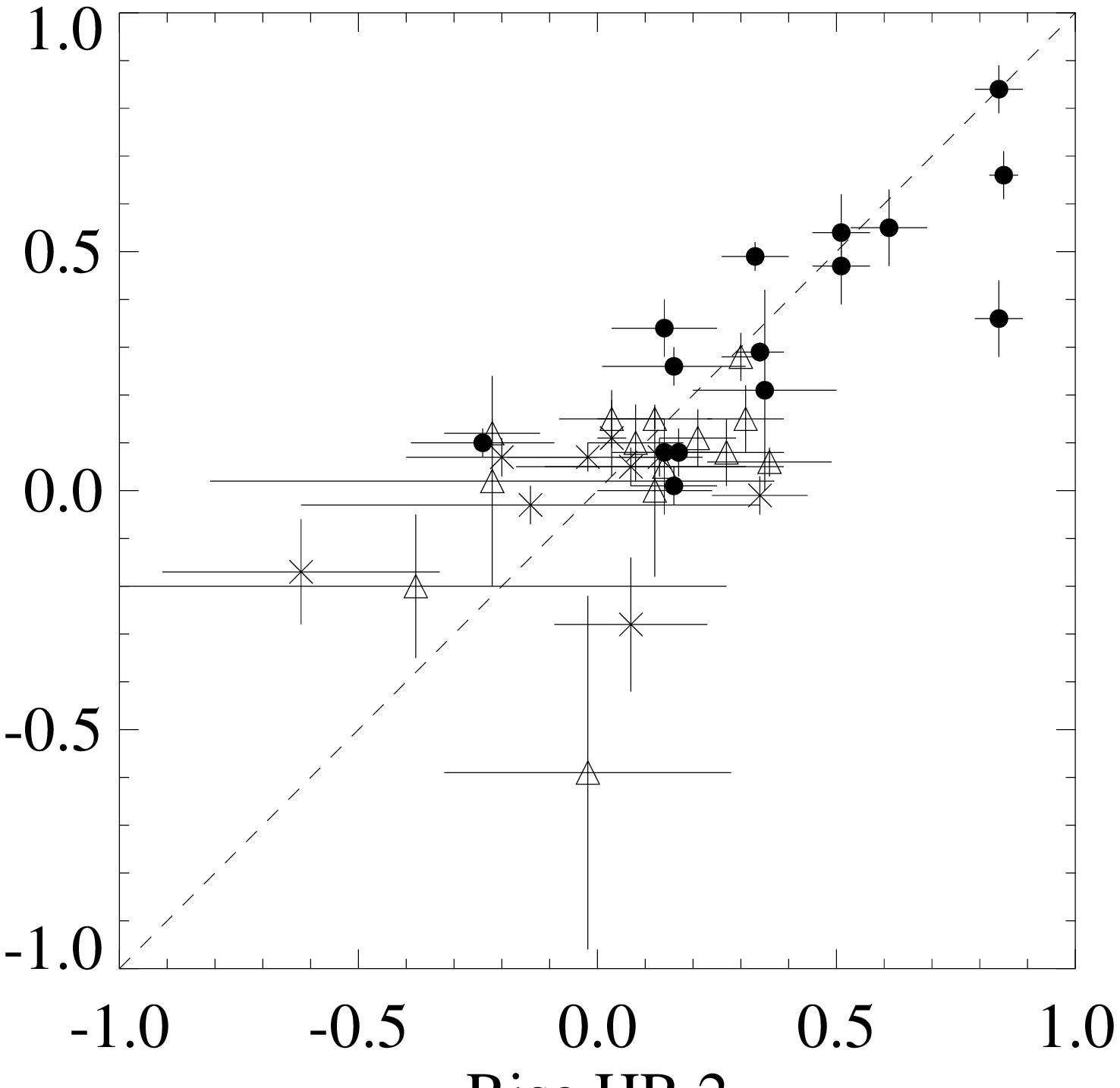}}}
\parbox{1cm}{\hspace*{1.cm}}
\parbox{4.5cm}{\resizebox{5.2cm}{!}{\includegraphics{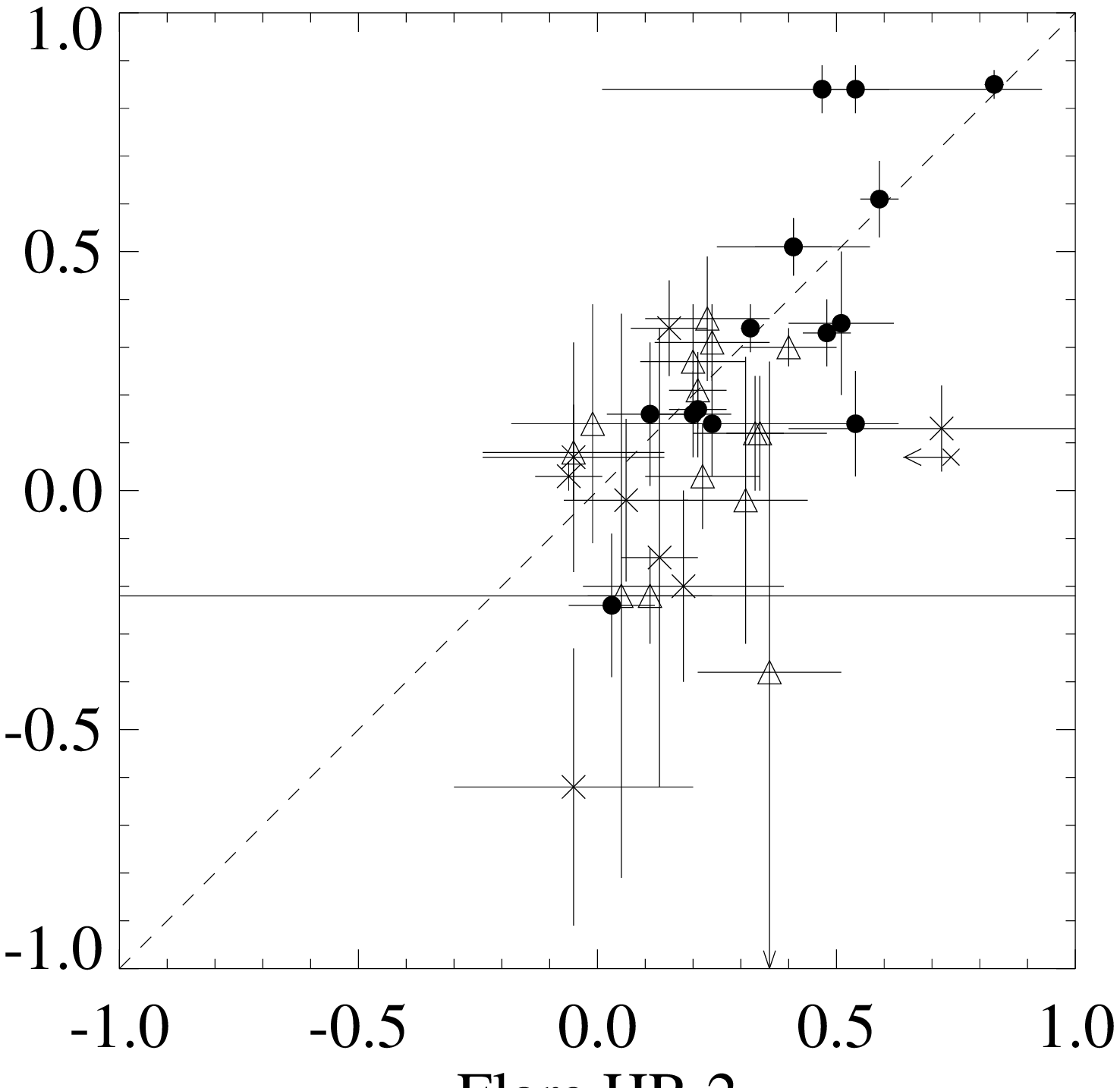}}}
}
\caption{Comparison of {\em ROSAT} PSPC hardness ratios for quiescent state,
flare rise and flare decay of the late-type stars from
Tables~\protect\ref{tab:det_flares_TTS},~\protect\ref{tab:det_flares_Ple},
and~\protect\ref{tab:det_flares_Hya}. {\em top} - $HR\,1$, {\em bottom} -
$HR\,2$. The meaning of the plotting symbols is the same as in Fig.~\ref{fig:Lf_Lq}. Stars located on the dashed line show no change in hardness between quiescent and flare state.}
\label{fig:hrs}
\end{center}
\end{figure*}

The observed hardness ratios, $HR\,1$ and $HR\,2$, 
are plotted in Fig.~\ref{fig:hrs}. 
The plots comparing quiescent and flare state
show marginal evidence that most of the stars lie
below the diagonal in the hardness plot (see lower left panel of 
Fig.~\ref{fig:hrs}) and thus are harder during the
flare intervals as compared to their quiescence. 
No significant difference in hardness is observed between flare rise
and flare decay. 
When impulsive heating takes place before the outburst the plasma cools 
quickly by radiation and conduction to the chromosphere. Therefore,
the similar hardness observed during rise and decay phase suggests
that heating takes place throughout the decay.

\begin{figure}
\begin{center}
\resizebox{8cm}{!}{\includegraphics{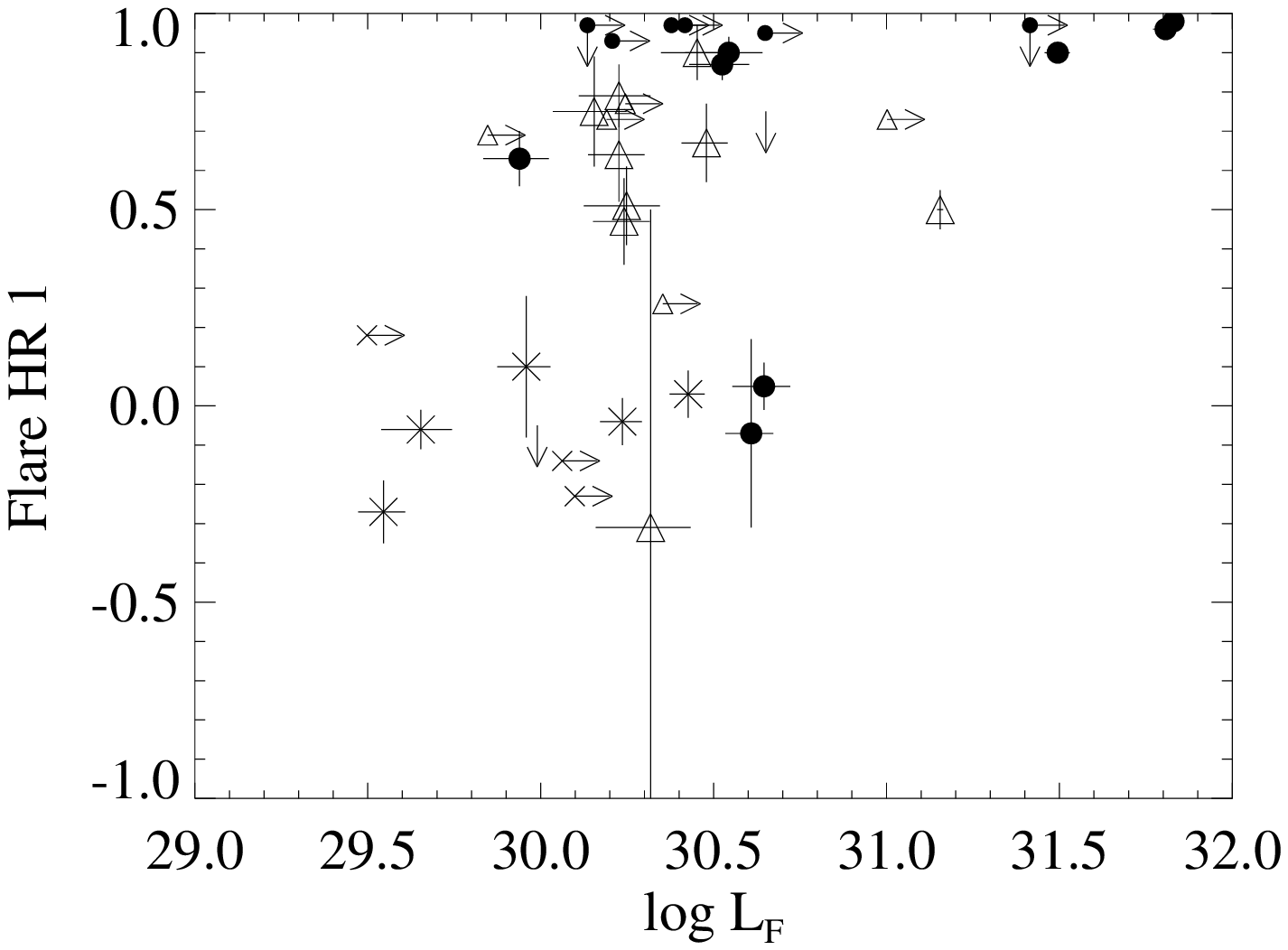}}
\resizebox{8cm}{!}{\includegraphics{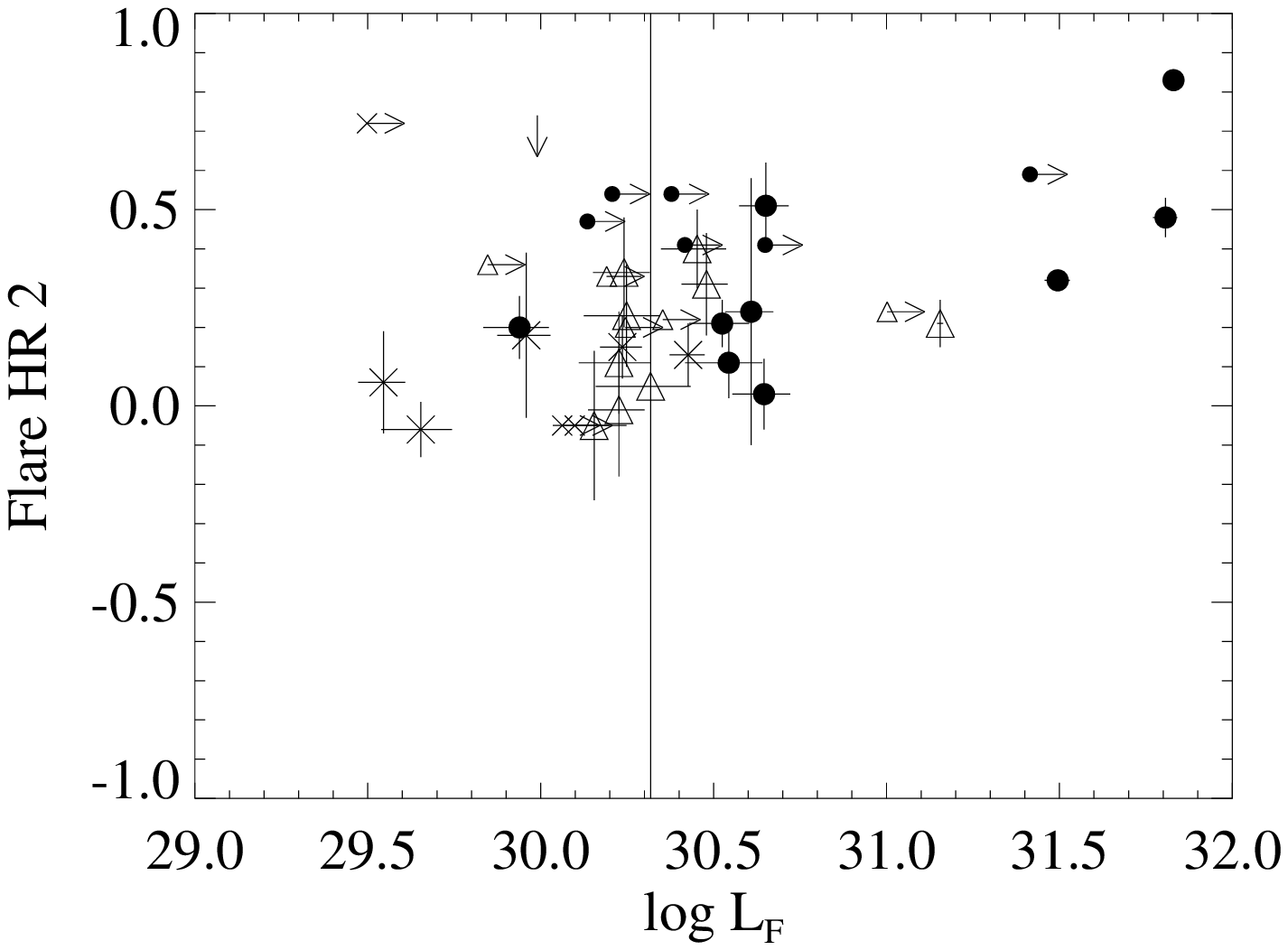}}
\caption{Correlation between {\em ROSAT} PSPC hardness ratios and flare
luminosity $L_{\rm F}$. The meaning of the plotting symbols is the same as
in Fig.~\ref{fig:hrs}. Lower limits for $L_{\rm F}$ are shown when a 
data gap proceeds the observed maximum of the lightcurve (see Tables~\ref{tab:det_flares_TTS},~\ref{tab:det_flares_Ple}~and~\ref{tab:det_flares_Hya})}
\label{fig:hr_Lx}
\end{center}
\end{figure}

To quantify the differences in hardness between different flare
stages we have computed mean hardness ratios for each of the stellar
groups. In Table~\ref{tab:hrs} we show the mean hardness 
for each activity stage (quiescence, rise, and decay) and each
sample of stars (TTSs, Pleiads, and Hyads).
The hardness changes systematically when 
the three groups are compared to each other:
TTSs display the hardest spectra, followed by Pleiads, 
which in turn are characterized by higher hardness ratios than the
Hyades stars. 
This is also manifest in the hardness plots of Fig.~\ref{fig:hrs} 
where the three samples occupy different regions.
In Sect.~\ref{subsect:lumfunct} it was shown that the flare luminosity
declines with stellar age. As a consequence, the spectral hardness 
and the flare luminosity are correlated. The relation between hardness
ratios and $L_{\rm F}$ is displayed in  
Fig.~\ref{fig:hr_Lx} and suggests that the more luminous flares are
associated with hotter plasma. 

\begin{table}
\begin{center}
\begin{tabular}{lrrr} \hline
        & \multicolumn{3}{c}{$HR\,1$} \\
	& \multicolumn{1}{c}{Quiescence}  & \multicolumn{1}{c}{Rise} & \multicolumn{1}{c}{Decay} \\ \hline      
TTS     & $0.75 \pm 0.38$ & $0.54 \pm 0.47$  & $0.70 \pm 0.44$ \\     
Pleiads & $0.54 \pm 0.26$ & $0.50 \pm 0.29$ & $0.58 \pm 0.30$ \\     
Hyads   & $-0.09 \pm 0.24$ & $-0.16 \pm 0.15$ & $-0.07 \pm 0.16$ \\ \hline    
        & \multicolumn{3}{c}{$HR\,2$} \\
	& \multicolumn{1}{c}{Quiescence}  & \multicolumn{1}{c}{Rise} & \multicolumn{1}{c}{Decay} \\ \hline      
TTS     & $0.36 \pm 0.24$ & $0.38 \pm 0.31$ & $0.39 \pm 0.21$ \\
Pleiads & $0.07 \pm 0.20$ & $0.08 \pm 0.22$ & $0.21 \pm 0.14$ \\
Hyads   & $-0.01 \pm 0.12$ & $-0.04 \pm 0.27$ & $0.14 \pm 0.27$ \\ \hline
\end{tabular}
\caption{Evolution of the {\em ROSAT} PSPC hardness ratios $HR\,1$ and $HR\,2$
during stellar flares. Given are the weighted means of flares on TTSs,
Pleiads, and Hyads measured during three different phases selected from the
X-ray lightcurve: Quiescence (pre- and post-flare), Rise, and Decay. 
A slight increase in hardness is observed between quiescence and rise
phase, but is not significant due to the large spread of the data within
one sample. During the decay the spectrum retains its hardness.
This indicates additional heating
taking place during the decay phase. For each of the flare stages the spectral hardness decreases from TTSs over Pleiads to Hyads.}
\label{tab:hrs}
\end{center}
\end{table}

\section{Discussion}\label{sect:discussion}

\subsection{Methods for flare detection}

Using binned data to detect flares introduces observational 
restrictions. The sensitivity for detection
of small flares is lower and 
very short flares remain unobserved due to the time binning.
Apart from these limitations, our flare detection produces reliable
results as verified by comparison to both an alternative approach 
using Bayesian statistics and, where possible, previous detections of
flares by visual inspection stated in the literature. 

The importance of Bayesian statistics to astronomical time series analysis 
has been described by 
\citey{Scargle98.1} and first applied to {\em ROSAT} observations 
of flare stars by \citey{Hambaryan99.1}. This approach, unlike the 
`classical' method used here, 
works on the raw, unbinned data and therefore has a time resolution
which is only limited by the instrument clock.

We have performed a detailed comparison of the events recognized by 
the two methods.
For the flare detection with the Bayesian algorithm the prior odds ratio, 
$O_{\rm pri}$, was set to 1. This means that at
the beginning one-rate and two-rate Poisson processes are assumed to 
have the same probability for being the correct description of the data set.
The significance of any detection of variability 
is then given by the value of the
posterior odds ratio, $O_{\rm 21}$. 
Applied to our data, 62 events are found at the $5\,\sigma$ level,
and 95 events have $O_{\rm 21}$ corresponding to at least $3\,\sigma$.
All but 5 of the flares discussed in this paper were among the $5\,\sigma$
detections. The remaining ones are detected at $>3\,\sigma$. But note 
that with the Bayesian method we find variability in 182 lightcurves
(in contrast to our 52 flares).

Although the Bayesian approach is sensitive to 
short events, we have persisted on the criteria explained in 
Sect.~\ref{subsect:method_det} for two reasons:
(i) While Bayesian statistics are sensitive to all kinds of temporal
variability, we are here interested in large flare events only. This makes
an additional selection process necessary. (ii) Comparison with the
classical flare search used in this paper has shown that the Bayesian
method needs further refinement. E.g. the outcome of the 
present algorithm used to search for flares depends 
sensitively on the value of the prior odds ratio.

\subsection{Interpretation of the results}\label{subsect:results}

Before flares on different stellar groups are compared, it must be
checked whether the composition of these samples is similar. 
The X-ray luminosity of MS stars depends on their spectral type. 
Therefore it would be desirable to separately 
investigate the flare activity from stars of different spectral types.
However this is hindered by the low flare statistics. 
We have performed statistical tests where the flaring TTSs, Pleiads, 
and Hyads have been compared regarding to their $T_{\rm eff}$ 
and hence spectral types. These tests have
shown that it is justified to jointly analyse flares on all late-type stars,
i.e. stars with spectral type G, K, and M.

We have shown that the relative number of flares increases 
when going from spectral types G to K 
(see Fig.~\ref{fig:detsens_spt}). Hereby, we have taken into
account that the detection sensitivity for
flares depends on the level of measured quiescent emission and hence on the
spectral type. 
An interpretation is that deeper convection zones are
favorable to the occurrence of surface flares.

\subsubsection{Age}\label{subsubsect:compare_age}

We found that
in terms of absolute flare luminosity and energy output TTSs surpass 
both Pleiads and Hyads. The mean flare luminosity of TTSs 
($L_{\rm F,TTS}=1.13~10^{31}\,{\rm erg/s}$) is 
almost an order of magnitude higher than that for Hyads 
($L_{\rm F,Hya}=1.15~10^{30}\,{\rm erg/s}$).
The mean 
Pleiades flare luminosity is intermediate between that for TTSs 
and Hyades stars  
with $L_{\rm F,Ple}=3.26~10^{30}\,{\rm erg/s}$
(see also Fig.~\ref{fig:Lf_Lq} and Table~\ref{tab:L_gkm_mean}).
This is partly due to the different distances of our stellar samples 
which result in different 
detection sensitivities for flares. Note, however, that
this effect can explain only why no events with small $L_{\rm F}$ are
observed on TTSs. But the lack of large events on Hyades stars is real.
In Sect.~\ref{sect:rate} flare rates for TTSs, Pleiads, 
and Hyads have been established 
from an evaluation of the observed flare durations and the total
observing time. 
Both, flare rate and
mean flare luminosity decline with increasing stellar age.

The quiescent luminosity of Hyades stars which showed a flare is larger than
the average $L_{\rm qui}$ of Hyads (see Sect.~\ref{subsect:lumfunct}). 
More than $90$\% of the detected Hyades stars are bright enough 
for detection of an average Hyads flare. 
Therefore, this result is not a selection effect, and we can 
conclude that only the most X-ray luminous Hyades stars exhibit X-ray 
flares.
The interesting question whether the enhanced
X-ray luminosity of flaring Hyades stars can be explained by their rotation
rate can not be pursued with this set of data, because 
only for half of the flaring Hyades stars measurements of $v\,\sin{i}$ are
available.

\subsubsection{Circumstellar Envelope}\label{subsubsect:compare_cw}

If magnetic interactions between star and disk take place, the field lines
will constantly become twisted by differential rotation 
(\cite{Montmerle00.1}). 
This may provide an environment favorable for magnetic reconnection
and related flare activity.

Six of the observed flare events 
can be attributed to cTTSs and 8 events to wTTSs. One of the flares
on TTSs occurred either on DD\,Tau, a cTTS, or on CZ\,Tau, a wTTS, both of
which are not resolved in the {\em ROSAT} PSPC observations. 
Two-sample tests show clear indications that flares
on cTTSs are more X-ray luminous than those on wTTSs 
(see Table~\ref{tab:L_gkm_2s}). 
This holds no matter on which side the ambiguous event is counted.
The flare rate is also slightly 
higher for cTTSs than for wTTSs, however with low significance. 
Given the fact that quiescent
X-ray emission of wTTSs is stronger than in cTTSs, this observation is 
surprising. A possible interpretation is that 
the stronger flare events on cTTSs may be due to violent
interaction with their disks.

\subsubsection{Multiple Flares}\label{subsubsect:multiflares}

During four observations a second flare followed the first one
(see lightcurves of VA\,334, VB\,141, RXJ\,0437.5+1851B, and T\,Tau in 
Figs.~\ref{fig:lcs_TTS}~and~\ref{fig:lcs_Hya}). 
From the number of observed flares and the total observing
time the average duration between two flare events is estimated to be
$> 100\,{\rm h}$. Therefore, from a statistical point of view it is very
unlikely to observe so many unrelated `double events'. We note, that double
flares have been reported in the optical. And 
\citey{Guenther99.1} have presented two flares that occurred 
within a few hours from the wTTS V819\,Tau.

A possible interpretation of multiple flares is the star-disk scenario
proposed by \citey{Montmerle00.1} and mentioned in the previous subsection.
However, this model does not seem to be accurate for our objects, which are
more evolved and in part are known not to possess disks.

\subsubsection{Projected Rotational Velocity}\label{subsubsect:compare_sf}

The statistical tests we have performed to discriminate between slow and
fast rotators (with boundary drawn at 20\,km/s) reveal no dependence of
individual flare parameters $L_{\rm F}$, $L_{\rm qui}$, 
$\tau_{\rm dec}$, and $L_{\rm F}/L_{\rm qui}$ on the rotation rate. However,
the flare frequency is about three times 
higher for fast rotators as compared to slow rotators: 
$F_{\rm slow} = 0.55 \pm 0.10$\% and $F_{\rm fast} = 1.55 \pm 0.38$\%.

\subsubsection {Binary Interactions}\label{subsubsect:compare_sm}
 
We have searched for evidence of binary interaction during X-ray flares
by dividing our sample of flares into 
spectroscopic binaries and all other systems, i.e. wide (or visual)
multiples and single stars, in which such interactions can not take place. 
The comparison of flare rates $F$ showed that large X-ray flares are
significantly more frequent on spectroscopic binaries:
$F_{\rm SB} = 1.43 \pm 0.25$\%
and $F_{\rm non-SB} = 0.64 \pm 0.12$\%.
We have taken account of all components in multiple systems when evaluating
the flare rate. Therefore, the difference in $F$ between close binaries and
other stars seems indeed to indicate that magnetic 
interactions within close binaries leads to increased flare activity.
But note, that interbinary events are expected to have longer durations
because of the larger scale of the magnetic configuration. Our statistical
observations did not show an increase of the time scales for
 spectroscopic binaries.

\subsubsection{Spectral signatures during flares}

From the lower panels of Fig.~\ref{fig:hrs} it can be concluded that
for most of the observed events the spectral hardness has increased during the 
flare. Due to the large uncertainties, however, the changes in the
mean hardness are only marginal. But, note, that the uncertainties
represent the standard deviation (computed by taking into account 
upper/lower limits to the hardness) and thus reproduce the spread in the
data.

We think that
the X-ray emission of TTSs is harder than that of Pleiades and Hyades
stars (see Table~\ref{tab:hrs}) for two reasons: (i) Because of their
circumstellar envelope TTSs suffer from 
much stronger absorption than Pleiads and Hyads, and absorption is stronger
for Pleiads than for Hyads due to the larger distance of the former, (ii)
the younger the stars, the stronger the activity, and therefore the harder
the spectrum.

\section{Conclusions}\label{sect:conclusions}

We have determined flare rates for PMS stars, Pleiades and Hyades
on a large data set and found that all stars are observed during flares
for less than 1\% of the observing time. 
Both frequency and strength of large X-ray flares
decline after the PMS phase.

To probe whether the activity changes in the
presence of a circumstellar disk, e.g. as a result of magnetic interactions
between the star and the disk, we have compared flares on cTTSs and wTTSs. 
We find that flares on cTTSs are stronger and more frequent.

A comparison of flares on spectroscopic binaries to 
flares on all other stars of our sample shows that the flare rate is
by a factor of $\sim 2$ higher for the close binaries. 

The flare rate of fast rotators is enhanced by a factor of $\sim$ 3 as
compared to slowly rotating stars. 

To summarize, our analysis 
confirms that age and rotation influence the magnetic activity of
late-type stars. All previous studies in this field 
have focused on the quiescent X-ray emission. Now, 
for the first time the rotation-activity-age connection has been 
examined for X-ray flares. 
Furthermore, from the sample of flares investigated here we find evidence 
that magnetic activity goes beyond solar-type coronal activity: On
young stars interactions between the star and a circumstellar disk
or the magnetic fields of close binary stars may play a role.

\begin{acknowledgements}
We made use of the Open Cluster Database, compiled by 
C.F. Prosser and J.R. Stauffer. We thank S. Wolk and W. Brandner
for useful discussions and an anonymous referee for valuable comments. RN acknowledges grants from the Deutsche
Forschungsgemeinschaft (Schwerpunktprogramm `Physics of star formation').
The {\em ROSAT} project is supported by the Max-Planck-Gesellschaft and
Germany's federal government (BMBF/DLR).
\end{acknowledgements}

\end{document}